%% file: 1a1744.tex
\documentclass[twocolumn]{aastex63}
\usepackage{amsmath}
\usepackage{booktabs}
\usepackage{comment}
\usepackage{xspace}
\usepackage{enumitem}
\usepackage{hyperref}
\usepackage{chronosys}
\usepackage{longtable}
\usepackage{latexsym}
\usepackage{amssymb}
\usepackage{longtable}
\usepackage{graphicx}
\usepackage{epsf}
\usepackage{url}

\received{}
\revised{}
\accepted{}

\shorttitle{\src X-ray and Radio Monitoring}
\shortauthors{Ng et al.}

\newcommand{\src}{1A~1744$-$361\xspace}

\graphicspath{{./}{figures/}}

\begin{document}

\title{X-ray and Radio Monitoring of the Neutron Star Low Mass X-ray Binary \src: Quasi Periodic Oscillations, Transient Ejections, and a Disk Atmosphere}

\correspondingauthor{Mason Ng}
\email{masonng@mit.edu}

\author[0000-0002-0940-6563]{Mason Ng}
\affiliation{MIT Kavli Institute for Astrophysics and Space Research, Massachusetts Institute of Technology, Cambridge, MA 02139, USA}

\author[0000-0003-0764-0687]{Andrew K. Hughes}
\affiliation{Department of Physics, University of Alberta, CCIS 4-181, Edmonton, AB T6G 2E1, Canada}

\author[0000-0001-8371-2713]{Jeroen Homan}
\affiliation{Eureka Scientific, Inc., 2452 Delmer Street, Oakland, CA 94602, USA}

\author[0000-0003-2869-7682]{Jon M. Miller}
\affiliation{Department of Astronomy, University of Michigan, 1085 South University Avenue, Ann Arbor, MI 48109, USA}

\author[0000-0002-8403-0041]{Sean N. Pike}
\affiliation{Department of Astronomy and Astrophysics, University of California, San Diego, CA 92093, USA}

\author[0000-0002-3422-0074]{Diego Altamirano}\affiliation{Physics \& Astronomy, University of Southampton, Southampton, Hampshire SO17 1BJ, UK}

\author[0000-0002-7252-0991]{Peter Bult} \affiliation{National Institute for Public Health and the Environment, P.O. Box 1, 3720 BA, Bilthoven, The Netherlands}

\author[0000-0001-8804-8946]{Deepto Chakrabarty}
\affiliation{MIT Kavli Institute for Astrophysics and Space Research, Massachusetts Institute of Technology, Cambridge, MA 02139, USA}

\author[0000-0002-5341-6929]{D. J. K. Buisson}\affiliation{Independent Researcher}

\author[0000-0003-0870-6465]{Benjamin M. Coughenour}\affiliation{Department of Physics, Utah Valley University, 800 W. University Parkway, MS 179, Orem, UT 84058, USA}

\author[0000-0002-5654-2744]{Rob Fender}\affiliation{Department of Physics, University of Oxford, Denys Wilkinson Building, Keble Road, Oxford OX1 3RH, UK}

\author[0000-0002-6449-106X]{Sebastien Guillot}\affiliation{Institut de Recherche en Astrophysique et Planétologie, UPS-OMP, CNRS, CNES, 9 avenue du Colonel Roche, BP 44346, F-31028 Toulouse Cedex 4, France}

\author[0000-0002-3531-9842]{Tolga G\"uver}
\affiliation{Istanbul University, Science Faculty, Department of Astronomy and Space Sciences, Beyaz\i t, 34119, \.Istanbul, T\"urkiye}
\affiliation{Istanbul University Observatory Research and Application Center, Istanbul University 34119, \.Istanbul, T\"urkiye}

\author[0000-0002-6789-2723]{Gaurava K. Jaisawal}\affiliation{DTU Space, Technical University of Denmark, Elektrovej 327-328, DK-2800 Lyngby, Denmark}

\author[0000-0002-3850-6651]{Amruta D. Jaodand}\affiliation{Cahill Center for Astronomy and Astrophysics, California Institute of Technology, 1200 E California Boulevard, Pasadena, CA 91125, USA}

\author[0000-0002-0380-0041]{Christian Malacaria} \affiliation{International Space Science Institute, Hallerstrasse 6, 3012 Bern, Switzerland}

\author[0000-0003-3124-2814]{James C. A. Miller-Jones}\affiliation{International Centre for Radio Astronomy Research - Curtin University, Perth, Western Australia
6845, Australia}

\author[0000-0002-0118-2649]{Andrea Sanna}\affiliation{Dipartimento di Fisica, Università degli Studi di Cagliari, SP Monserrato-Sestu km 0.7, I-09042 Monserrato, Italy}

\author[0000-0001-6682-916X]{Gregory R. Sivakoff}\affiliation{Department of Physics, University of Alberta, CCIS 4-181, Edmonton, AB T6G 2E1, Canada}

\author[0000-0001-7681-5845]{Tod E. Strohmayer}\affiliation{Astrophysics Science Division, NASA Goddard Space Flight Center, Greenbelt, MD 20771, USA}

\author[0000-0001-5506-9855]{John A. Tomsick}\affiliation{Space Sciences Laboratory, 7 Gauss Way, University of California, Berkeley, CA 94720-7450, USA}

\author[0000-0002-5686-0611]{Jakob van den Eijnden}\affiliation{Department of Physics, University of Warwick, Coventry CV4 7AL, UK}

\begin{abstract}

We report on X-ray (NICER/NuSTAR/MAXI/Swift) and radio (MeerKAT) timing and spectroscopic analysis from a three-month monitoring campaign in 2022 of a high-intensity outburst of the dipping neutron star low-mass X-ray binary \src.~The 0.5--6.8~keV NICER X-ray hardness-intensity and color-color diagrams of the observations throughout the outburst suggests that \src spent most of its outburst in an atoll-state, but we show that the source exhibited Z-state-like properties at the peak of the outburst, similar to a small sample of other atoll-state sources. A timing analysis with NICER data revealed several instances of an $\approx8$ Hz quasi-periodic oscillation (QPO; fractional rms amplitudes of $\sim5\%$) around the peak of the outburst, the first from this source, which we connect to the normal branch QPOs (NBOs) seen in the Z-state. Our observations of \src are fully consistent with the idea of the mass accretion rate being the main distinguishing parameter between atoll- and Z-states. Radio monitoring data by MeerKAT suggests that the source was at its radio-brightest during the outburst peak, and that the source transitioned from the `island' spectral state to the `banana' state within $\sim3$ days of the outburst onset, launching transient jet ejecta. The observations present the strongest evidence for radio flaring, including jet ejecta, during the island-to-banana spectral state transition at low accretion rates (atoll-state). The source also exhibited Fe XXV, Fe XXVI K$\alpha$, and K$\beta$ X-ray absorption lines, whose origins likely lie in an accretion disk atmosphere.

\end{abstract}


a\section{Introduction} \label{sec:intro}

\subsection{Neutron Star Low-Mass X-ray Binaries} \label{sec:nslmxbs}

Low-mass X-ray binaries (LMXBs) are binary systems comprising a compact object (i.e., a black hole or neutron star) accreting matter from a low-mass companion star \citep[$M_c < 1 M_\odot$; see][for a review]{bahramian22}. Neutron star (NS) LMXBs have traditionally been broken down into two classes based on their correlated spectral and timing properties and on the patterns they trace in the X-ray color-color diagram (CCD): atoll sources trace out atoll-shaped regions in the CCD, tend to have lower luminosities ($\sim0.001-0.5\,L_{\rm Edd}$), and often exhibit sporadic transient outbursts followed by extended periods of quiescence. Z sources trace a `Z' shape in the CCD and tend to be highly luminous and persistent, accreting at close to the Eddington limit for a $1.4M_\odot$ NS \citep[$>0.5\,L_{\rm Edd}$;][]{hasinger89,vdk95,belloni02,vdk04}.

Atoll sources have been observed in multiple spectral states, namely the `island' and `banana branch' states, historically defined by their positions in the CCD \citep{hasinger89,wijnands17}. Typically, the island states have low accretion rates, and therefore low X-ray luminosities. Furthermore, the power spectrum is characterized by strong band-limited noise ($\gtrsim10-20\%$ in fractional rms amplitude). At high(er) accretion rates, an atoll source occupies the banana state, and the power spectrum shows a weak ($\sim3\%$) red noise component \citep{wijnands99a}. The island X-ray spectra are hard (i.e., dominated by high-energy X-ray photons), and the spectra usually include a significant nonthermal component often parameterized by a power law (or broken power law). In contrast, the banana state X-ray spectra are dominated by soft, thermal X-ray photons from a blackbody (of the NS surface), multi-colored disk blackbody \citep[from the accretion disk;][]{lin07}, or the boundary layer between the NS surface and accretion disk. Some atolls have not been observed to show state transitions, existing either only in the island state (e.g., 4U~1812$-$12: \citealt{tarana06,gladstone07}; XTE~J1814$-$338: \citealt{vanstraaten05}) or banana state (e.g., GX~9$+$9: \citealt{hasinger89,kong06,fridriksson11}).

The X-ray spectral states are correlated with the properties of the relativistic jets fueled by the accretion flow \citep[e.g.,][]{migliari06,munoz14}. In the island state most dominated by hard X-ray photons \cite[i.e., the `extreme island' state; see][]{mendez97,vdk06,wijnands17}, the accretion flow fuels a continuous, compact jet that exhibits a flat or inverted radio spectral index ($F_\nu\propto\nu^\alpha$; $\alpha\gtrsim0$) up to a break frequency (often at sub-mm wavelengths) where the radio emission is optically thin \citep[$\alpha\,{\sim}\,-0.7$; e.g.,][]{migliari2010,russell2013,trigo2018}. In some atolls, the compact jet is quenched in the banana states, becoming undetectable after the transition \citep[e.g., Aql~X$-$1;][]{millerjones2010}. This behavior is analogous to the jet quenching observed in all black hole (BH) LMXBs during the transition from the hard to soft state \citep[e.g.,][]{coriat2011} and motivated the typical comparisons between the extreme island and banana states with the hard and soft states of BH LMXBs. However, some atolls have exhibited banana state radio detections \citep[e.g., 4U~1820$-$30, Ser~X$-$1;][]{migliari2004}. Many BH LMXBs exhibit bright radio flaring during a hard to soft state transition \citep{fender2006}. In contrast to compact jet emission, radio flares show temporally evolving spectra that rapidly transition to optically thin (i.e., steep) spectral indices at radio frequencies. The flaring is thought to result from ejection events, and thus are de-coupled from the accretion flow and do not follow the $L_R$--$L_X$ relation. Although a similar ejection process has been suggested for atolls \citep{migliari06}, there are no observations of bright flaring or spatially resolved jet ejecta during the `island-to-banana' transition. Therefore, there may be a weak quenching mechanism or possibly a different jet launching process in atoll NS LMXBs \citep[see,][for a summary of the differences between jet quenching in NS and BH LMXBs]{vandeneijnden2021}. 

For Z sources, the CCD spectral states alternate between the `horizontal branch' (HB; top of the Z), the `normal branch' (NB; the diagonal), and the `flaring branch' (FB; bottom). The X-ray spectra of Z sources tend to be softer than atoll sources \citep{muno02a}. Like some atolls, Z sources experience quenching of the compact jet in the FB and flaring during NB--HB transitions \citep{migliari06}. The Z source radio flaring is significantly more common than observed in atolls, thought to be the result of frequent transitions between the NB and HB. These flares have been spatially resolved as discrete ejection events in a number of Z source LMXBs (e.g., Cyg~X$-$2: \citealt{spencer2013}; Sco~X$-$1: \citealt{fomalont01, motta19}). 

\subsection{Quasi-Periodic Oscillations} \label{sec:qpos}

Quasi-periodic oscillations (QPOs) are broad, ephemeral features in the power spectra of NS and BH LMXBs usually described with Lorentzians \citep{vdk00}. QPOs are observed throughout the Z-track in NS LMXBs, where the HB QPOs (HBOs) are roughly between 15--60~Hz, the NB QPOs (NBOs) are between 5--7~Hz, and the FB QPOs (FBOs) are between 7--20~Hz \citep[e.g.,][]{middleditch86,vdk89,motta17}. The HBOs are thought to be linked to the disk truncation radius \citep[e.g.,][]{stella98,ingram10,motta19}, while NBOs are thought to be connected to subsonic density (radial) oscillations in a thick disk \citep{fortner89,alpar92} or relativistic and transient ejections of plasma \citep[e.g.,][]{fomalont01,migliari06}.

Atoll sources show a variety of QPO phenomena, including mHz QPOs \citep[e.g.,][]{revnivtsev01,heger07,mancuso23}, $\sim1{\rm\,Hz}$ QPOs seen in some atolls exhibiting absorption dips \citep[e.g.][]{jonker99,homan99}, and hecto-Hz QPOs \citep[e.g.,][]{ford98,altamirano08,bult15}.
A class of atoll sources also show NBO-like QPOs near the highest source luminosities; they include XTE~J1806$-$246, the ultracompact X-ray binary 4U~1820--30, and the transient source Aql~X$-$1 \citep{wijnands99a,wijnands99b,reig04}. These QPOs only lasted for hundreds of seconds, and were found in the tip of the upper banana branch in the CCD; they resembled the FBOs and NBOs seen in Z sources when the source accretion rate was close to the Eddington limit \citep{vdk89}. Both atolls and Z sources also exhibit kHz QPOs \citep[e.g.,][]{vdk96,motta17}. The power spectra of the NS and BH LMXBs are additionally characterized by several broadband and higher frequency noise components \citep{vdk00}.

\subsection{Z and Atoll States}

Some NS LMXBs have transitioned between atoll and Z-like behavior at high mass accretion rates \citep{lin09}.~Detailed X-ray timing and spectral studies with the archetypal transitional Z/atoll source, XTE~J1701$-$462, suggested that the transitions between the two spectral states are driven by changes in the mass accretion rate \citep{homan07b,homan07c,lin09,homan10,fridriksson15}. The transient NS LMXB IGR~J17480$-$2446 had also been observed to show transitions between Z-source behavior and atoll source behavior \citep{altamirano10,chakraborty11}. While the transitions themselves were not directly observed, Cir~X$-$1 \citep{oosterbroek95,shirey98,shirey99} also showed Z-source-like properties at the higher luminosities ($>0.5\,L_{\rm Edd}$) and atoll source-like properties at lower luminosities ($\sim0.001-0.5\,L_{\rm Edd}$). Finally, the NBO-like QPOs observed from atolls XTE~J1806$-$246, 4U~1820$-$30, and Aql~X$-$1 \citep[mentioned above;][]{wijnands99a,wijnands99b,reig04} at the highest source luminosities suggest a luminosity threshold at which the NBO production mechanism is activated. \emph{In this work, we will discuss our results in the framework of \cite{lin09} and \cite{homan10}, i.e., that atoll sources and Z sources are not distinct subpopulations of NS LMXBs but, instead, represent two states on a continuum defined by the mass accretion rate (atoll-state at low luminosities, Z-state at high luminosities). We will refer to them as atoll-state sources and Z-state sources from here on.}

\subsection{Geometric Properties of NS LMXBs} \label{sec:nslmxb_geometry}

Some LMXBs show absorption dips in their X-ray light curves (e.g., 4U~1624$-$490: \citealt{smale01}; 4U~1254$-$690: \citealt{smale02}). Given the expected viewing geometry of these phenomena, the binary inclination angle of non-eclipsing dippers is expected to be $i\approx60-75^\circ$ \citep{frank87}. The absorption dips are thought to be due to intervening cold and/or partially ionized absorbers, around the interaction point where the accretion stream hits the accretion disk, along the line-of-sight \citep{diaztrigo06}, or the partial covering by an accretion disk corona from an extended absorber \citep{church98,church05,balucinskachurch11}. These absorption dips are observed to be energy-dependent, and are accompanied by an increase in spectral hardness due to scattering of softer photons by the gas cloud. Some sources exhibit both eclipses and absorption dips (e.g., Swift J1858.6$-$0814:  \citealt{buisson21}; EXO~0748$-$676: \citealt{parmar86}), and so the binary inclination angle of such systems is expected to be $i\approx75-80^\circ$ \citep{frank87}.



\subsection{\src} \label{sec:source}

\cite{tobrej23} recently reported on joint NICER/NuSTAR spectroscopic analysis of \src, showing that a phenomenological model comprising a power law with a high energy cutoff described the spectrum well, and hinted at existence of a reflection component. They also showed prominent absorption features at $6.92$~keV and $7.98$~keV, which they attributed to H-like iron and a blend of Fe XXV and Ni XXVII lines, respectively. \cite{mondal23} also recently reported spectral analysis with NuSTAR data which found that the observations showed relativistic disk reflection and similar absorption features. The spectral analysis in both work identified the source to be in the banana branch state.

In this work, we present the X-ray (NICER, NuSTAR, MAXI, and Swift) and radio (MeerKAT) monitoring campaign of the dipping NS LMXB \src, which was first discovered by the Ariel 5 satellite in 1976 due to an outburst \citep{davison76,carpenter77}. \src has exhibited several known outbursts between 1989 and the most recent outburst in 2022. Observations of the high-intensity 2003 outburst reported in \cite{bhattacharyya06a} revealed two sets of energy-dependent absorption dips in two successive RXTE orbits, which led the authors to infer an orbital period of $P_{\rm orb} = 97 \pm 22$ min, though they did not rule out periods twice or half as long. Observations of the low-intensity 2004 outburst revealed the presence of a significant ($4.3\sigma$) $\nu\approx3.5$ Hz QPO with a quality factor $Q\approx2$ and a fractional rms amplitude of $\approx5\%$ \citep{bhattacharyya06a}. They also reported a low significance ($2.3\sigma$) lower kHz QPO with $\nu\approx800$ Hz, high $Q$ value of $Q\approx62.5$, and a fractional rms amplitude of $\sim6\%$ \citep{bhattacharyya06a}. The QPOs were connected to the island state \citep{bhattacharyya06b}, similar to 4U~1746$-$37 and EXO~0748$-$676 \citep{homan99,jonker00,homan12}, while the energy-dependent absorption dips were only observed in the high-intensity outbursts \citep{bhattacharyya06b}. \cite{bhattacharyya06a} have also reported the NS in \src to have a spin frequency of $\nu\approx530$ Hz, based on the discovery of thermonuclear burst oscillations in a type I X-ray burst seen in the high-intensity 2005 outburst.

\src went into outburst in June 2022 for about three months in its longest known outburst yet. Radio observations taken by MeerKAT detected bright radio emission during the early stages of the outburst \citep{hughes22}. In this paper, we present X-ray and radio observations of \src taken over the span of almost three months to characterize the (aperiodic and periodic) timing and spectral evolution of the source. In \S \ref{sec:observations}, we outline the observations and data analysis procedures. In \S \ref{sec:results}, we present the results from the timing and spectral analyses. We will discuss the results in \S \ref{sec:discussion} and finally conclude in \S \ref{sec:conclusion}. All uncertainties reported represent $1\sigma$ confidence intervals unless otherwise noted.

\section{Observations and Data Analysis} \label{sec:observations}

\subsection{Neutron star Interior Composition Explorer (NICER)}

NICER, an external payload on the International Space Station, consists of 52 operating co-aligned X-ray concentrator optics and silicon drift detectors in focal plane modules (FPMs). NICER has fast-timing capabilities in the 0.2--12.0~keV energy range, allowing for a GPS time-tagging accuracy of 100 ns \citep{gendreau16,lamarr16,prigozhin16}. 

NICER observed \src starting from 2022 June 3 through 2022 August 31, with ObsIDs starting with 5202 and 5406. We processed the NICER observations with \textsc{HEASoft} version 6.31.1 and the NICER Data Analysis Software (\textsc{NICERDAS}) version 10a (2022-12-16\_V010a) with calibration version \texttt{xti20221001}. Our data processing criteria included the following: a source angular offset of $\texttt{ANG\_DIST}<0\fdg015$; elevation angle from the Earth limb $\texttt{ELV}>20^\circ$; NICER being outside the South Atlantic anomaly; bright Earth limb angle $\texttt{BR\_EARTH}>30^\circ$; undershoot rate (dark current; per FPM) range of $\texttt{underonly\_range}=$~0--500; overshoot rate (charged particle saturation; per FPM) range of $\texttt{overonly\_range}=$~0--1.5. We also applied $\texttt{COR\_SAX}$ (magnetic cut-off rigidity in GeV/c) filtering of $\texttt{COR\_SAX}>1.5$ to filter out background flares. This resulted in 53.5 ks of filtered exposure for scientific analysis out of 95.3 ks of unfiltered exposure. 

For the timing analysis, we applied Solar system barycenter corrections in the ICRS reference frame (so that times are in Barycentric Dynamical Time; TDB), with source coordinates ${\rm R.A.}=267\fdg0548, {\rm\, Decl.}=-36\fdg13251$, using \texttt{barycorr} in FTOOLS with the JPL DE421 solar system ephemeris \citep{folkner09}. The spectral analysis was conducted with \texttt{XSPEC} 12.13.0c \citep{arnaud96}. The spectra were first grouped with the optimal binning scheme and rebinned such that each bin had at least 25 counts \citep{kaastra16}. Background spectra were constructed with the nibackgen3C50 model \citep{remillard22} available through \texttt{nicerl3-spect}. The associated response matrices were generated with \texttt{nicerarf} and \texttt{nicerrmf} for each observation.

\subsection{Nuclear Spectroscopic Telescope Array (NuSTAR)}

NuSTAR is the first hard X-ray focusing telescope and consists of two co-aligned grazing incidence telescopes \citep{harrison13}. NuSTAR observed \src between 2022 June 8 19:36 UTC (MJD 59738.81747; TT) and 2022 June 9 18:11 UTC (MJD 59739.75854; TT) for a total filtered exposure time of approximately 34.3 ks. The NuSTAR data reduction was performed with the standard pipeline tools (\texttt{nupipeline} and \texttt{nuproducts}) in NuSTARDAS v.1.9.7 and CALDB 20221229. We generated the final source and background event files with 100\arcsec\ regions around the source, and in a source-free region near the source, respectively. 

Initial analysis of \src revealed relativistic reflection and strong emission and absorption features from the $\sim6.96$~keV and $\sim8.0$~keV Fe XXVI K$\alpha$ and K$\beta$ lines \citep{pike22}. In this work, we will focus on characterizing the timing properties of the source from the NuSTAR observation and defer detailed analysis of the spectroscopic properties to another publication (Pike et al. 2024, in prep.)

\subsection{MeerKAT Radio Telescope} \label{sec:data_meerkat}

The MeerKAT radio telescope is an interferometric array located in South Africa. We observed \src with MeerKAT as a part of the large survey project ThunderKAT \citep{2016mks..confE..13F}. Our observing began with a single rapid response observation on 2022 May 31 (MJD 59730), $\sim$ 2~days after the first MAXI/GSC X-ray detection \citep[and $\sim8$~hours after the outburst's initial reporting;][]{2022ATel15407....1K}. Following this rapid response, we began a monitoring campaign on 2022 June 3 (MJD 59733), observing the source every $\sim$ 7~days until 2022 August 27 (MJD 59818). Each observation consisted of a single scan with 15 minutes on-source flanked by two 2-minute scans of a nearby gain calibrator (J1830$-$3602). Each epoch also included a 5-minute scan of PKS B1934$-$638 (J1939$-$6342), for flux and bandpass calibration. Our observations used MeerKAT's L-band receiver, with a central frequency of 1.28~GHz and a total (un-flagged) bandwidth of 856~MHz. 

We performed flagging, calibration, and imaging using a modified version of the semi-automated routine \textsc{OxKAT}\footnote{\url{https://github.com/IanHeywood/oxkat}}  \citep{2020ascl.soft09003H}, originally developed for the MeerKAT International GHz Tiered Extragalactic Exploration (MIGHTEE) Survey \citep[see][for a detailed description of the routine]{heywood22}. The first step (1GC) uses \textsc{casa} \citep[v5.6;][]{casa22} to remove radio frequency interference (RFI) and apply the standard flux density, bandpass, and complex gain calibrations. The second step (FLAG) applies a second round of flagging using \textsc{tricolour} \citep{hugo22}, followed by preliminary imaging of the source field using \textsc{wsclean} \citep[v2.9;][]{offringa14}. The preliminary image is then used to create an imaging mask. To maximize imaging sensitivity, we modified \textsc{OxKAT}, changing the Briggs' robustness weight \citep{briggs95} from -0.3 (the default) to 0. We cannot increase the robustness further (i.e., ${>}\,0$) as the MeerKAT synthesized beam becomes significantly non-Gaussian. The final step (2GC) begins with a masked deconvolution before using the model image for direction-independent (DI) self-calibration with \textsc{CubiCal} \citep{kenyon18}. We then performed a second round of masked deconvolution using the DI self-calibrated visibilities. Given the high image fidelity of the 2GC images, we did not perform any direction-dependent self-calibration and adopted the 2GC images as our final data products. 

We measured the radio properties of the source in each image using the \textsc{casa} task \texttt{imfit}. The task fits an elliptical Gaussian component in a small sub-region around the source, measuring the position and flux density. As the source was unresolved, we fixed the component shape to be the synthesized beam of each image. We measured the uncertainty on the flux measurement using the local root-mean-square (rms) noise. For each epoch, we extracted the rms from an annular region using the \textsc{casa} task \texttt{imstat}. Each annulus was centered on the position of the Gaussian component. The inner radius was fixed at the major axis of the synthesized beam, and we scaled the outer radius such that the annular area was equal to 100 synthesized beams. 

\subsection{Swift X-ray Telescope}
As a part of our ThunderKAT monitoring, for each radio epoch, we acquired quasi-simultaneous (within ${\sim}\,$days) X-ray observations using the X-ray telescope aboard the Neil Gehrels Swift Observatory \citep[henceforth Swift/XRT;][]{gehrels04,burrows05}. The weekly-cadence monitoring began on 2022 June 1 (MJD 59731) and continued until 2022 August 8 (MJD 59798), totaling 10 individual epochs (target ID: 31222). We increased the cadence of our Swift/XRT monitoring (one observation every $\leq$ 3~days) after the source's X-ray flux began decaying. We acquired an additional 12 epochs of this `high-cadence' monitoring that began on 2022 August 10 (MJD 59801) and ended on 2022 August 24 (MJD 59815). There was no further Swift/XRT monitoring. 

	\begin{figure*}[htbp]
		\includegraphics[width=1\linewidth]{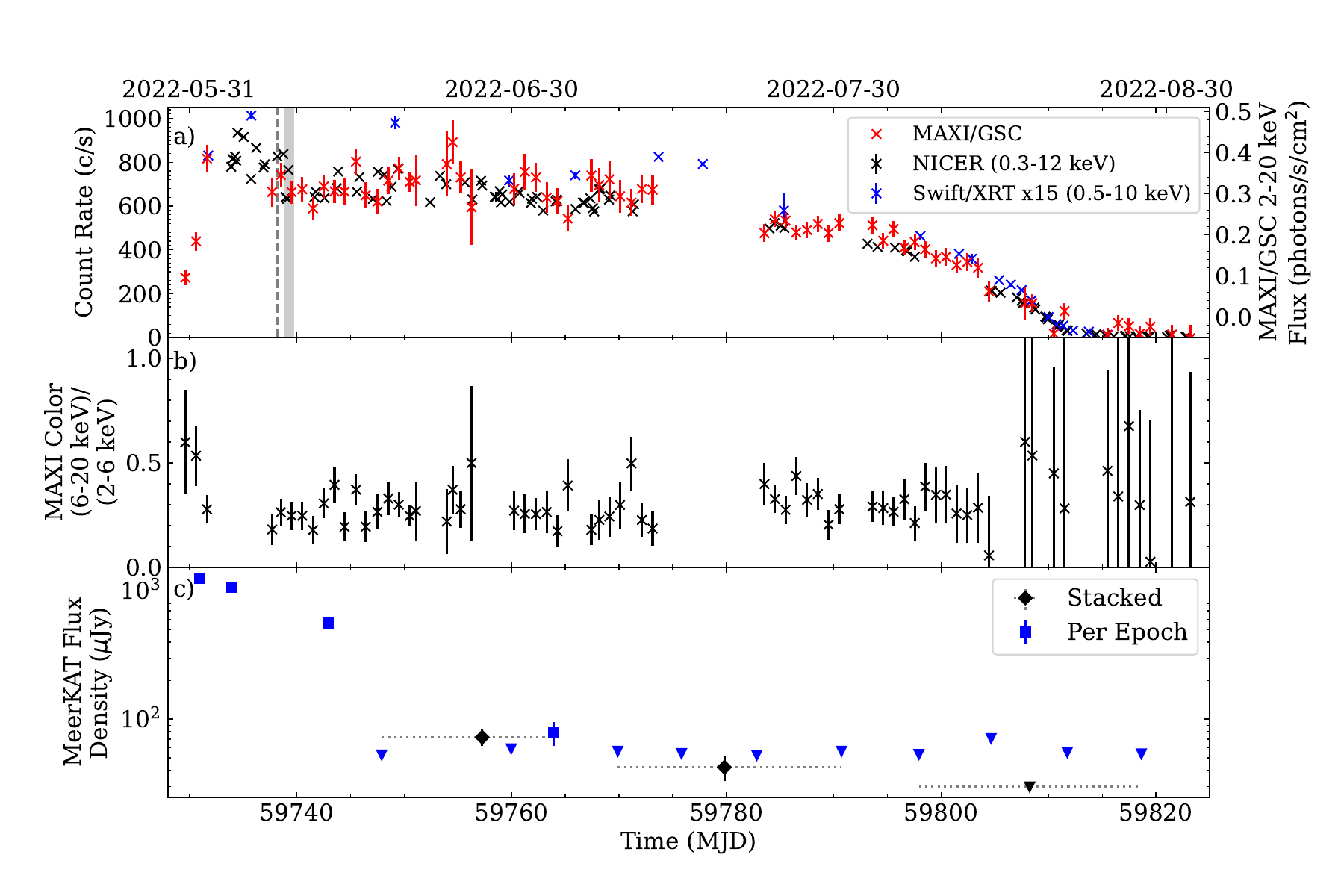}
		\caption{a) Count rate evolution of the rise and decay of the outburst of \src from NICER (black; 10 ks bins; 0.3--12.0~keV), Swift/XRT (blue; rescaled by a factor of 15; 0.5--10~keV), and MAXI/GSC (red; fluxes in 24~h bins; 2--20~keV). MAXI/GSC data were extracted from the MAXI On-demand service \citep{matsuoka09}. The vertical dotted line corresponds to the absorption dip (started on MJD 59738.19985, lasted for $\sim 100$ s) that is shown in Figure~\ref{fig:simple_dip_lc}. The shaded gray region corresponds to the NuSTAR observation interval. b) Temporal evolution of the X-ray color as defined by the ratio of 6--20~keV and 2--6~keV fluxes from MAXI/GSC; c) 1.3 GHz flux densities ($\mu$Jy) from MeerKAT. Blue squares and filled triangles are the flux density per observation and the upper limits, respectively. Black diamonds denote stacked observations  to increase the S/N of the detection (the corresponding observation epochs included are indicated by the dotted lines). The last data point for stacked observations (black filled triangle) is the 3$\sigma$ upper limit at 30~$\mu$Jy. The values for the flux densities are given in Table \ref{tab:meerkat_flux} in the Appendix.}
		\label{fig:all_lc} 
	\end{figure*}

We extracted 0.5--10~keV light curves using the Python API version of the Swift/XRT pipeline, \texttt{swifttools} \citep{evans07,evans09}, extracting a single data point for each observation, shown in panel a of Figure~\ref{fig:all_lc}. Although our X-ray spectroscopy results presented in this work largely derive from the extensive and more sensitive NICER coverage, we also extracted the spectrum from the first Swift/XRT observation on 2022 June 1 (MJD 59731) with the online Swift/XRT product generator, since this observation was two days prior to the first NICER observation \citep{evans09}.

\section{Results} \label{sec:results}

\begin{figure*}[t]
		\centering
		\includegraphics[width=\linewidth]{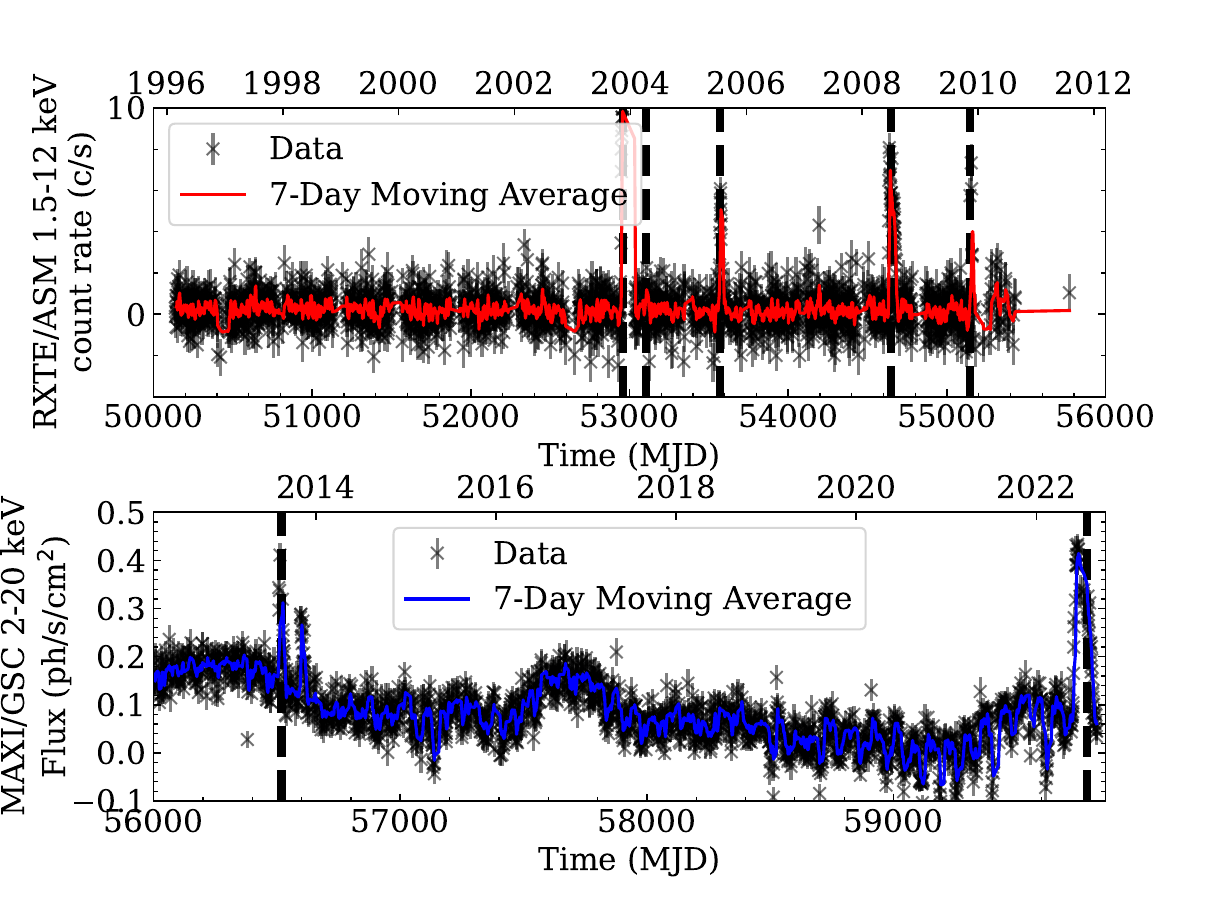}
		\caption{Long-term monitoring of \src with RXTE/ASM (top; 1.5--12~keV count rate) and MAXI/GSC (bottom; 2--20~keV flux); black crosses and solid lines show the data points and the 7-day moving average, respectively. The secondary top x-axes for each panel displays the date in calendar years, with each tick corresponding to January 1 of the labeled year. It is worth noting that there is also another X-ray source (4U~1746$-$37) in the same MAXI/GSC field of view ($0\fdg9$ away) as \src. The black dashed vertical lines indicate rough dates of the outbursts of \src that have been reported in the literature - they correspond to MJDs 52960 (2003-11-17), 53104 (2004-09-04), 53570 (2005-07-07), 54647 (2008-06-30), 55145 (2009-11-10), and 56507 (2013-08-03).}
		\label{fig:monitoringlc}
	\end{figure*}

We first show the evolution of the 2022 outburst of \src in panel~a) of Figure~\ref{fig:all_lc} with NICER, Swift/XRT coverage, and MAXI/GSC monitoring. The NuSTAR observation interval is represented by the gray shaded region. For context, we also show light curves from long-term monitoring of the field by RXTE/ASM and MAXI/GSC in Figure~\ref{fig:monitoringlc}. The 2022 outburst of \src had the longest known outburst duration from the source, and had a similar brightness level to the 2003 outburst.


	\begin{figure*}[t]
		\centering
		\includegraphics[width=1.1\linewidth]{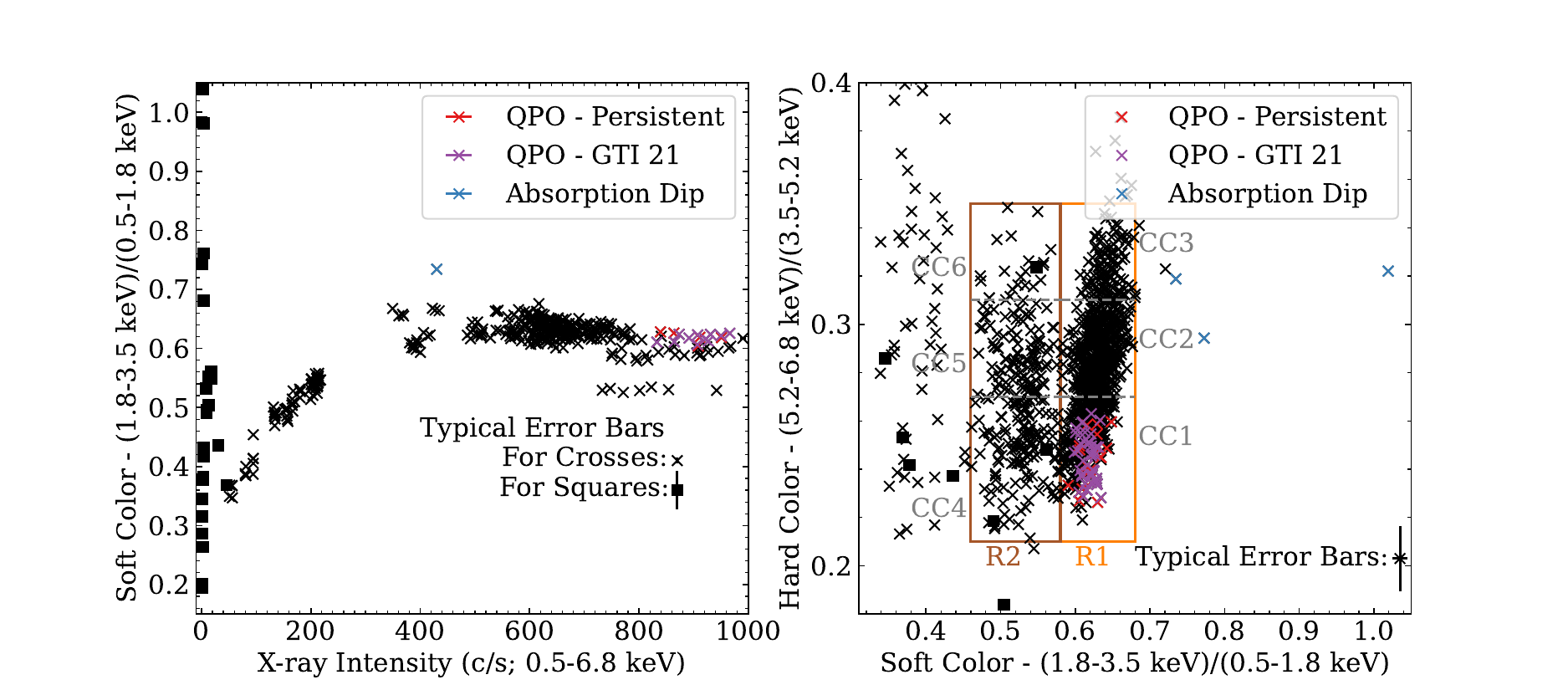}
		\caption{Left: Hardness-intensity diagram of \src throughout the outburst, which is typical of the atoll-state. Crosses correspond to 128~s bins for the data, and squares correspond to 256~s bins (for bins with count rates $\leq 50$~c/s, after MJD~$\sim59811$). The red crosses correspond to the time interval containing the persistent emission post-absorption dip; the blue cross corresponds to the absorption dip; purple crosses correspond to the GTI that showed the second instance of the QPO (see text for details). Right: Color-color diagram of \src throughout the outburst. We display a typical error bar for the 32~s bins; we note the typical error bar for the 256~s bins was 6 times larger on average (not shown). Crosses correspond to 32~s bins for the data, and squares correspond to 256~s bins (for bins with count rates $\leq 50$~c/s, after MJD~$\sim59811$). We also show two regions (denoted R1 and R2) in the CCD for which we conducted pulsation searches (see \S\ref{sec:coherent}); we also broke up the two spectral regions into six smaller regions (CC1 to CC6) for further power spectral analysis. Background subtraction has been taken into account for all data points.}
		\label{fig:hid_ccd}
	\end{figure*}

To gain an understanding of the spectral variability of \src, using NICER data, we constructed the HID which is shown in Figure~\ref{fig:hid_ccd}, where the soft color was defined as the ratio of background-subtracted count rates in the 1.8--3.5~keV and 0.5--1.8~keV bands, and the X-ray intensity is represented by background-subtracted count rates in the 0.5--6.8~keV band. We also generated the CCD, shown in Figure \ref{fig:hid_ccd}. We used the same definition for the soft color, and for the hard color we had the ratio of background-subtracted count rates in the 5.2--6.8~keV and 3.5--5.2~keV bands. In order to increase the signal-to-noise (S/N) in each data point, we increased the bin size to 256~s (from 128~s) after the broadband (0.3--12.0~keV) count rate fell below 50 c/s (around MJD 59811). The background-subtracted light curves were generated with \texttt{nicerl3-lc} using the `spaceweather' (sw) background model. Figure~\ref{fig:hid_ccd} clearly shows the source to be in the atoll-state throughout the outburst \citep{hasinger89,vdk04}. 

	\begin{figure}[t]
		\centering
		\includegraphics[width=1.1\linewidth]{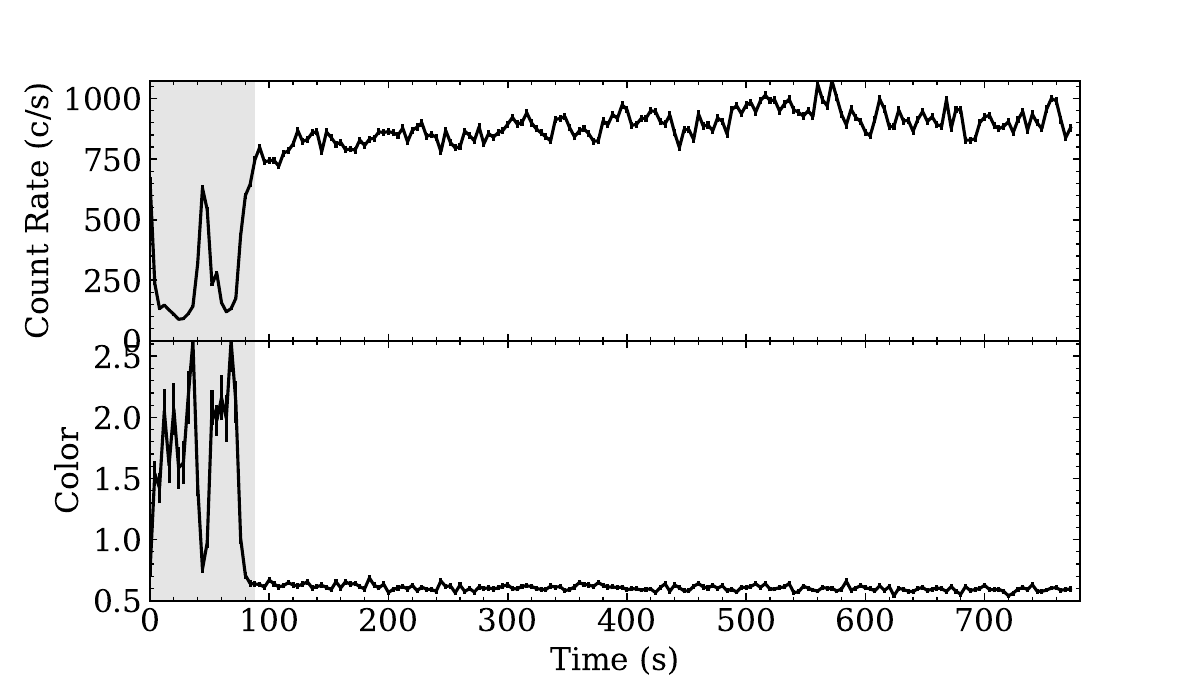}
		\caption{Top: 0.3--10.0~keV 4 s-binned NICER light curve corresponding to the absorption dip ($t=0$ corresponds to MJD 59738.19431 (TT)). Bottom: The color is defined as the ratio of 2--10~keV and 0.3--2~keV background-subtracted count rates. The gray shaded region corresponds to the absorption dip; the rest of the GTI is referred to as the persistent emission post-absorption dip. The absorption dip corresponds to an increased spectral color/hardness.}
		\label{fig:simple_dip_lc}
	\end{figure}

In the ObsID 5202800106, we found a $\sim100$ s absorption dip, starting at MJD~59738.19431 (TT), in the light curve. The corresponding light curve and spectral color evolution from the absorption dip and the persistent emission immediately after are shown in Figure \ref{fig:simple_dip_lc}, where the absorption dip is accompanied by an increase in the color, presumably due to the scattering of softer photons by an intervening gas cloud. The time intervals of the persistent emission post-absorption dip are indicated by the red crosses in the HID (Figure~\ref{fig:hid_ccd}).

We also note that to increase the S/N of the observations at the tail end of the outburst, we attempted to combine several observations. The ObsID 5202800160 (see Table~\ref{tab:obsprops}) was generated by combining (using the \texttt{ftmerge} tool) all housekeeping, auxiliary, and event files from ObsIDs 5202800149 and 5202800150. We were not able to combine subsequent observations as either the combined data exhibited a very large spread of colors (e.g., because too many observations were combined), or the background dominated the source spectrum well below 10~keV. Thus, we ended up not using data from ObsIDs 5202800151 to 5202800157 (2022 August 25 to 2022 August 31) in our final analysis.

Finally, to track the outburst rise, we calculated the temporal evolution of the X-ray color from MAXI/GSC data (first three data points in panel b of Figure~\ref{fig:all_lc}), and corroborated the evolution with the Swift/XRT spectrum of the first observation on 2022 June 1. We used the Windowed Timing observation (since \src is bright) which had 1.4~ks of exposure. The 0.5--10.0~keV spectrum was well described by an absorbed cutoff power law with $n_H = 0.44\times10^{22}{\rm\,cm^{-2}}$ (see \S\ref{sec:continuum_evol}), power law index $\Gamma_{\rm co} = 0.61_{-0.04}^{+0.04}$, e-folding energy $E_f = 3.21_{-0.16}^{+0.18}{\rm\,keV}$, and normalization $0.479_{-0.007}^{+0.007}{\rm\,photons\,s^{-1}\,cm^{-2}\,keV^{-1}}$. The 0.6--10.0~keV absorbed flux was approximately $F\approx2.6\times10^{-9}{\rm\,erg\,s^{-1}\,cm^{-2}}$ (uncertainties were reported at 90\% confidence for this spectral fit).

\subsection{X-ray Timing} \label{sec:timing}

\subsubsection{Aperiodic Analysis} \label{sec:aperiodic} 

In order to look for QPOs, we first created an averaged Leahy-normalized power spectrum for each individual GTI longer than 50 s, with 32~s segments. We made use of the \texttt{AveragedPowerspectrum} class in \texttt{Stingray} to perform the calculation \citep{huppenkothen19a,huppenkothen19b}, where we also logarithmically binned the averaged power spectra with $f=1.05$. Almost all of the power spectra were well-fit with a power law plus a constant; we show the results in Table \ref{tab:obsprops} in the Appendix. However, the power spectra corresponding to GTIs 12 and 21 (see Table~\ref{tab:obsprops}), observations that contained the absorption dip (see Figure~\ref{fig:simple_dip_lc}) and an observation one day after, respectively, exhibited a residual feature at around $\nu\simeq8$ Hz. 

For GTI 12 (in ObsID 5202800106), we investigated the origins of the residual feature by isolating the dip emission and the persistent emission immediately after the dip (within the GTI). We found that the residual feature was seen only in the persistent emission and not in the absorption dip, though with a $2\sigma$ upper limit of $6.8\%$, so we cannot rule out the QPO being absent in the dip emission. Thus in what follows, we focused the analysis on the persistent emission immediately following the dip. We proceeded to model the residual by adding a Lorentzian on top of the initial continuum model (power law plus constant) and found a significant improvement. To quantify the improvement of the additional model component, we considered the difference in the Akaike information criterion \citep[AIC;][]{akaike74}, with $\Delta {\rm AIC}_{12} = {\rm AIC}_1 - {\rm AIC}_2$, where 

\begin{equation} \label{eq:aic}
    {\rm AIC} = -2\text{ln}\mathcal{L} + 2d,
\end{equation}

\noindent $\mathcal{L}$ is the maximum likelihood of any given model ($\mathcal{L}\propto e^{-\chi^2/2}$), $\chi^2$ is the chi-squared of the fit, and $d$ is the number of degrees of freedom \citep{liddle07,tan12,arcodia18}. To evaluate the two models, we can define a rejection probability at a confidence level defined by 

\begin{equation} \label{eq:prob}
    P_1/P_2 \approx e^{-\Delta{\rm AIC}_{12}/2}.
\end{equation}

 	\begin{figure*}[htbp!]
		\centering
		\includegraphics[width=1.1\linewidth]{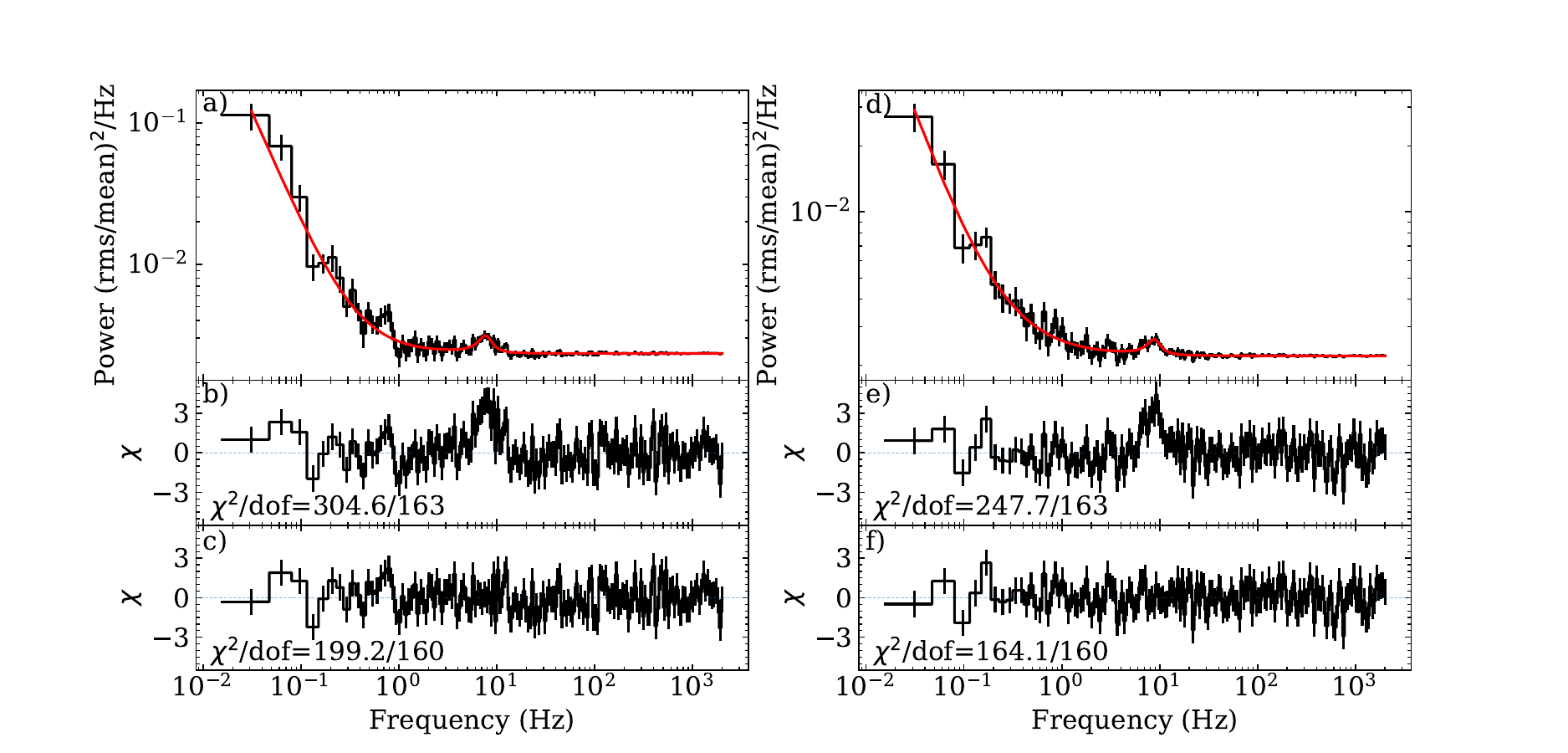}
		\caption{a) Power spectrum corresponding to the persistent emission post absorption-dip. The best-fit model, comprising a power law (and a constant for the Poisson level) and a Lorentzian for the QPO, is in red. b) Residuals [$\chi=$ (data-model)/error] when fitting the data to only a power law (plus constant) - the reduced $\chi^2$ value is given by $\chi^2$/d.o.f. = $304.6/163$. c) Residuals when the Lorentzian is included in the fit - the reduced $\chi^2$ value is given by $\chi^2$/d.o.f. = $199.2/160$. d), e), and f) are similar plots, but they correspond to GTI 21 (in ObsID 5202800107), the segment of data showing the second QPO instance.}
		\label{fig:qpo_ps_final}
	\end{figure*}

In Figure~\ref{fig:qpo_ps_final}, we show the fractional rms-squared-normalized power spectra, the residuals when fitting with an initial power law plus constant continuum, and the residuals when including an additional Lorentzian component. We found that the $\chi^2$ values decreased from $\chi^2=304.6$ to $\chi^2=199.2$ (from 163 to 160 d.o.f.), thus $\Delta {\rm AIC}_{12} = 111.4$, and so $P_1/P_2\approx6.5\times10^{-25}$. We found the Lorentzian (QPO) parameters to be $\nu_0 = 7.64\pm0.19$~Hz with ${\rm FWHM}=3.0\pm0.6$~Hz, which implied a quality factor, $Q = \nu_0/{\rm FWHM} = 2.5\pm0.5$. The fractional rms amplitude of the QPO (from the normalization of the Lorentzian) was found to be $f_{\rm rms,1} = 6.0\pm0.4\%$ over 0.3--12.0~keV. 

We have found an additional instance of the QPO in one GTI of an observation about one day later, in GTI 21 (in ObsID 5202800107). The QPO had Lorentzian parameters $\nu_0 = 8.75\pm0.25$ Hz with ${\rm FWHM} = 3.4\pm0.7$ Hz, which implied $Q = 2.6\pm0.5$ ($\chi^2=247.7$ to $\chi^2=164.1$; $P_1/P_2 \approx 3.5\times10^{-20}$). The fractional rms amplitude of the QPO was $f_{\rm rms,2} = 4.6\pm0.4\%$ over 0.3--12.0~keV. We also show the corresponding power spectrum and residuals in the right panels of Figure~\ref{fig:qpo_ps_final}. 

When we visually inspected the power spectral fits for all GTIs, we noticed a weak ($\sim2\sigma$) indication of the QPO in GTI 6 (in ObsID 5202800103). To investigate this further, we ran \texttt{nicerclean} on the unfiltered event file (ufa file) and generated the cleaned event file. We found another instance of the QPO in the GTI with the highest count rates (MJDs 59735.31189 to 59735.31792 (TDB)), where we needed a power law and an additional Lorentzian ($\chi^2=257.9$ to $\chi^2=148.2; P_1/P_2 \approx 7.5\times10^{-26}$) to describe the power spectrum. The Lorentzian (QPO) parameters were $\nu_0 = 7.34\pm0.14{\rm\,Hz}, {\rm FWHM} = 2.3\pm0.4{\rm\,Hz}$, thus $Q = 3.1\pm0.6$. Finally, the fractional rms amplitude was $f_{\rm rms,3} = 6.2\pm0.4\%$ over 0.3--12.0~keV (note that we did not plot the power spectrum given the non-standard filtering).

We also looked for energy dependence of the QPO, where we fixed the centroid frequency and FWHM, and subdivided the data into energy bands of 0.3--1.0, 1.0--2.0, 2.0--3.0, 3.0--4.0, and 4.0--12.0~keV, tabulating the results in Table~\ref{tab:qpo_Edep}. We did not find evidence for the energy dependence of the fractional rms amplitude for the QPO, which is consistent with the behavior of 1 Hz QPOs from dipping/eclipsing LMXBs \citep{jonker99,homan12} but not with sources that show NBO-like oscillations \citep{wijnands99a,wijnands99b}. However, we note that the energy coverage and sensitivities differ between RXTE and NICER (and uncertainties here were large). 

We characterized the power spectra from the different spectral regions in the CCD (based on the soft color) as shown in Figure~\ref{fig:hid_ccd}, summarizing these results in Table~\ref{tab:cc_timing}. In order to understand the variability within each spectral region, we further subdivided each region into three spectral subregions based on the hard color and repeated the power spectral characterization (see Table~\ref{tab:cc_timing}). The 0.03--50 Hz variability (characterized by the fractional rms amplitude) in R1 turned out to be primarily due to the spectral subregion occupied by the QPO in the lower region of the soft state; the variability decreased going up the soft state (R1) from $7.3\pm1.4\%$ to $2.0\pm1.5\%$. For R2, there were no discernible differences between the spectral subregions within uncertainties. It is worth noting that the source moved erratically through the CCD throughout the outburst, so we cannot ascertain any smooth temporal evolution of the power spectrum. 

\begin{figure}[htbp!]
		\centering
		\includegraphics[width=1.1\linewidth]{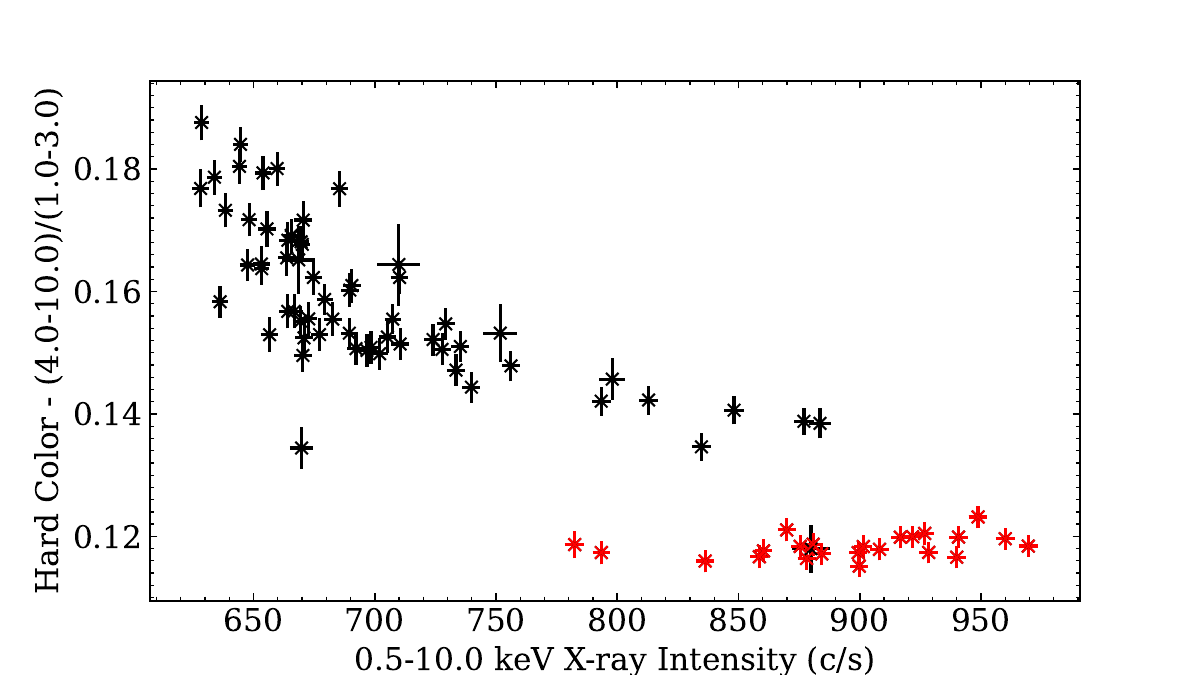}
		\caption{HID with data points from NICER ObsID 5202800107, where the third QPO instance appeared. The hardness is defined as the hard color, with the count rate ratio of 4--10~keV over 1--3~keV; the intensity is the NICER 0.5--10.0~keV count rate. The red and black points are data points corresponding to GTIs that do and do not exhibit the QPO, respectively. The HID shows two separate tracks, which only Z-state sources have shown before on short timescales, but given the limited data, we cannot confidently connect these branches to the Z-state tracks.}
		\label{fig:hid_z}
	\end{figure}

We have also plotted the HID using data points from ObsID 5202800107, with an observation span of five hours, in Figure~\ref{fig:hid_z}. The HID shows two separate tracks, and only Z-state sources are known to trace out multi-branched tracks on such short timescales (e.g., GX~13$+$1: \citealt{schnerr03,homan04,fridriksson15}; Cyg~X$-$2: \citealt{fridriksson15}). However, given the limited data, we caution over-interpretation of these tracks as directly related to that of the branches in the Z-state. 

\input{spectralregions_timing}
\input{qpo_Edep.tex}


\subsubsection{Pulsation Searches} \label{sec:coherent}

Given the discovery of coherent burst oscillations at $\sim530$ Hz from \src \citep{bhattacharyya06a}, the pulsation frequency must be around the same frequency \citep{chakrabarty03,watts12}. Thus we conducted a coherent pulsation search with two independent techniques --- stacked power spectra and acceleration searches (only for NICER data).

For the stacked power spectra, we first identified search regions in the CCD (see Figure~\ref{fig:hid_ccd}), as the appearance of pulsations could be associated with different spectral states of the source \citep{vaughan94}. For each region (R1 and R2), the corresponding time windows of the 32-s bins were collated and joined, with new GTI files generated for each region. We then used \texttt{ftselect} to extract the events given the filtered GTI file. The final averaged power spectra were calculated using a segment size of 256~s and a bin size of $\Delta t = 2^{-11}{\rm\,s}$ using the \texttt{AveragedPowerspectrum} routine in \texttt{Stingray} \citep{huppenkothen19a,huppenkothen19b}.

\begin{figure*}[t]
		\centering
		\includegraphics[width=\linewidth]{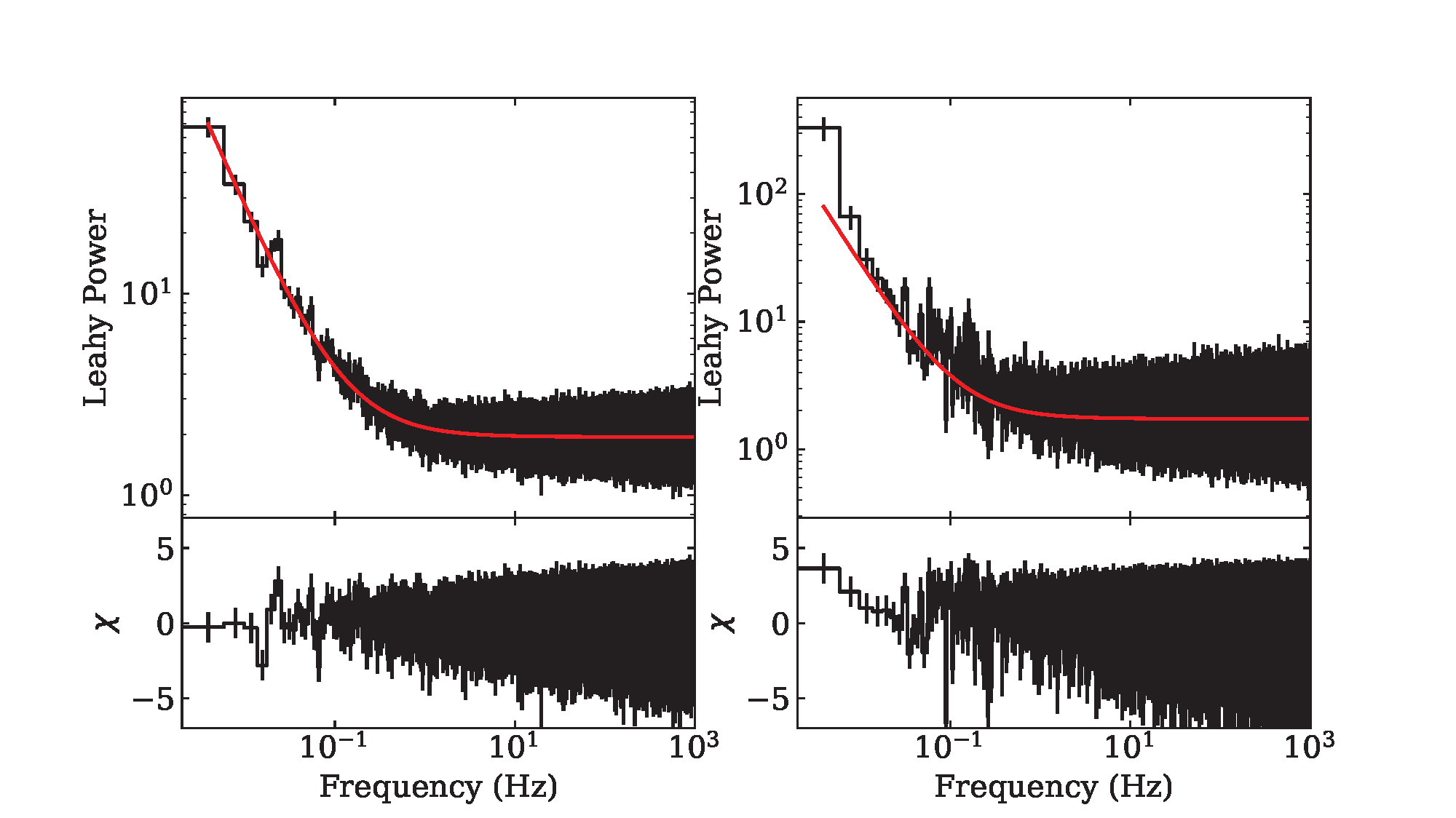}
		\caption{Averaged power spectra (256~s segments) using the events in R1 (left) and R2 (right), where $m=76$ and $N_\gamma = 12897259$, and $m=23$ and $N_\gamma = 1542304$, respectively. The red solid lines in the top panel are the best-fit power law (plus constant) model. The panels below show the residuals after subtracting the best-fit model from the data.}
		\label{fig:r1r2}
	\end{figure*}

The averaged power spectrum from each of the two search regions are shown in Figure~\ref{fig:r1r2}. We did not find any significant coherent periodicity in any of the search regions, but we can place upper limits on the fractional rms amplitude. Since we did not know the precise spin frequency of the pulsar, and given the relatively low significance ($\sim4\sigma$) of the burst oscillation \citep{bhattacharyya06a}, we calculated the $3\sigma$ upper limits on the fractional rms amplitude over 1--1000 Hz. We found that for R1 and R2, the upper limits were roughly $0.04\%$ and $0.13\%$, respectively. We also found that for the 3--25~keV NuSTAR data (using FPMA), the $3\sigma$ upper limit was $0.10\%$. We derived these upper limits by generating 1000 realizations of the event lists and GTIs corresponding to each of the regions (for NuSTAR, we used the entire FPMA event list). Across each frequency bin, we generate a cumulative distribution function (CDF) of the simulated powers, and took the power value where the CDF was equal to 0.9973 ($3\sigma$ upper limit), and calculated the corresponding fractional rms amplitude \citep{vaughan94,muno02b}. Operationally, the simulations were carried out with \texttt{simulate\_times} in \texttt{Stingray} \citep{huppenkothen19a,huppenkothen19b}. 

We used acceleration searches to probe for coherent periodicity while accounting for the Doppler modulation of the pulsation frequency due to an orbit \citep{ransom02}. Acceleration searches are most sensitive with segments (with length $T$) where $T \lesssim P_{\rm orb}/10$, such that the pulsar acceleration is roughly constant and so the frequency evolution is linear within the time segment \citep{ransom02}. We carried out the acceleration searches on the individual GTIs throughout the outburst of \src, with a minimum of $100{\rm\,s}$ in length (to have enough photons). A total of 108 GTIs were searched and the median segment length was $442{\rm\,s}$, with a standard deviation of $299{\rm\,s}$. We searched over the frequency range 525--535~Hz, energy ranges 0.3--2.0~keV and 2.0--10.0~keV. We found no significant coherent periodicity candidates from the acceleration searches.


\subsubsection{NuSTAR} 

\begin{figure}[htbp!]
        \centering
        \includegraphics[width=1.15\linewidth]{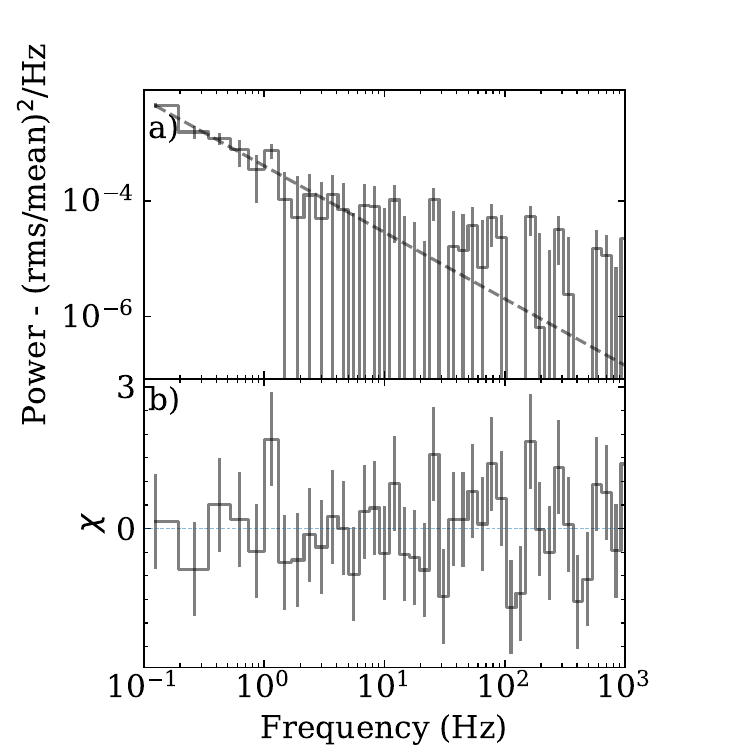}
        \caption{a) The 3--25~keV (temporal) cospectrum for the entire NuSTAR observation (constructed from 8~s segments); b) the residuals after fitting the observed cospectrum with a power law with index $1.15\pm0.10$.}
        \label{fig:nustar_ps}
    \end{figure}

We calculated the averaged (temporal) cospectrum from the entire NuSTAR observation with 8~s segments and fit a simple power law to the observed cospectrum; the cospectrum and the corresponding residuals are shown in Figure~\ref{fig:nustar_ps}. The cospectrum is well-described by a power law, $f(\nu) = A\nu^{-\alpha}$, with amplitude $A=(4.1\pm0.7)\times10^{-4}$, spectral index $\alpha=1.15\pm0.10$, with the reduced $\chi^2$ being 0.74 (69 d.o.f.) The 0.1--50 Hz fractional rms amplitude was $2.7\pm2.4\%$. The cospectra were calculated using \texttt{Stingray} with data from the two independent FPMA and FPMB detectors, such that any periodic (or quasi-periodic) signals in phase between the two detectors can be detected and any unrelated variability is eliminated \citep[e.g., dead time effects;][]{huppenkothen18,huppenkothen19a,huppenkothen19b}. The cospectrum is the real part of the cross power density spectrum \citep{bendat11}.

\begin{figure}[htbp!]
        \centering
        \includegraphics[width=1.15\linewidth]{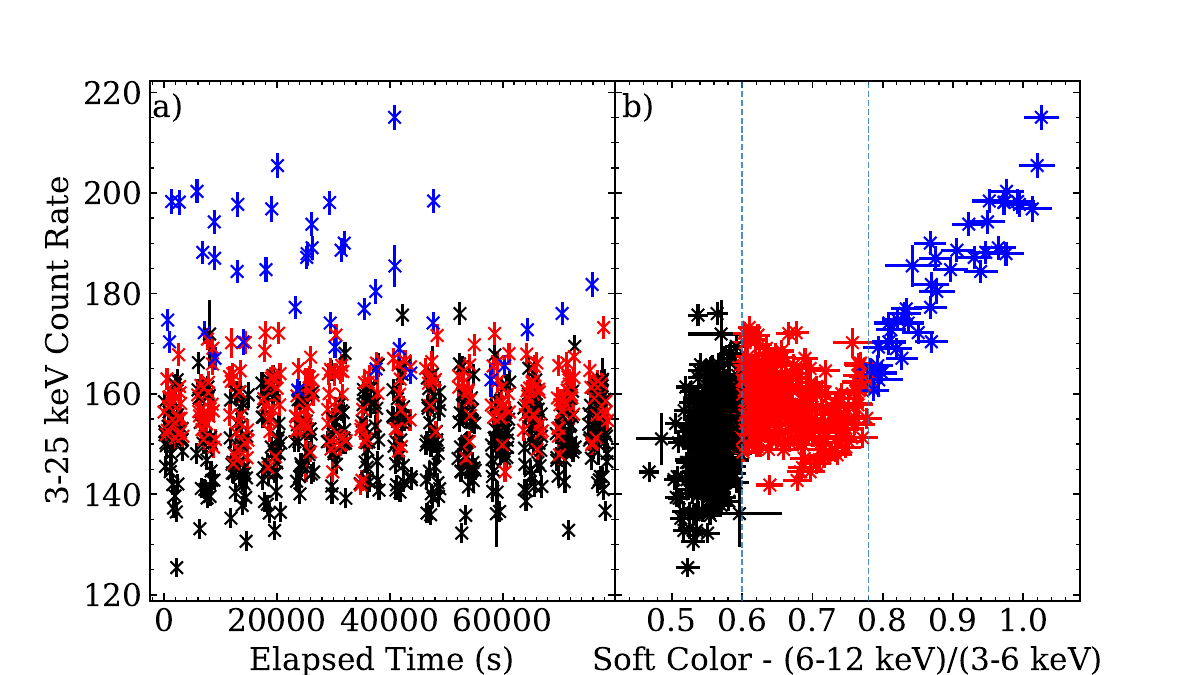}
        \caption{a) 3--25~keV NuSTAR light curve (combined FPMA and FPMB); b) 3--25~keV count rate vs. soft color, defined as the ratio of counts in the 6--12~keV and 3--6~keV bands. All data points are constructed from 64~s bins. The black, red, and blue data points corresponding to different spectral branches for which we investigated the cospectral properties - see Table~\ref{tab:nustar_cospec} for the best fit parameters to the (temporal) cospectra.}
        \label{fig:nustar_lc}
    \end{figure}

To further investigate the cospectral properties of \src, we first constructed the HID, which is shown in Figure~\ref{fig:nustar_lc}. The HID (on the right) clearly shows three distinct spectral states which have been color-coded, and the corresponding points are shown in the light curve (on the left). In particular, the blue points corresponding to flaring activity in the light curve (Figure~\ref{fig:nustar_lc}a) show up as a Z-state-like flaring branch in the HID (Figure~\ref{fig:nustar_lc}b). For each of the three spectral regions, we calculated the averaged cospectrum (also 8~s segments) and we found that the power law amplitudes and spectral indices were roughly consistent with each other. There is weak evidence of a slightly shallower power law at lower count rates of the source (lower soft colors). We did not see evidence of any QPOs in all of the cospectra. The best fit parameters are summarized in Table~\ref{tab:nustar_cospec}.

\input{nustar_cospec}

\subsection{X-ray Spectroscopy} \label{sec:spectroscopy}

\subsubsection{Evolution of the Continuum Components} \label{sec:continuum_evol}

The broadband 0.6--10.0~keV (see below for the choice of energy range) spectra were generally well-fit by several continuum models that were a function of the spectral states (R1, R2) as identified in Figure~\ref{fig:hid_ccd}. We found that adopting a single spectral model for the entire outburst did not suffice. In particular, the source transitioned from R1 to R2 between MJDs 59797 and 59804 (NICER data gap). Thus we performed spectral fits within the following framework -- for GTIs 0 to 93 (MJDs 59733 to 59797, inclusive), we adopted an underlying continuum consisting of a disk blackbody (for the accretion disk) and a thermal blackbody (possibly from the boundary layer at the NS surface). For subsequent observations, we performed spectral fitting on an individual ObsID basis to maximize the S/N. We tried a combination of a thermal blackbody (for the NS surface) and either a power law or a cutoff power law for the Comptonization component. We found that the cutoff power law worked best ($E_{\rm cut}\approx7-11{\rm\,keV}$) between ObsIDs 5202800140 and 5202800144 (MJDs 59804 to 59808), when the source was in R2, and the high energy cutoff after that was being pegged at 200~keV by the spectral fitting process, which indicated that a simple power law sufficed. Thus from ObsID 5202800145 onwards (MJD 59809), we employed a thermal blackbody and a power law. We note that we ultimately did not use Comptonization models (e.g., \texttt{thcomp}) as NICER did not have sufficient high energy coverage to provide meaningful constraints. We also tried including a partially ionizing absorber component \texttt{zxipcf} to the fit, but the data lacked constraining power. The fitting strategy described above was similarly adopted for NS LMXBs Aql~X$-$1 and 4U~1608$-$52 \citep{lin07}.

\begin{figure*}[t]    
        \setlength{\belowcaptionskip}{-50pt}
        \setlength{\abovecaptionskip}{-35pt}
        \setlength{\intextsep}{5pt}    
        \vspace{0cm}
        \centering
        \includegraphics[width=0.9\linewidth]{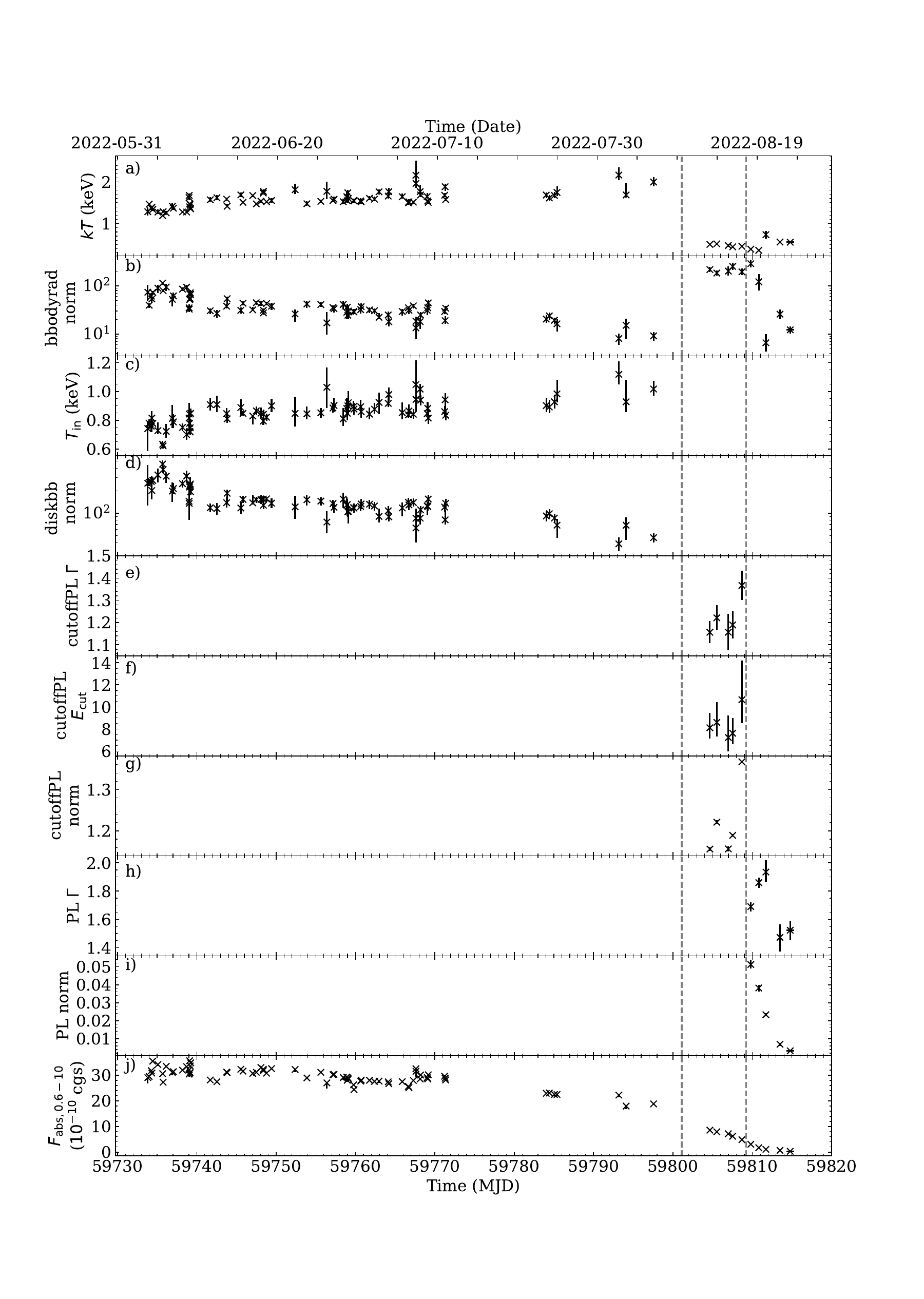}
        \vspace{-\belowcaptionskip}
        \caption{0.6--10.0~keV spectral fit with \texttt{XSPEC} of individual NICER GTIs and observations (see text for details) throughout the outburst. The uncertainties reported here are $90\%$ confidence limits. Panels a)-d) show results from R1 with the model \texttt{tbabs(diskbb+bbodyrad+gauss+gauss)}; the results of the Gaussian lines are reported elsewhere in the text (see \S\ref{sec:spectrallines} and Table~\ref{tab:lines_eqw}). Panels~e)-g) show results from spectral fits in R2, with the model \texttt{tbabs(bbodyrad+cutoffpl)}. The results corresponding to R2 are in between the two vertical dashed lines. Panels h) and i) show results from fitting the rest of the spectra (after the second vertical dashed line; after MJD 59809) with \texttt{tbabs(bbodyrad+powerlaw)}. The power law photon index and normalization are defined similarly to the cutoff power law. Finally, panel~j) corresponds to the 0.6--10.0~keV absorbed flux in units of $10^{-10}{\rm\,erg\,s^{-1}\,cm^{-2}}$. The spectroscopic fits to the NICER data show that increasing blackbody (\texttt{bbodyrad}) and disk blackbody (\texttt{diskbb}) temperatures with decreasing blackbody normalizations (both components) in R1 while the outburst flux decayed. There were no identifiable trends in R2 and beyond given the small number of data points.}
        \label{fig:spectroscopy}
    \end{figure*}

The full results from the spectroscopic analysis are presented in Figure~\ref{fig:spectroscopy}. For spectral region R1, we fit the spectra with an absorbed disk blackbody model and a thermal blackbody component. Some of the observations in R1 exhibited clear residuals around 1.0~keV (an emission feature) and around 7.0~keV (an absorption feature), so we added two Gaussian components (see \S\ref{sec:spectrallines}). We restricted the energy range of the $\sim1$~keV ($\sim7$~keV) component to be 0.85--1.15~keV (6.9--7.1~keV), and the line width to be between 0.001--0.2~keV. For the $\sim7$~keV absorption line, we allowed the normalization of the Gaussian to be negative. In \texttt{XSPEC} parlance, the full model was expressed as \texttt{tbabs(diskbb+bbodyrad+gauss+gauss)}. We tried fitting the absorption line with the multiplicative \texttt{gabs} model but we recovered very high (unphysical) values of the line depth.

In all spectral fits, the line of sight hydrogen column density was fixed at $n_H = 0.44\times10^{22}{\rm\,cm^{-2}}$, roughly consistent with the HI4PI value of $0.26\times10^{22}{\rm\,cm^{-2}}$ \citep{hi4pi16}, which was determined through the following. We initially performed the spectral fit (with the absorbed disk blackbody and thermal blackbody model) to all observations on an individual GTI basis by letting all spectral parameters be free. The value of $n_H$ typically ranged between (0.39--0.45)$\times10^{22}{\rm\,cm^{-2}}$ with typical errors of (0.01--0.03)$\times10^{22}{\rm\,cm^{-2}}$, so we calculated a weighted average of the $n_H$ values, which resulted in the derived value above. Hereafter, we fixed $n_H$ and repeated the spectral fitting. The spectral fitting was restricted to 0.6--10.0~keV as the column density absorption dominated below 0.6~keV, and the background dominated above 10~keV.

Returning to the results, we see that in R1, there is a strong anticorrelation between the blackbody temperature and normalization (\texttt{bbodyrad} in \texttt{XSPEC}), as well as between the temperature at the inner (accretion) disk radius and the disk blackbody normalization (\texttt{diskbb} in \texttt{XSPEC}). In this period, both temperatures were rising (and normalizations decreasing) while the overall 0.6--10.0~keV (absorbed) flux was gradually decreasing. In R2, where the absorbed blackbody and cutoff power law model was used, the flux was rapidly decreasing, while the blackbody temperature and associated normalization were roughly constant. There was a slight increase in the power law index and the normalization. Finally, after R2, fit with an absorbed blackbody and power law model, the flux was also decreasing but there was no monotonic evolution in the spectral parameters. We note that the thermal blackbody values changed dramatically across R1 to R2; however, the continua employed in both states were different so we caution any physical interpretation here.


\subsubsection{Evolution of the Spectral Lines} \label{sec:spectrallines}

\begin{figure}[htbp!]
        \centering
        \includegraphics[width=0.95\linewidth]{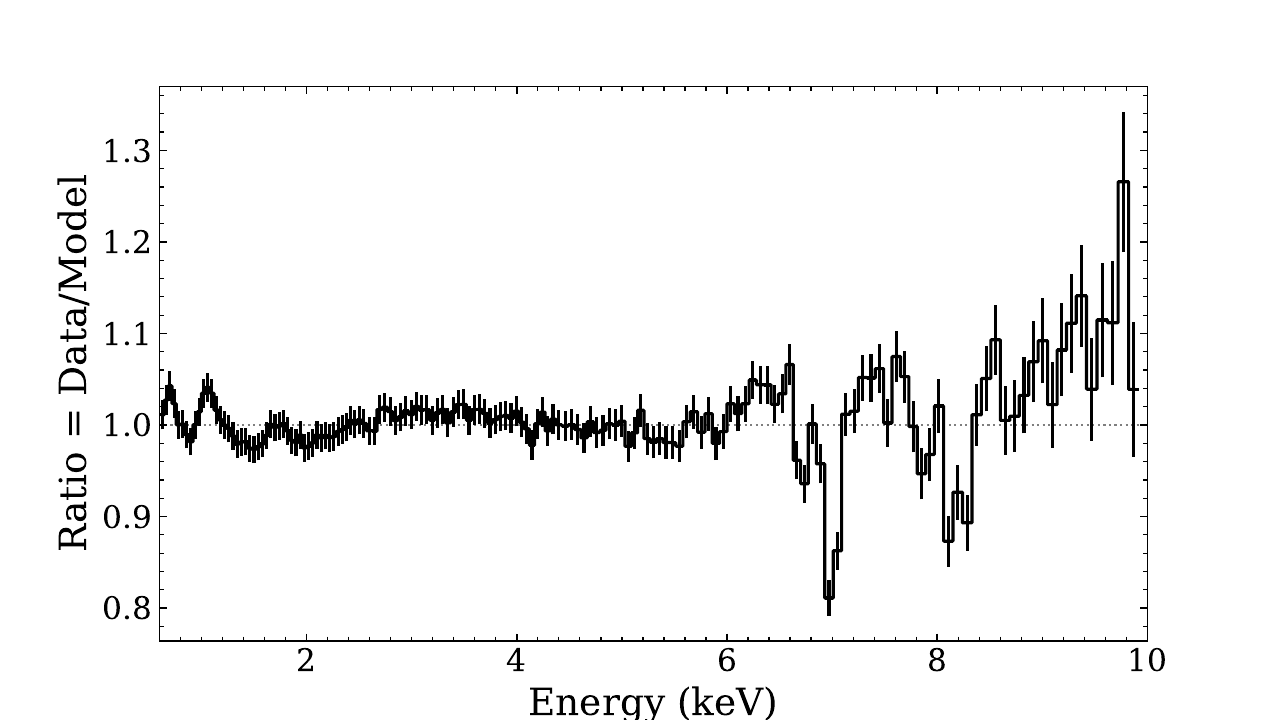}
        \caption{Ratio of the observed spectrum to the underlying model (\texttt{tbabs(diskbb+bbodyrad)}) as a function of energy for the spectral region CC1. Strong negative residuals, identified with Fe XXV, Fe K$\alpha$, and Fe K$\beta$ absorption features, are seen as unmodeled features in the ratio plot.}
        \label{fig:cc1_abs}
    \end{figure}

As mentioned in \S\ref{sec:continuum_evol}, some of the observations in R1 exhibited clear residuals around 1.0~keV and 7.0~keV. To understand the spectral lines in more detail, we combined all of the observations that appeared in R1 and R2, as well as in CC1-CC6. As an example, we show the ratio of the data to the baseline model (absorbed disk blackbody and thermal blackbody) in Figure~\ref{fig:cc1_abs} for spectral region CC1 to illustrate the strong absorption features. For the combined spectra in R1 and R2, the best-fit parameters are given in Table~\ref{tab:R1R2_spec}, and the corresponding equivalent widths (and upper limits) of the lines are given in Table~\ref{tab:lines_eqw}. We noticed that in the combined data in R1, there was a narrow 6.7~keV absorption feature as well as a potentially broad 8~keV absorption feature, which are likely due to the Fe XXV and Fe XXVI K$\beta$ absorption lines, respectively. 

\input{R1_R2_spectra}

In order to look for any finer evolution of the 1~keV and 7~keV spectral lines in the spectral regions, we ran all the spectral fits without both Gaussians, with one Gaussian (each line individually), and with both Gaussians. If the fit improved by adding a line, whether an individual line (1 or 7~keV) or adding a second line on top of the first fit, at the 99.73\% confidence level as assessed by the difference in AIC values (see Equations \ref{eq:aic} and \ref{eq:prob}), the spectral line was considered a detection. There was no obvious correlation between the presence (or absence) of the spectral lines with the position in R1. The equivalent widths of both lines also did not suggest an obvious correlation.


The 1.0~keV emission line feature could possibly be due to either incomplete modeling of the interstellar medium absorption\footnote{\url{https://heasarc.gsfc.nasa.gov/docs/nicer/data_analysis/workshops/NICER-CalStatus-Markwardt-2021.pdf}}, or a pseudo-continuum of weak narrow lines \citep[e.g., IGR J17062-6143;][]{vandeneijnden17}, or due to faint Fe L or Ne X emission \citep{degenaar13,bult21,ngmason22}. 


\input{lines_eqw}

\subsubsection{Dip Spectrum vs Non-Dip Spectrum} \label{subsubsec:dip_results}

\begin{figure}[t]
		\centering
		\includegraphics[width=1.1\linewidth]{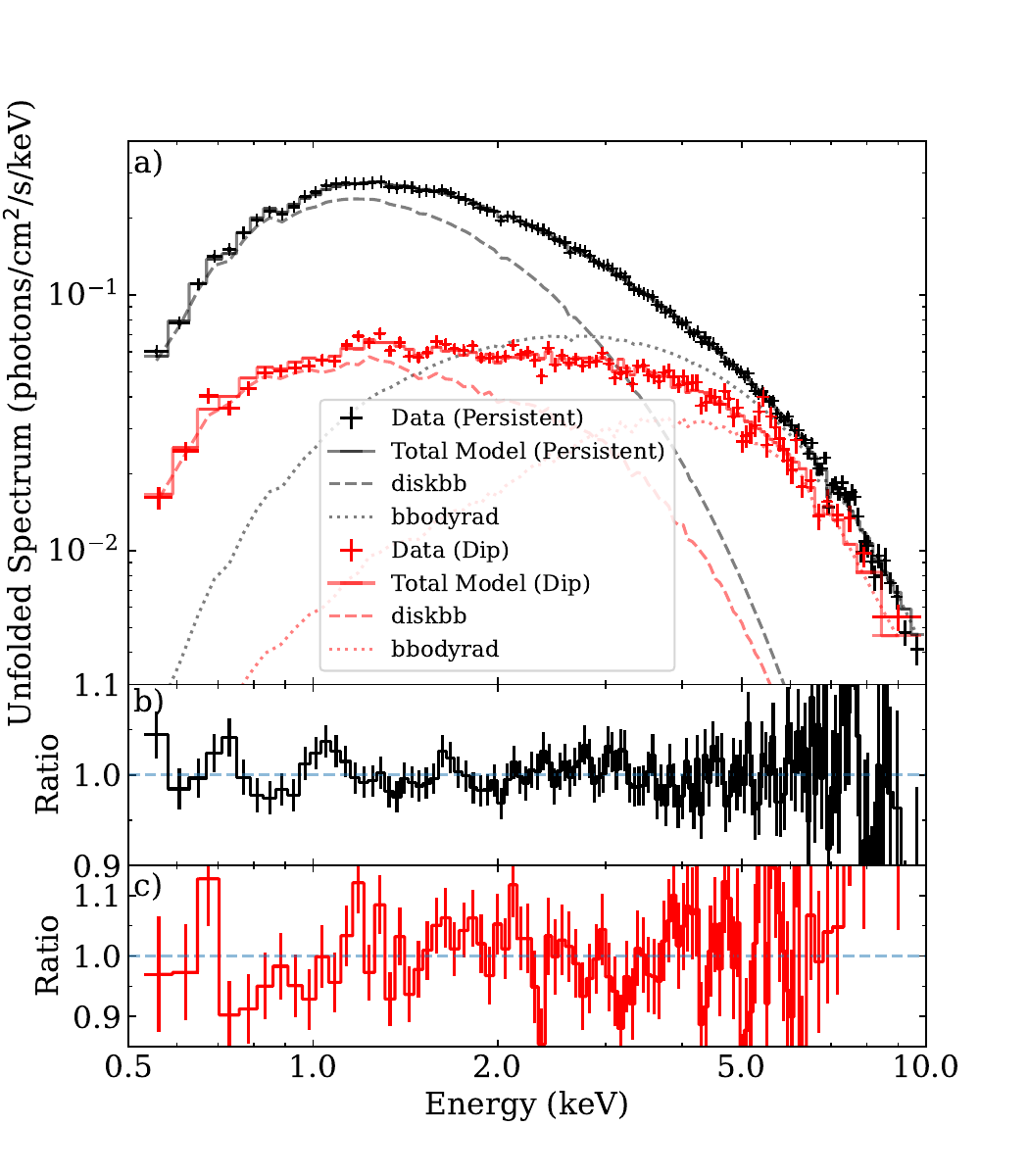}
		\caption{a) Unfolded spectra corresponding to the source emission during the absorption dip (red) and in the persistent emission immediately following the dip (black). The spectra were jointly fit with an absorbed disk blackbody (\texttt{diskbb}), a thermal blackbody (\texttt{bbodyrad}), a neutral partial covering component (\texttt{tbpcf}) and a partial covering partially ionized absorber component (\texttt{zxipcf}). In \texttt{XSPEC} parlance, the model is given by \texttt{tbabs(tbpcf(zxipcf(diskbb+bbodyrad)))}. The best-fit spectral parameters are given in Table~\ref{tab:dip_persist_spec}; b) Ratio of the spectrum from the persistent emission to the model (specified above); c) Similar to b), but for the emission during the absorption dip. The spectra have been rebinned for visual purposes.}
		\label{fig:dippluspersistent}
	\end{figure}

We investigated the spectral variation of \src during the absorption dip and in the persistent emission immediately following the dip. The uncertainties reported in this subsection are $90\%$ confidence limits. We performed a simultaneous fit of the two spectra with an absorbed disk blackbody, thermal blackbody, a neutral partial covering component, and a partial covering partially ionized absorber. We found that the column density for the partially ionized absorber, $N_H$, was unconstrained, so we set it to $N_H=10^{23}{\rm\,cm^{-2}}$. The corresponding spectral parameters for both blackbody components (\texttt{diskbb} and \texttt{bbodyrad}) were tied across the dip and persistent emission spectra. The best-fit spectral parameters are shown in Table~\ref{tab:dip_persist_spec}. The unfolded spectra corresponding to the absorption dip (in red pluses) and persistent emission (in black pluses) are shown in Figure~\ref{fig:dippluspersistent}, along with the disk blackbody (dashed lines) and thermal blackbody (dotted lines) model components.  

We found that the best-fit values for both blackbody components are consistent with spectra around the GTI containing the absorption dip (see Figure~\ref{fig:spectroscopy}). For the neutral partial covering component, we found that the column density decreased from $8.7_{-0.9}^{+0.8}\times10^{22}{\rm\,cm^{-2}}$ to $4.8_{-1.4}^{+1.3}\times10^{22}{\rm\,cm^{-2}}$ as the neutral absorber presumably moved away from the line of sight, though not highly significant. However, the partial covering fraction significantly dropped from $0.714_{-0.015}^{+0.027}$ to $0.16_{-0.03}^{+0.06}$. For the partial covering partially ionized absorber, the column density was $N_H = 100_{-18}^{+39}\times10^{22}{\rm\,cm^{-2}}$ during the dip, and the ionization parameter decreased from ${\rm log}(\xi) = 4.37_{-0.11}^{+0.30}$ to ${\rm log}(\xi) = 3.20_{-0.15}^{+0.12}$ as the neutral absorber was presumably in the line of sight. There were no significant changes in the covering fraction.

The presence of a partial covering partially ionized component is common among dipping NS LMXBs \citep{diaztrigo06,raman18,gambino19}, but an additional neutral partial covering component is not usually required. To test the necessity of the additional component, we performed a similar fit to that presented in Table~\ref{tab:dip_persist_spec}, but without the component (\texttt{tbpcf}). The column density of the partial covering partially ionized component remained fixed at $N_H = 10\times10^{22}{\rm\,cm^{-2}}$ in the persistent emission (unconstrained), but we found that $N_H=14.0_{-1.1}^{+1.3}\times10^{22}{\rm\,cm^{-2}}$ during the absorption dip. We also found that the ionization parameter decreased from ${\rm log}(\xi) = 4.39_{-0.15}^{+0.56}$ to ${\rm log}(\xi) = 1.55_{-0.15}^{+0.18}$ as the neutral absorber moved presumably in the line of sight. We did not see any statistically significant change in the covering fraction, with a lower limit of $>0.53$ for the persistent emission and a value of $0.774_{-0.004}^{+0.004}$ during the absorption dip. Finally, for the blackbody components, we found that $T{\rm in} = 0.78_{-0.02}^{+0.02}{\rm\,keV}$, ${\rm norm}_{\rm diskbb} = 250_{-20}^{+30}$ (scaling defined in Table~\ref{tab:R1R2_spec}), $kT = 1.287_{-0.021}^{+0.017}{\rm\,keV}$, and ${\rm norm}_{\rm bbodyrad} = 83_{-6}^{+7}$. The fit statistic from this fit is $\chi^2/{\rm d.o.f.}=281/236$; the fit presented in Table~\ref{tab:dip_persist_spec} is a statistically significant improvement with a rejection probability (see Equation~\ref{eq:prob}) of $5.0\times10^{-7}$. 

To investigate whether the varying absorption alone can account for the absorption dip, we performed an alternative fit to what was presented in Table~\ref{tab:dip_persist_spec} --- we tied the absorption components across the dip and persistent spectra ($N_H$ still constrained, so it was fixed at $10\times10^{22}{\rm\,cm^{-2}}$), but allowed the blackbody parameters (both \texttt{diskbb} and \texttt{bbodyrad}) to vary. For the absorption components, we found $n_{\rm H,n} = 4.4_{-0.8}^{+1.4}\times10^{22}{\rm\,cm^{-2}}$, a covering fraction for the neutral absorber of $0.21_{-0.06}^{+0.06}$, and an upper limit on the covering fraction for the ionized absorber of $>0.81$. We found that for the spectrum during the absorption dip (persistent emission), $T_{\rm in}=0.44_{-0.03}^{+0.04}{\rm\,keV}$ ($T_{\rm in}=0.78_{-0.03}^{+0.04}{\rm\,keV}$), ${\rm norm}_{\rm diskbb}=430_{-110}^{+120}$ (${\rm norm}_{\rm diskbb}=320_{-50}^{+50}$; scaling defined in Table~\ref{tab:R1R2_spec}), $kT=1.36_{-0.03}^{+0.03}$ keV ($kT=1.32_{-0.03}^{+0.03}$ keV), and ${\rm norm}_{\rm bbodyrad}=51_{-4}^{+5}$ (${\rm norm}_{\rm bbodyrad}=72_{-6}^{+9}$). We see no evolution in the disk blackbody normalization and blackbody temperature ($kT$) within uncertainties, a small difference in the blackbody normalization, and a significant evolution in the inner disk temperature. We found the fit to describe the persistent and dip spectra well, with $\chi^2/{\rm d.o.f.}=252/233$. The rejection probability between the two models is $P_1/P_2=0.05$, so both fits are statistically consistent with the two spectra.

If we assume an upper limit of $D_{10} < 0.9$, and $i = 60-75^\circ$, we find that the upper limit of the apparent inner disk radius from the dip (persistent) spectrum is about $r_{\rm in} < 22.8_{-1.9}^{+1.7}$ km ($r_{\rm in} < 26_{-4}^{+3}$ km). The realistic disk radius is related to the apparent inner disk radius, $r_{\rm in}$, through $R_{\rm in} = \xi\kappa^2r_{\rm in}$, where $\kappa$ is the spectral hardening factor and $\xi = \sqrt{3/7}(6/7)^3$ \citep{kubota98}. In this case, for $L\approx0.1L_{\rm Edd}$, we adopted $\kappa=1.7$ \citep{shimura95}, thus for the dip (persistent) portion we obtain $R_{\rm in} < 27.2_{-2.3}^{+2.0}{\rm\,km}$ ($R_{\rm in} < 31_{-5}^{+4}{\rm\,km}$), or $R_{\rm in} < 13.2_{-1.2}^{+0.9}R_G$ ($R_{\rm in} < 15.0_{-2.4}^{+1.9}R_G$), where $R_G=GM/c^2$ is the gravitational radius with $M=1.4M_\odot$. These limits are consistent with those derived from reflection modeling of other NS LMXBs \citep{ludlam22} as well as recent reflection modeling with NuSTAR observations \cite{mondal23}. Additional detailed reflection modeling of \src is outside of the scope of this work and will be reported in Pike et al. 2024, in prep. We will discuss the results in \S\ref{sec:dip}.


The 2008 outburst of \src was about 10--20 times fainter than the latest outburst, so making fair comparisons is difficult, but Fe XXVI absorption lines were observed from the outburst with Chandra and \cite{gavriil12} found an upper limit of a redshift flow to be $v < 221{\rm\,km\,s^{-1}}$. Therefore, a static disk atmosphere cannot be ruled out. Unfortunately, the spectral resolution of the NICER observations is too low for us to ascertain any useful limits from the spectral fit of the absorption dip and the persistent emission post dip. Letting the redshift parameter freely vary yielded a best fit value of $z = 0.0003_{-0.0049}^{+0.0064}$, which is consistent with zero. High-resolution X-ray spectroscopc studies with the recently launched X-ray Imaging and Spectroscopy Mission (XRISM) of NS LMXBs will be able to ascertain the nature of the disk atmosphere \citep{tashiro18,trueba20,trueba22}.


\input{dip_persist_spectral_fit.tex}
 
\subsection{Correlated X-ray Behavior} \label{sec:correlations}

We can combine the insights from the X-ray timing and spectroscopic analyses we have taken by looking at correlations between the different timing and spectroscopic variables. To do so, we calculated the Spearman correlation coefficient (SCC) and the corresponding p-values. There were some correlations that can already be seen from the overall spectroscopic evolution of the source (see Figure~\ref{fig:spectroscopy}) --- the blackbody temperature ($kT$) and normalization (from \texttt{bbodyrad} in \texttt{XSPEC}) had an SCC of $-0.95$, which translated to a p-value of $\sim10^{-35}$, which is unsurprising as they are intrinsically correlated for a given flux. For the correlation between the 0.6--10.0~keV absorbed flux and blackbody components, the SCC and p-values were $-0.53$ and $1.1\times10^{-6}$ ($T_{\rm in}$), $0.61$ and $6.9\times10^{-9}$ (disk blackbody normalization), $-0.39$ and $6.0\times10^{-4}$ (blackbody temperature), and $0.61$ and $1.1\times10^{-8}$ (blackbody normalization). 

The potentially more interesting correlations come from comparing quantities derived from spectroscopy and from timing for each data segment. For example, we found a strong correlation between the blackbody temperature $(kT)$ and the fractional rms amplitude of the power spectrum over 0.03--50 Hz with an SCC of $-0.64$ and p-value of $4.0\times10^{-9}$. The corresponding SCC and p-values for the correlation between the fractional rms amplitude and the blackbody normalization were $0.55$ and $1.8\times10^{-6}$, respectively. 

\subsection{MeerKAT} \label{sec:meerkat_results}

We have shown the MeerKAT 1.3 GHz flux densities in the bottom panel of Figure~\ref{fig:all_lc} and Table~\ref{tab:meerkat_flux} in the Appendix. The measured flux densities in each of the 15-minute observations are shown in blue, where we plot both ${>}\,4\sigma$ detections (squares) and $3\sigma$ upper limits (triangles). During the first three observations, the source was radio-bright with a flux density ${\gtrsim}\,500\,\mu$Jy. As a result, we were able to measure the (intra-band) spectral index in the bright epochs by breaking the bandwidth in half (i.e., 856--1284 and 1284--1712$\,$MHz) and measuring the flux density in each sub-band. We measured radio spectral indices ($S_\nu \propto \nu^\alpha$) of $\alpha=-0.2\pm0.1$, $-0.3\pm0.1$, and $-0.6\pm0.2$ during the observations on MJD 59731, 59733, and 59742, respectively. Following the bright detections, the source flux density rapidly decayed over $\sim$~7~days, dropping below our single observation detection threshold of ${\sim}\,{80}\,\mu$Jy for all but one of our remaining observations. We performed image stacking to increase the S/N, grouping the last eleven epochs into three stacked images consisting of three, four, and four independent observations. The stacked flux densities are shown in black, where we detected the source at ${\gtrsim}\,4.5\sigma$ in the first two stacked images (diamonds). However, even with the improved S/N we still did not detect the source in the final stacked image, and thus represent it as a $3\sigma$ upper limit. We did not observe any correlated evolution (in time) when comparing the X-ray and radio light curves, as would be expected from hard-state jet emission.

To constrain the position of \src on the $L_R$--$L_X$ plane, we calculated the radio luminosity at 5 GHz ($L_{R,5})$ through 

\begin{equation}
    L_{R,5} = 4\pi S_\nu \nu d^2 \approx 3.9\times10^{26} {\rm\,erg\,s^{-1}} \left(\frac{S_\nu}{1{\rm\,\mu Jy}}\right) \left(\frac{d}{8{\rm\,kpc}}\right)^2, 
\end{equation}

\noindent where $1{\rm\,Jy} = 10^{-23}{\rm\,erg\,s^{-1}\,cm^{-2}\,Hz^{-1}}$, $\nu=5{\rm\,GHz}$, and adopting the standard assumption of a flat radio spectral index ($\alpha=0$) in $S_\nu \propto \nu^\alpha$. We find an upper limit on $L_{R,5}$ for the last stacked data point to be $L_{R,5} < 1.2\times10^{28}{\rm\,erg\,s^{-1}}$, assuming the source is at a distance of $8\,$kpc.

\section{Discussion} \label{sec:discussion}

We have reported on the X-ray timing and spectroscopic analysis with NICER as well as radio monitoring observations with MeerKAT, throughout the nearly three-month long outburst of the dipping NS LMXB \src. We also presented supplementary NuSTAR observations (timing only), and monitoring data with MAXI and Swift. With NICER we have discovered, for the first time, several instances of a $\nu_0\approx8$ Hz QPO in the high intensity, banana spectral state of the source. The QPO had fractional rms amplitudes of $\sim5\%$. We identify this QPO as an NBO-like QPO, and argue that through \src, we are probing the lower accretion rate boundary at which NBOs (can) occur. Additionally, we argue that since: a) the source exhibited two separate tracks in a very short timescale, only seen in the Z-state of other sources, in a NICER observation (Figure~\ref{fig:hid_z}); b) we observed a flaring branch in the 3--25 keV NuSTAR HID (Figure~\ref{fig:nustar_lc}); c) we saw NBOs at the peak of the outburst, the source is exhibiting Z-state-behavior at the outburst peak. Finally, MeerKAT observations confirmed that at the peak of the outburst, \src was at its radio-brightest. We discuss below that the results are consistent with the picture of a transient ejection of material accompanying the island-to-banana spectral state transition of the NS LMXB in the atoll-state. The observations provided the strongest observational evidence for radio flaring (and jet ejecta) during spectral state transitions in the atoll-state. The spectroscopic results are possibly consistent with the presence of a disk atmosphere and a contracting boundary layer of the NS during the banana state, where the source spent most of the outburst in.

\subsection{X-ray and Radio Evolution} \label{sec:outburst_summary}

The X-ray spectral and timing properties of \src showed clear signatures of a spectral state transition. When we began our NICER monitoring (MJD 59734), the X-ray spectra were best-fit with thermal components (i.e., a blackbody and multi-color disk blackbody), characteristic of a banana state \citep{lin07}. At later times, for GTI segments after MJD 59800, the X-ray spectra favored power law models, which are often used to describe X-ray spectra dominated by high energy Comptonized photons, and thus an island state. The change in the best-fit X-ray spectral model coincided with the evolution of the timing parameters, namely the transition from the R1 to the R2 spectral states (highlighted in Figure~\ref{fig:hid_ccd}). Moreover, an increase in the average (continuum) fractional rms (from $3\%$ to $10\%$) is typical of atoll-state sources transitioning from banana to island spectral states \citep{hasinger89,vdk04}. The changes in the X-ray spectral continua, position on the CCD, and average fractional rms strongly suggest that a physical change in the accretion flow occurred around MJD 59800. 

The radio evolution consisted of a bright initial detection on our first day of observing (MJD 59731), followed by a rapid, monotonic decrease. Initially, the radio detections were thought to be associated with a steady compact jet \citep{hughes22}. However, the best-fit model for the X-ray spectra (i.e., the source was in a banana state) and the temporal evolution of the radio is more consistent with emission originating from a transient jet ejection \citep[otherwise known as an optically-thick flare;][]{fender2019}. Moreover, the steepening of the radio spectral index ($\alpha=0.2\pm0.1$ to $-0.6\pm0.2$; optically thick-to-thin synchrotron emission) is also suggestive of a transient jet ejection. There is an alternative explanation for optically thin radio emission in NS LMXBs; \citet{Russell2021} observed optically thin radio emission in the ``high X-ray mode'' (equivalent to the banana state) for the ultracompact NS LMXB 4U~1820$-$30. In this system, the emission was attributed to a heavily quenched compact jet with a break frequency that had moved into the radio regime. However, it is difficult to apply a similar explanation to our observations of \src. The optically thin radio emission in 4U~1820$-$30 is persistent and modestly variable (a factor of $\sim$\,2), whereas \src showed a rapid decay corresponding to, at a minimum, a factor of $\sim$\,100 decrease in the radio flux. These decay properties, combined with the temporal coincidence between the radio flare and the X-ray state transition, lead us to favor the jet ejection scenario. 

We looked for evidence of ballistic motion as ejecta have been spatially resolved in several BH \citep[e.g.,][]{hjellming1995,bright2020} and NS LMXBs \citep[e.g.,][]{fomalont01,millerjones2012,motta19}, although for NSs the ejections have only been resolved for Z-state sources. We saw no evidence of movement of the radio emission in \src. However, this was expected as the MeerKAT beam size (${\sim}\,6^{\prime\prime}$) and decay timescale (undetectable after ${\sim}\,40\,$days) would make proper motion undetectable for distances ${>}\,1\,$kpc. We use the maximum flux density ($1.2\,$mJy) and observing frequency ($1.3\,$GHz) to derive a distance-scaling relation for the minimum energy ($E_\text{min}$) of the jet ejection \citep[following the method described in][]{fender2019}, 

\begin{equation}
    E_\text{min} \approx 2.2\times10^{37}\,\text{erg} \left(\frac{d}{8\,\text{kpc}}\right)^{40/17}
\end{equation}

A minimum energy of $\sim{10}^{37}\,$erg is consistent with past observations of flaring NS LMXBs \citep[e.g., Swift J1858.6-0814;][]{rhodes22}. Measuring an accurate distance \citep[e.g., through the photospheric radial expansion of a type I X-ray burst;][]{Kuulkers2003} will remove the distance ambiguities and confirm (or refute) the apparent consistency between the minimum energy calculations.

\cite{motta19} investigated simultaneous X-ray and radio observations of the NS LMXB Z-state source Sco~X$-$1 and found that ultrarelativistic outflows were connected to the simultaneous manifestation of an NBO/HBO pair. For \src, we did not observe any HBOs, and the NBO-like QPOs were not continuously observed --- the three GTIs for which NBO-like QPOs were observed had multiple GTIs in between where the QPOs were not observed. Additionally, simultaneous X-ray and radio observations of another Z-state source, GX 17$+$2, associated the occurrences of NBO with jet formation \citep{penninx88,migliari07}. Unlike the simultaneous X-ray and radio observations that \cite{motta19} analyzed, we only observed the QPO instances days after the ejecta would have been launched. Our X-ray and radio monitoring, although not simultaneous, is consistent with the picture of NBO-like QPOs and transient ejections occurring around the outburst peak. In order to confirm (or refute) this suggestive picture, more dense monitoring of future outbursts of similar sources in the radio and X-rays at the onset of the outburst are required, which could be achieved by coordinated all-sky X-ray monitoring and rapid follow-up observations\footnote{\url{https://heasarc.gsfc.nasa.gov/docs/nicer/science_nuggets/20201203.html}}.


\subsection{Observational Evidence for Radio Flaring During Spectral State Transitions in Atoll-State Sources} \label{sec:atoll_flaring}

Using the behavior of BH LMXBs as a reference, ejection events in NS LMXBs have been predicted to occur during a transition from a low to high mass accretion rate \citep[i.e., `island-to-banana';][]{migliari06,munoz14}. Given that we did not observe the rise of the radio flare and that our early-time NICER X-ray observations were already consistent with banana state emission, it is likely that we missed the initial island-to-banana state transition. Looking at the MAXI/GSC light curves (see Figure~\ref{fig:all_lc}), the onset of the outburst began on MJD 59728, six days before our first NICER observations. During this time, the MAXI hardness ratio showed a drop from ${\sim}\,0.6$ on MJD~59729 (the highest hardness ratio of the entire outburst) to ${\sim}\,0.3$ on MJD~59731 (the same day as the initial radio detection). It is plausible that the observed hardness ratio evolution was a signature of a state transition that coincided with the launching of a jet ejection, and the source rapidly transitioned (${\lesssim}\,3\,$days) during the rise of the X-ray flux. This is corroborated by the Swift/XRT observation on 2022 June 1 (see \S\ref{sec:results}), where the 0.5--10.0~keV spectrum was well described with an absorbed cutoff power law with a photon index of $\Gamma_{\rm co}\approx0.6$ and e-folding energy $E_f\approx3.2{\rm\,keV}$, suggesting that \src was still in the island state. The behavior of \src constitutes some of the strongest observational evidence for radio flaring (and jet ejecta) during state transitions in atoll-state LMXBs.  

\begin{figure}[htbp!]
	\centering
	\includegraphics[width=\linewidth]{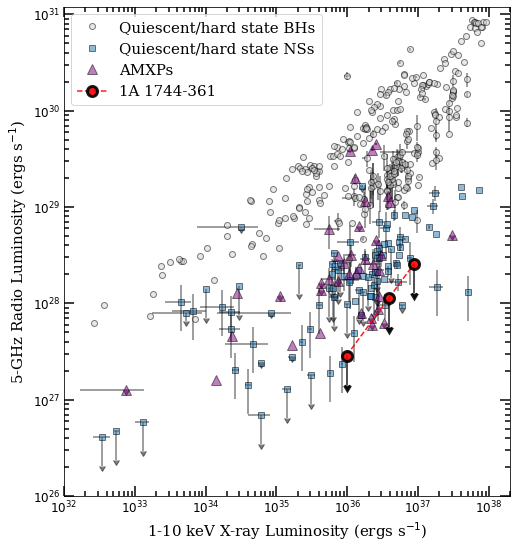}
	\caption{1--10~keV X-ray and 5~GHz radio luminosity plane (i.e., $L_R$--$L_X$ plane). The plot includes archival hard state BH LMXBs (gray circles), hard state NS LMXBs (blue squares), and accreting millisecond X-ray pulsars (AMXPs; purple triangles) as seen in Figure~4 of \citealt{vde2022}, which contains a large number of sources from \cite{bahramian22_lrlx}. As \src was never detected in the hard state, its position on the $L_R$--$L_X$ plane is limited to upper limits (red circles). Furthermore, due to the unknown distance of \src, the plot includes three data points at distances of 4, 8, and 12~kpc. The upper limits reside within the lower end of radio luminosities for known hard-state NS LMXBs. Future, higher sensitivity observations (and an accurate distance measurement) are critical for determining the source's exact position on the $L_R$--$L_X$ plane, and whether it has an anomalously radio-quiet compact jet.}
		\label{fig:lrlx}
    \end{figure}

We did not detect any radio re-brightening after the source transitioned back to the island state (towards the end of the outburst). Following the transition, assuming \src did, in fact, re-form the compact jet, we used the final stacked image upper limit ($\sim30{\rm\,\mu Jy}$) to place constraints on the position of the source on the $L_R$--$L_X$ diagram as this observation coincided with the hard state X-ray emission. We present these results for three assumed distances (4, 8, and 12$\,$kpc) in Figure~\ref{fig:lrlx}. The luminosities at all three distances put \src among the lower-end of radio luminosities for NS LMXBs at our measured X-ray luminosity. Higher sensitivity radio monitoring of future outbursts will allow us to directly detect (or better constrain) the radio flux density of the compact jet, thereby determining whether \src is anomalously radio-quiet for an NS LMXB.  

\subsection{QPO Phenomenology} \label{sec:qpo_discussion}

In the most recent high-intensity outburst in 2022, we discovered instances of a $\nu\approx8$~Hz QPO over 1.0--3.0~keV (see Table \ref{tab:qpo_Edep}) around the peak of the outburst. This is similar to the NBO-like QPOs observed in Aql~X$-$1 \citep{reig04}, XTE~J1806$-$246 \citep{wijnands99b}, and 4U~1820$-$30 \citep{wijnands99a}. While the QPOs (including from \src) manifested near the peak luminosities of the outburst, the luminosities were below $0.5\,L_{\rm Edd}$ (except XTE~J1806$-$246), unlike the NBOs in Z-state sources \citep[$0.5-1.0 L_{\rm Edd}$;][]{vanparadijs88,wijnands96,piraino02}. For 4U~1820$-$30, the peak luminosity was approximately $0.2-0.4{\rm\,L_{\rm Edd}}$, and for Aql~X$-$1 it was approximately $0.05{\rm\,L_{\rm Edd}}$. In the other high-intensity outbursts of \src in 2005 and 2008 (see Figure~\ref{fig:monitoringlc}) with RXTE, no QPOs were observed (except for a low significance $3\sigma$ indication at 29 Hz in 2008) though the bolometric luminosity was on the order of $0.01 L_{\rm Edd}$. Additionally, the high-intensity outburst in 2003 did not show any QPOs in the data. 

The QPOs from this outburst were found in the banana state in the CCD. The QPOs found in XTE~J1806$-$246 and 4U~1820$-$30 exhibited a clear energy dependence below 12~keV, whereas we found no such relation in the NICER data. However, this could just be due to the differences in sensitivities in the different energy bands. Detections of NBOs/NBO-like QPOs below 3~keV are rare. At the higher accretion rates exhibited in Z-state sources, NBOs have been observed in the soft energy range with Z-state sources: over 2--3.1~keV from Cyg~X$-$2 \citep{wijnands01} with RXTE and down to $\sim0.4$~keV from Cyg~X$-$2 and Sco~X$-$1 with NICER \citep{jia23}. We have seen in various atoll-state and Z-state sources that the appearance of the different QPOs is a function of the spectral state of the source \citep{motta17}. The fact that we have observed NBO-like QPOs from \src implies that at the peak of the outburst, the source was a Z-state source, and the QPOs were undetectable by the time the source was below a luminosity threshold. Thus we suggest that the `NBO-like QPOs' we have observed from other atoll-state sources \citep{wijnands99a,wijnands99b} and NBOs observed from Z-state sources are simply the same phenomena (NBOs) observed at different mass accretion rates. With \src, we are probing the lower accretion rate boundary at which NBOs are observed. Observing atoll-state sources at higher luminosities will confirm this picture.

We also note from the low-intensity 2004 outburst from \src that a $\sim3.5$~Hz QPO was discovered \citep{bhattacharyya06a}. Seven sources in total have also exhibited $\sim1$~Hz QPOs which are empirically observed in the island spectral states and is thought to originate from a geometrically modulated precessing inner accretion flow \citep{homan99,jonker00,homan12,homan15}.

\subsection{Absorption Dip} \label{sec:dip}

The source light curve exhibited an absorption dip during the outburst, as shown in Figure~\ref{fig:simple_dip_lc}, a known feature of the source \citep{bhattacharyya06a,gavriil12}. We note that during the roughly 90 s duration of the absorption dip, the source flux temporarily increased to near-persistent levels (approximately 16--20 s; 18--22\% of the dip duration), and the color correspondingly fell. 

We also investigated the spectra during the absorption dip and the persistent emission immediately following the dip (Table~\ref{tab:dip_persist_spec}). The spectral results, as well as the lone absorption dip observed, are consistent with the idea of an intervening orbital partial covering absorber along the line of sight appearing for about 100~s \citep{jimenezgarate02,diaztrigo09}. In particular, we required a partial covering partially ionized absorber and a neutral partial covering absorber (see \S\ref{subsubsec:dip_results}) whose parameters varied across the dip and persistent emission spectra. We also tested for the possibility that the different observed spectra were due to changes in the blackbody parameters. We found a significant increase in the inner disk temperature from $T_{\rm in}=0.44_{-0.03}^{+0.04}{\rm\,keV}$ to $T_{\rm in}=0.78_{-0.03}^{+0.04}{\rm\,keV}$ and a small increase in the blackbody normalization (${\rm norm}_{\rm bbodyrad}=51_{-4}^{+5}$ to $72_{-6}^{+9}$; scaling in Table~\ref{tab:R1R2_spec}) as the intervening absorber left the line of sight. From the light curve shown in Figure~\ref{fig:simple_dip_lc}, these spectroscopic results imply that the inner disk temperature increased by almost a factor of two in a matter of seconds, which seems implausible. It is more likely that the spectral changes during the dip can be explained by the presence of a neutral partial covering absorber. In fact, the partial covering fraction of $0.16_{-0.03}^{+0.06}$ is similar to the fraction of time that the source flux temporarily increased during the absorption dip, which is consistent with the idea of the covering fraction being the fraction of time that the source was fully obscured by the neutral absorber.

The best-fit value for the ionization parameter, ${\rm log}(\xi) \approx 4.4$, is consistent with the lower limit of ${\rm log}(\xi) > 3.6$, derived from the absorption dip in the July 2008 outburst \citep{gavriil12}. Absorption dips observed in other dipping LMXBs also revealed ionization parameters in the persistent emission ${\rm log}(\xi) > 4.0$ \citep{iaria07b,iaria07a,marino22}. However, we do not rule out ionized absorption during the dip. 

\subsection{Accretion Disk Atmosphere} \label{sec:diskatmosphere}

Looking at the combined spectra from spectral regions R1 and R2 (see Table~\ref{tab:R1R2_spec}), the blackbody normalization from R1 of $26.5$ (scaled by $(R_{\rm km}/D_{10})^2$) implied an emitting region of size $R\approx4.6{\rm\,km}$. In fact, for R1, the normalization took on values over $10-100$ ($R\approx2.8-9{\rm\,km}$), though note these are all upper limits as we adopt $d<9{\rm\,kpc}$. The decreasing blackbody normalizations could be a result of a contracting boundary layer. 

The presence of significant absorption lines from Fe XXV K$\alpha$, Fe XXVI K$\alpha$, and Fe XXVI K$\beta$ in R1 suggested the presence of a highly ionized disk atmosphere (plasma with an inferred equatorial geometry), which have been observed in several NS LMXBs \citep{parmar02,iaria07b,iaria07a,hyodo09,marino22}. They are also a common feature in NS LMXBs exhibiting absorption dips \citep{diaztrigo06}. On the other hand, in R2, we saw significantly weaker Fe XXVI K$\alpha$ and a non-detection of the Fe XXVI K$\beta$ line (albeit with a high upper limit), which possibly implies a cooler disk atmosphere. This could also be due to the photoionization of the disk atmosphere from the corona \citep{ko91,neilsen09}. We tried to investigate individual segments within the spectral regions (R1 and R2), but within uncertainties and due to the high upper limits, we could not ascertain any differences between the spectral subregions. In addition, the spectrum during the persistent emission after the absorption dip (Table~\ref{tab:dip_persist_spec}) exhibited a high ionization parameter ($\text{log}(\xi)\approx4.4$) and a large covering fraction ($>0.7$), suggesting the presence of an ionized medium possibly in the form of an accretion disk atmosphere. Thus, with the expected $60-75\degr$ orbital inclination of the system (observed dip, but no eclipses), we are likely probing into the highly ionized accretion disk atmosphere, with the absorption lines originating in the disk atmosphere.  

We also note that this outburst was the first time that the Fe XXV K$\alpha$ and Fe XXVI $K\beta$ lines were detected from this source (\citealt{tobrej23}, \citealt{mondal23}, Pike et al. 2024, in prep.). Higher spectral resolution observations, such as that of XRISM, will be crucial in revealing the dynamic (or static) nature of the accretion disk atmosphere or wind from NS LMXBs \citep{trueba20,trueba22,gandhi22}


\section{Conclusions} \label{sec:conclusion}

In this work, we have reported on the X-ray timing and spectroscopic analysis with NICER (timing only for NuSTAR; and monitoring with MAXI and Swift) of the dipping atoll-state NS LMXB \src, as well as radio monitoring observations with MeerKAT, throughout its nearly three-month long outburst in 2022. The overall outburst proceeded as follows:
$ $ \\

\begin{enumerate}[nolistsep]
    \item On ${\sim}\,$MJD 59728, \src entered into an outbursting state, its X-ray flux rapidly increased, and it likely transitioned from the island to the banana spectral states within ${\sim}\,3\,$days of the onset of the outburst, launching transient jet ejecta.
    \item The source remained at a high accretion rate (i.e., in the banana state) from ${\sim}\,$MJD 59728 to 59800, where any radio detection was the result of the fading jet ejecta.
    \item At around ${\sim}\,$MJD~59800, the source transitioned back to the island state. We detected no radio emission, and, therefore, if a compact jet formed, its radio emission was below our luminosity detection threshold of $L_{R,5}<1.2\times10^{28}\,(d/8{\rm\,kpc})^2{\rm\,erg\,s^{-1}}$
    \item For the remainder of our monitoring, the X-ray luminosity decreased, dipping below NICER's sensitivity limit on ${\sim}\,$MJD~59820, as the source returned to quiescence.
\end{enumerate}
The results presented in this work support the notion that atoll sources and Z sources are not distinct source classes, but instead are two states on a continuum defined by the mass accretion rate, as previously suggested in detailed spectral and timing studies of XTE~J1701$-$462 \citep{lin09,homan10} and other NS LMXBs \citep{fridriksson15}. Thus, it is more appropriate to refer to atoll-state sources and Z-state sources. We have discovered several instances of a $\nu_0\approx8$ Hz QPO in the high intensity, banana state outburst of the source with fractional rms amplitudes of $\sim5\%$ for the first time. The HID of a NICER observation around the outburst peak (Figure~\ref{fig:hid_z}) suggests that the source exhibited two distinct tracks on a very short time scale ($\lesssim2{\rm\,hr}$, which is typically seen in Z-state sources). The HID for the NuSTAR observation, taken around the peak of the outburst (Figure~\ref{fig:nustar_lc}), shows a flaring branch. While the HID/CCD corresponding to the outburst with the NICER data suggested an atoll-state source, the aperiodic timing features (along with the HIDs) observed by NICER and NuSTAR of \src resemble that of other atoll-state sources that exhibit Z-state-like behavior around the peak of the outbursts, such as XTE~J1806$-$246 \citep{wijnands99a}, 4U~1820$-$30 \citep{wijnands99b}, and Aql~X$-$1 \citep{reig04}. In fact, with the growing understanding that the spectral evolution of atoll-state sources and Z-state sources differ only in the mass accretion rate \citep{lin09,homan10,fridriksson15}, we posit that the `NBO-like QPOs' observed in several atoll-state sources \citep{wijnands99a,wijnands99b,reig04} are in fact NBOs manifesting at lower accretion rates. The NBO production mechanism is likely activated at a luminosity threshold below that of the Eddington luminosity, as suggested from observations of 4U~1820$-$30 and XTE~1806$-$246 \citep{wijnands99a,wijnands99b}. To probe this further, dense multiwavelength monitoring and rapid follow-up of future outbursts of similar sources across different accretion states are required. 

Around the peak of the outburst of \src, we detected NBOs and inferred the transient ejection of material accompanying the island-to-banana state transition with quasi-simultaneous X-ray and radio monitoring. This observed scenario is possibly consistent with the simultaneous X-ray and radio observations of Sco~X$-$1 which found that ultrarelativistic outflows were connected to the appearance of an NBO/HBO pair \citep{motta19}. However, the robust confirmation of such a scenario will require dense monitoring of future outbursts of such sources in the X-rays and radio wavelengths with sensitive instruments. 

We also observed an absorption dip in the light curve of one observation throughout the outburst. The X-ray spectroscopy results are consistent with the idea of a non-orbiting neutral partial covering absorber with a non-trivial interior structure along the line of sight. The large covering fraction ($>70\%$) of the ionized absorber during the persistent emission suggests the presence of a highly ionized medium surrounding the NS. We have also investigated the evolution of the 6.7~keV, 6.96~keV, and 8.0~keV Fe XXV, Fe XXVI K$\alpha$ and K$\beta$ absorption lines, respectively, where the presence of these absorption lines also suggest the presence of a highly ionized medium, with origins likely in the disk atmosphere of the source. 






\facilities{NICER, MAXI, NuSTAR, Swift, MeerKAT}

\software{Astropy \citep{astropy:2013, astropy:2018}, NumPy and SciPy \citep{virtanen20}, Matplotlib \citep{hunter07}, IPython \citep{perez07}, tqdm \citep{dacostaluis22}, HEASoft 6.31.1\footnote{http://heasarc.gsfc.nasa.gov/ftools} \citep{heasoft}}

\acknowledgments

We thank the anonymous referee for comments that improved the clarity of the presentation and description of the results. We thank Thomas Dauser and Javier Garcia for their persistence to get \texttt{relxill} to be able to compile on MacOS systems with the M1 chip. MN also thank Megan Masterson and Jingyi Wang for discussions about aperiodic timing and NICERDAS tools. AKH and GRS are supported by NSERC Discovery Grant RGPIN-2021-0400. JvdE acknowledges a Warwick Astrophysics prize post-doctoral fellowship made possible thanks to a generous philanthropic donation. NICER work at NRL is also supported by NASA. This research has made use of data and/or software provided by the High Energy Astrophysics Science Archive Research Center (HEASARC), which is a service of the Astrophysics Science Division at NASA/GSFC and the High Energy Astrophysics Division of the Smithsonian Astrophysical Observatory. This research has made use of MAXI data provided by RIKEN, JAXA and the MAXI team. This research has made use of data from the NuSTAR mission, a project led by the California Institute of Technology, managed by the Jet Propulsion Laboratory, and funded by the National Aeronautics and Space Administration. Data analysis was performed using the NuSTAR Data Analysis Software (NuSTARDAS), jointly developed by the ASI Science Data Center (SSDC, Italy) and the California Institute of Technology (USA). We acknowledge the use of public data from the Swift data archive. This work made use of data supplied by the UK Swift Science Data Centre at the University of Leicester.

AKH and GRS respectfully acknowledge that they perform the majority of their research from Treaty 6 territory, a traditional gathering place for diverse Indigenous peoples, including the Cree, Blackfoot, Métis, Nakota Sioux, Iroquois, Dene, Ojibway/Saulteaux/Anishinaabe, Inuit, and many others whose histories, languages, and cultures continue to influence our vibrant community.

\bibliography{1a1744.bib}
\bibliographystyle{aasjournal}

\appendix

\restartappendixnumbering
\section{Tables}
\input{meerkat_flux}
\input{obsid_props_final}

\end{document}

%% file: spectralregions_timing.tex
\begin{deluxetable*}{cccccccc}[htbp!]
\tablecaption{Best-fit parameters from fitting the power spectra with a power law plus constant continuum for each spectral region defined in Figure \ref{fig:hid_ccd}.
\label{tab:cc_timing}}
\tablehead{
 \colhead{Spectral Region} &
 \colhead{$f_{\rm rms,0.03-50}\ (\%)$} &
 \colhead{$m$} & 
 \colhead{No. photons} & 
 \colhead{$\chi^2$/d.o.f.} & 
 \colhead{PL Amplitude} & 
 \colhead{PL $\alpha$} & 
 \colhead{PL Constant} 
}
\startdata
R1 & $5.6 \pm 1.4$ & 411 & 17510405 & $474/360$ & $(3.56 \pm 0.11)\times10^{-4}$ & $1.061 \pm 0.012$ & $(3.0027 \pm 0.0006)\times10^{-3}$ \\
R2 & $10.5 \pm 2.4$ & 123 & 1995884 & $683/360$ & $(1.08 \pm 0.06)\times10^{-3}$ & $1.13 \pm 0.02$ & $(7.883 \pm 0.003)\times10^{-3}$ \\
CC1 & $7.3 \pm 1.4$ & 95 & 4919162 & $464/360$ & $(5.7 \pm 0.2)\times10^{-4}$ & $1.106 \pm 0.018$ & $(2.4706 \pm 0.0010)\times10^{-3}$ \\
CC2 & $4.5 \pm 1.4$ & 128 & 5126233 & $369/360$ & $(2.10 \pm 0.19)\times10^{-4}$ & $1.02 \pm 0.03$ & $(3.1937 \pm 0.0011)\times10^{-3}$ \\
CC3 & $2.0 \pm 1.5$ & 22 & 848125 & $425/360$ & $(2.6 \pm 2.1)\times10^{-5}$ & $1.5 \pm 0.2$ & $(3.316 \pm 0.003)\times10^{-3}$ \\
CC4 & $11.5 \pm 2.1$ & 34 & 847789 & $658/360$ & $(1.12 \pm 0.08)\times10^{-3}$ & $1.12 \pm 0.03$ & $(5.126 \pm 0.004)\times10^{-3}$ \\
CC5 & $10.1 \pm 2.7$ & 14 & 211120 & $551/360$ & $(1.2 \pm 0.7)\times10^{-4}$ & $1.74 \pm 0.19$ & $(8.492 \pm 0.009)\times10^{-3}$ \\
CC6 & $13.5 \pm 9.5$ & 3 & 3584 & $303/360$ & $-0.021 \pm 0.003$ & $0.43 \pm 0.06$ & $0.1085 \pm 0.0008$ \\
\enddata
\tablecomments{We present the fractional rms amplitude ($\%$) over 0.03--50 Hz, the number of averaged segments for the power spectrum ($m$), the number of photons, the $\chi^2$ values, the power law amplitudes, spectral indices ($\alpha$), and the constant value.}
\end{deluxetable*}

%% file: qpo_Edep.tex
\begin{deluxetable}{cccc}[ht]
\tablecaption{Energy dependence of the QPO. \label{tab:qpo_Edep}}
\tablehead{
 \colhead{Energy (keV)} &
 \colhead{$f_{\rm rms,1} (\%)$} &
 \colhead{$f_{\rm rms,2} (\%)$} &
 \colhead{$f_{\rm rms,3} (\%)$}
}
\startdata
0.3--1.0 & \nodata & $3.9\pm1.8$ & \nodata \\ 
1.0--2.0 & $6.3\pm0.5$ & $4.4\pm0.5$ & $5.3\pm0.5$ \\ 
2.0--3.0 & $5.8\pm1.3$ & $5.8\pm0.9$ & $6.8\pm0.9$ \\ 
3.0--4.0 & $4.2\pm3.6$ & $<9.8\ (2\sigma)$ & \nodata \\ 
4.0--12.0 & $<45.9\ (2\sigma)$ & $<11.9\ (2\sigma)$ & $7.7\pm2.1$ \\ 
0.3--12.0 & $6.0\pm0.4$ & $4.6\pm0.4$ & $6.2\pm0.4$
\enddata
\tablecomments{We explored the energy dependence of the QPO by fixing the centroid energy and FWHM for each respective QPO instance, and fitting for the normalization. We calculated $2\sigma$ upper limits if the uncertainty was larger than the best estimate. Energy bands for which no data are listed indicate non-convergent fits. The QPO instances occurred in GTIs 12, 21, and 6, respectively (see Table~\ref{tab:obsprops} for corresponding ObsID and MJDs).}
\end{deluxetable}

%% file: nustar_cospec.tex
\begin{deluxetable}{ccccc}[htbp!]
\tablecaption{Best-fit parameters from the power law fit to the NuSTAR cospectra defined from Figure \ref{fig:nustar_ps}. \label{tab:nustar_cospec}}
\tablehead{
 \colhead{Soft Color} &
 \colhead{PL Amplitude} &
 \colhead{$\alpha$} &
 \colhead{$f_{\rm rms}$ } &
 \colhead{$\chi^2$/d.o.f.} \\ 
 \colhead{} &
 \colhead{$10^{-4}$} &
 \colhead{} &
 \colhead{$\%$ } & 
 \colhead{} 
}
\startdata
All & $4.1\pm0.7$ & $1.15\pm0.10$ & $2.7\pm2.4$ & $51/69$ \\ 
$<0.6$ & $5.0\pm0.9$ & $0.87\pm0.10$ & $2.6\pm1.9$ & $94/69$ \\ 
$[0.6, 0.78]$ & $3.6\pm1.4$ & $1.26\pm0.22$ & $2.8\pm2.6$ & $62/69$ \\ 
$>0.78$ & $2.1\pm3.0$ & $1.7\pm0.7$ & $3.1\pm2.9$ & $82/69$  
\enddata
\tablecomments{The power law amplitudes and indices, the 0.1--50 Hz fractional RMS amplitudes ($\%$), and $\chi^2$ values are shown.}
\end{deluxetable}

%% file: R1_R2_spectra.tex
\begin{deluxetable}{cccc}[htbp!]
\tablecaption{0.6--10.0 keV spectral fits for the spectral regions R1 and R2, as defined in Figure \ref{fig:hid_ccd}. \label{tab:R1R2_spec}}
\tablehead{
 \colhead{Model} & 
 \colhead{Parameter} &
 \colhead{R1} &
 \colhead{R2} 
}

\startdata
\texttt{bbodyrad}  & $kT$ (keV) & $1.675_{-0.018}^{+0.032}$ & $0.468_{-0.016}^{+0.016}$ \\ 
                  & ${\rm norm}_{\rm bbodyrad}$ & $26.5_{-1.3}^{+1.3}$ & $178_{-23}^{+24}$ \\ 
\texttt{diskbb}  & $T_{\rm in}$ (keV) & $0.913_{-0.012}^{+0.019}$ & \nodata \\ 
                  & ${\rm norm}_{\rm diskbb}$ & $116_{-3}^{+6}$ & \nodata \\ 
\texttt{cutoffpl}  & $\Gamma$ & \nodata & $1.13_{-0.03}^{+0.03}$ \\ 
                  & $E_{\rm cut}$ (keV) & \nodata & $6.1_{-0.3}^{+0.4}$\\
                  & ${\rm norm}_{\rm cutoffpl}$ & \nodata & $0.1334_{-0.0013}^{+0.0021}$ \\
\texttt{gauss} (Fe XXV) & $E_{\rm XXV}$ (keV) & $6.72_{-0.04}^{+0.03}$ & \nodata \\ 
                  & $\sigma_{\rm XXV}$ (keV) & $0.001_{-0.001}^{+0.064}$ & \nodata \\ 
                  & ${\rm norm}_{\rm XXV}$ ($10^{-4}$) & $-1.8_{_-0.9}^{+0.9}$ & \nodata \\ 
\texttt{gauss} (K$\alpha$) & $E_{\rm K\alpha}$ (keV) & $6.985_{-0.012}^{+0.009}$ & $6.99_{-0.09}^{+0.07}$ \\ 
                  & $\sigma_{\rm K\alpha}$ (keV) & $0.02_{-0.02}^{+0.03}$ & Unconstrained  \\ 
                  & ${\rm norm}_{\rm K\alpha}$ ($10^{-4}$) & $-7.1_{-1.1}^{+0.9}$ & $-0.6_{-0.4}^{+0.4}$ \\ 
\texttt{gauss} (K$\beta$) & $E_{\rm K\beta}$ (keV) & $8.06_{-0.09}^{+0.08}$ & \nodata \\ 
                  & $\sigma_{\rm K\beta}$ (keV) & $0.5_{-0.3}^{+0.2}$ & \nodata \\ 
                  & ${\rm norm}_{\rm K\beta}$ ($10^{-3}$) & $-1.4_{-0.9}^{+0.8}$ & \nodata \\ 
\midrule 
\nodata & Flux\tablenotemark{a} & $2.949_{-0.009}^{+0.006}$ & $0.7453_{-0.0056}^{+0.0017}$ \\
\nodata & $\chi^2/{\rm d.o.f.}$ & 181/159 & 109/142 
\enddata
\tablenotetext{a}{Absorbed flux over 0.6--10.0 keV in units of $10^{-9}{\rm\,erg\,s^{-1}\,cm^{-2}}$}
\tablecomments{The uncertainties reported here are $90\%$ confidence limits. The spectral parameters are the blackbody temperature ($kT$, keV), blackbody normalization (scaled by $R_{\rm km}^2/D_{10}^2$, where $R_{\rm km}$ is the source radius in km and $D_{10}$ is the source distance in units of 10 kpc), temperature at the inner disk radius ($T_{\rm in}$, keV), and the disk blackbody normalization (scaled by $(R_{\rm in}/D_{10})^2\text{cos}(\theta)$; $R_{\rm in}$ is the apparent inner disk radius in km, and $\theta$ is the disk inclination angle (in degrees). We also have the power law photon index ($\Gamma$), the cutoff energy ($E_{\rm cut}$, keV), and the normalization (flux in photons/keV/cm$^2$/s at 1 keV). $E_{K_{\alpha}}$ and $E_{K_{\beta}}$ are the Gaussian centroid energies for the Fe XXVI K$\alpha$ and K$\beta$ absorption lines, respectively; $\sigma_{K_\alpha}$ and $\sigma_{K_\beta}$ refer to the same lines but describe the Gaussian width; ${\rm norm}_{K_\alpha}$ and ${\rm norm}_{K_\beta}$ refer to the same lines are the line normalizations in photons/s/cm$^2$.}
\end{deluxetable}

%% file: lines_eqw.tex
\begin{deluxetable}{cccccc}
  \tablecaption{Equivalent widths (in eV) for relevant lines.
  \label{tab:lines_eqw}}
  \tablewidth{0pt}
  \tablehead{
    \colhead{} &
    \colhead{} &
    \multicolumn{4}{c}{\uline{Line Energy}} \\
    \colhead{Region} &
    \colhead{Exposure (s)} &
    \colhead{1 keV} &
    \colhead{6.7 keV} &
    \colhead{7 keV} &
    \colhead{8 keV}
  }
  \startdata
    R1  & 34322 & $4.8_{-2.7}^{+2.7}$ & $-5.5_{-4.3}^{+5.3}$ & $-31_{-4}^{+6}$  & $-46_{-10}^{+14}$ \\
    R2  & 11451 & $<2.7$ & $>-20$ & $-14_{-9}^{+7}$ & $>-30$ \\
    CC1 & 9710 & $8.2_{-3.0}^{+4.0}$ & $-12_{-8}^{+6}$ & $-37_{-5}^{+6}$  & $-58_{-16}^{+16}$ \\
    CC2 & 20590 & $6.6_{-4.3}^{+3.5}$ & $>-7$ & $-29_{-4}^{+5}$  & $-36_{-13}^{+14}$ \\
    CC3 & 4022 & $2.6_{-2.5}^{+2.2}$ & $>-9$ & $-22_{-6}^{+6}$  & $>-40$ \\
    CC4 & 3323 & $4.9_{-2.5}^{+2.4}$ & $>-15$ & $-29_{-19}^{+14}$ & $>-55$ \\
    CC5 & 8128 & $<5$ & $>-25$ & $-12_{-12}^{+9}$ & $>-55$ \\
    CC6 & 1352  & $<20$  & \nodata  & \nodata & \nodata \\
  \enddata
\tablecomments{The uncertainties reported here are $90\%$ confidence limits. We report on the 1 keV emission line, the Fe XXV (6.7 keV), Fe XXVI K$\alpha$ (7 keV), and Fe XXVI K$\beta$ (8 keV) absorption lines for the different spectral regions defined in Figure \ref{fig:hid_ccd}. All upper limits shown are $2\sigma$ upper limits from fixing the line energy and width to the values from R1. There were no upper limits on the equivalent widths of the three absorption lines for spectral region CC6 as the background dominated past 5 keV.}
\end{deluxetable}

%% file: dip_persist_spectral_fit.tex
\begin{deluxetable}{cccc}[htbp!]
\tablecaption{0.6--10.0 keV spectral fits for the absorption dip interval and the persistent emission post-dip. \label{tab:dip_persist_spec}}
\tablehead{
 \colhead{Model} & 
 \colhead{Parameter} &
 \colhead{Dip} &
 \colhead{Persistent} 
}
\startdata
\texttt{tbpcf} & $n_{\rm{H,n}}$ ($10^{22}{\rm\,cm^{-2}}$) & $8.7_{-0.9}^{+0.8}$ & $4.8_{-1.4}^{+1.3}$ \\ 
& pcf & $0.714_{-0.015}^{+0.027}$ & $0.16_{-0.03}^{+0.06}$ \\ 
& Redshift & 0 & 0 \\
\texttt{zxipcf} & $N_H$ ($10^{22}{\rm\,cm^{-2}}$) & $100_{-18}^{+39}$ & $10$ (fixed) \\ 
                & \text{log}($\xi$) & $3.20_{-0.15}^{+0.12}$ & $4.37_{-0.11}^{+0.30}$ \\ 
                & $f_c$ & $0.62_{-0.13}^{+0.13}$ & $>0.74$ \\ 
                & $z$ & 0 & 0 \\ 
\texttt{diskbb}  & $T_{\rm in}$ (keV) & Tied & $0.81_{-0.04}^{+0.03}$ \\ 
                  & ${\rm norm}_{\rm diskbb}$ & Tied & $270_{-30}^{+60}$ \\ 
\texttt{bbodyrad}  & $kT$ (keV) & Tied & $1.33_{-0.03}^{+0.03}$  \\ 
                  & ${\rm norm}_{\rm bbodyrad}$ & Tied & $69_{-8}^{+10}$ \\ 
\midrule 
\nodata & $F_{0.6-10.0, \rm unabs}\ (10^{-9}$\ cgs) & $1.795_{-0.020}^{+0.019}$ & 
$3.870_{-0.013}^{+0.013}$ \\
$\chi^2/{\rm d.o.f.}$    &   \nodata      & \multicolumn{2}{c}{260/232} 
\enddata
\tablecomments{The uncertainties reported here are $90\%$ confidence limits. For interstellar absorption, we used the \texttt{tbabs} model \citep{wilms00} and fixed $n_H = 0.44\times10^{22}{\rm\,cm^{-2}}$ (see \S \ref{sec:continuum_evol} for more information). The variable $n_{\rm H,n}$ is the column density of the neutral absorber, ``pcf" is the partial covering fraction, $N_H$ is the column density of the (partial covering) partially ionized absorber, ${\rm log}(\xi)$ is the ionization parameter, $f_c$ is the covering fraction, and $z$ is the redshift. The unabsorbed 0.6--10.0~keV flux was determined using \texttt{cflux}.}
\end{deluxetable}


%% file: meerkat_flux.tex
\begin{deluxetable}{ccc}[ht]
\tablecaption{Measured MeerKAT flux densities ($\mu{\rm Jy}$) as a function of time for the stacked data points and per epoch (see description in \S \ref{sec:meerkat_results}). Upper limits are quoted to $3\sigma$. 
\label{tab:meerkat_flux}}
\tablehead{
 \colhead{Type} &
 \colhead{Date (MJD)} &
 \colhead{Flux ($\mu{\rm Jy}$)}
}
\startdata
Stacked & $59730.9693$ & $1248\pm23$ \\ 
& $59733.9204$ & $1066\pm23$ \\
& $59742.9526$ & $563\pm20$ \\
& $59757.2662$ & $72\pm11$ \\ 
& $59779.8157$ & $42\pm9$ \\ 
& $59808.2435$ & $<30$ \\
\hline
Epoch & $59730.9693$ & $1248\pm23$ \\ 
& $59733.9204$ & $1066\pm23$ \\
& $59742.9526$ & $563\pm20$ \\ 
& $59747.9011$ & $<53$ \\ 
& $59759.9651$ & $<59$ \\ 
& $59763.9325$ & $79\pm17$ \\ 
& $59769.8848$ & $<56$ \\ 
& $59775.8213$ & $<54$ \\ 
& $59782.8344$ & $<52$ \\ 
& $59790.7224$ & $<56$ \\ 
& $59797.9283$ & $<53$ \\
& $59804.6530$ & $<71$ \\
& $59811.7461$ & $<55$ \\ 
& $59818.6467$ & $<54$ 
\enddata
\end{deluxetable}

%% file: obsid_props_final.tex
\startlongtable
\begin{deluxetable}{ccccccccccc}
\centerwidetable
\tablecaption{Properties of the NICER observations throughout the outburst of \src. We presented the GTI numbers, corresponding ObsID, start time of the interval in MJD (TT units), the amount of filtered exposure (in s), the averaged total (source plus background) count rate (c/s), the background count rate (c/s) as modeled with the 3C50 model \citep{remillard22}, the number of averaged segments in the power spectra ($m$), the fractional RMS amplitude (in $\%$) over 0.03--50 Hz, the power law amplitude, the power law index, and the reduced $\chi^2$ between the model (power law plus constant) and the observed power spectrum. Entries with ellipses denote unconstrained values.  \label{tab:obsprops}}
\tablehead{
 \colhead{GTI} &
 \colhead{ObsID} &
 \colhead{MJD (TT)} &
 \colhead{Exposure} &
 \colhead{$r_{\rm tot}$} &
 \colhead{$r_{\rm bg}$} &
 \colhead{$m$} &
 \colhead{$f_{\rm rms}$} &
 \colhead{$A$} &
 \colhead{PL Index} &
 \colhead{$\chi^2$} \\
 \colhead{} &
 \colhead{} &
 \colhead{} &
 \colhead{s} &
 \colhead{c/s} &
 \colhead{c/s} &
 \colhead{} &
 \colhead{$\%$ (0.03--50 Hz)} &
 \colhead{} &
 \colhead{} &
 \colhead{}
}
\startdata
0 & 5202800101 & 59733.826 & 54.0 & $780.3\pm3.8$ & $1.5\pm0.17$ & 1 & $5.5\pm2.6$ & $-0.66\pm0.42$ & $0.15\pm0.22$ & 0.53 \\
1 & 5202800102 & 59733.993 & 862.0 & $811.5\pm1.0$ & $2.47\pm0.01$ & 26 & $9.1\pm2.0$ & $0.72\pm0.05$ & $1.11\pm0.04$ & 0.91 \\
2 & 5202800102 & 59734.267 & 277.0 & $828.3\pm1.8$ & $2.47\pm0.02$ & 8 & $8.2\pm2.2$ & $0.5\pm0.1$ & $1.3\pm0.1$ & 0.7 \\
3 & 5202800102 & 59734.332 & 547.0 & $805.0\pm1.2$ & $2.47\pm0.01$ & 17 & $9.1\pm2.1$ & $0.72\pm0.07$ & $1.17\pm0.05$ & 0.99 \\
4 & 5202800102 & 59734.445 & 1053.0 & $934.4\pm0.9$ & $2.48\pm0.01$ & 32 & $7.5\pm1.9$ & $0.62\pm0.05$ & $1.24\pm0.04$ & 0.72 \\
5 & 5202800103 & 59735.094 & 83.0 & $915.5\pm3.3$ & $2.3\pm0.01$ & 2 & $6.8\pm2.3$ & $-0.23\pm0.13$ & $0.52\pm0.36$ & 0.7 \\
6 & 5202800103 & 59735.696 & 306.0 & $809.9\pm1.6$ & $0.774\pm0.004$ & 9 & $5.6\pm2.0$ & $0.29\pm0.08$ & $1.28\pm0.13$ & 0.8 \\
7 & 5202800103 & 59735.759 & 406.0 & $656.8\pm1.3$ & $0.782\pm0.003$ & 12 & $5.7\pm2.1$ & $0.17\pm0.06$ & $1.38\pm0.14$ & 0.9 \\
8 & 5202800104 & 59736.15 & 121.0 & $866.3\pm2.7$ & $1.04\pm0.03$ & 3 & $7.8\pm2.1$ & $0.0\pm0.01$ & $2.86\pm1.61$ & 0.79 \\
9 & 5202800104 & 59736.922 & 38.0 & $783.3\pm4.5$ & $1.07\pm0.01$ & 1 & $6.5\pm2.6$ & $0.01\pm0.08$ & $1.94\pm1.98$ & 0.57 \\
10 & 5202800104 & 59736.924 & 140.0 & $773.7\pm2.4$ & $1.08\pm0.03$ & 4 & $6.7\pm2.2$ & $0.11\pm0.09$ & $1.6\pm0.34$ & 0.82 \\
11 & 5202800105 & 59737.05 & 165.0 & $790.5\pm2.2$ & $1.23\pm0.01$ & 5 & $6.7\pm2.1$ & $0.24\pm0.11$ & $1.28\pm0.2$ & 0.81 \\
12 & 5202800106 & 59738.194 & 779.0 & $795.3\pm1.0$ & $0.98\pm0.01$ & 24 & $12.9\pm4.5$ & $0.44\pm0.06$ & $1.62\pm0.08$ & 1.86 \\
13 & 5202800106 & 59738.215 & 41.0 & $993.2\pm4.9$ & $1.0\pm0.03$ & 1 & $3.4\pm2.3$ & $-0.82\pm0.13$ & $0.38\pm0.09$ & 0.48 \\
14 & 5202800106 & 59738.713 & 176.0 & $838.2\pm2.2$ & $2.57\pm0.08$ & 5 & $3.5\pm1.9$ & $0.08\pm0.09$ & $1.15\pm0.37$ & 0.93 \\
15 & 5202800107 & 59739.027 & 131.0 & $640.7\pm2.2$ & $0.88\pm0.01$ & 4 & $4.1\pm2.2$ & $-0.11\pm0.08$ & $0.33\pm0.43$ & 0.71 \\
16 & 5202800107 & 59739.041 & 1138.0 & $666.6\pm0.8$ & $0.877\pm0.004$ & 35 & $4.1\pm2.0$ & $0.12\pm0.03$ & $1.19\pm0.11$ & 0.91 \\
17 & 5202800107 & 59739.092 & 277.0 & $710.2\pm1.7$ & $0.74\pm0.01$ & 8 & $4.2\pm2.0$ & $0.01\pm0.03$ & $1.73\pm0.83$ & 0.81 \\
18 & 5202800107 & 59739.108 & 995.0 & $572.1\pm0.8$ & $0.741\pm0.004$ & 31 & $5.4\pm2.1$ & $0.15\pm0.04$ & $1.08\pm0.1$ & 0.96 \\
19 & 5202800107 & 59739.156 & 1030.0 & $580.8\pm0.8$ & $0.741\pm0.004$ & 32 & $5.2\pm2.1$ & $0.15\pm0.04$ & $1.06\pm0.1$ & 0.78 \\
20 & 5202800107 & 59739.182 & 158.0 & $678.0\pm2.1$ & $0.74\pm0.01$ & 4 & $4.2\pm2.2$ & $-0.15\pm0.08$ & $0.52\pm0.28$ & 0.78 \\
21 & 5202800107 & 59739.221 & 1470.0 & $901.5\pm0.8$ & $0.879\pm0.003$ & 45 & $7.3\pm1.8$ & $0.46\pm0.04$ & $1.07\pm0.04$ & 1.48 \\
22 & 5202800108 & 59741.693 & 579.0 & $656.4\pm1.1$ & $1.06\pm0.01$ & 18 & $4.3\pm2.0$ & $0.1\pm0.04$ & $1.22\pm0.16$ & 0.91 \\
23 & 5202800109 & 59742.533 & 517.0 & $636.1\pm1.1$ & $1.06\pm0.01$ & 16 & $4.9\pm2.1$ & $0.05\pm0.03$ & $1.52\pm0.24$ & 1.1 \\
24 & 5202800110 & 59743.756 & 775.0 & $680.9\pm0.9$ & $1.25\pm0.01$ & 24 & $2.9\pm2.0$ & $0.09\pm0.04$ & $1.17\pm0.16$ & 0.73 \\
25 & 5202800110 & 59743.821 & 773.0 & $757.4\pm1.0$ & $1.18\pm0.01$ & 24 & $5.0\pm1.9$ & $0.21\pm0.04$ & $1.04\pm0.09$ & 0.92 \\
26 & 5406620101 & 59745.544 & 599.0 & $664.6\pm1.1$ & $1.95\pm0.01$ & 18 & $2.9\pm2.0$ & $0.02\pm0.02$ & $1.51\pm0.43$ & 0.94 \\
27 & 5202800111 & 59745.819 & 900.0 & $732.3\pm0.9$ & $1.171\pm0.004$ & 28 & $3.7\pm1.9$ & $0.19\pm0.04$ & $0.96\pm0.08$ & 0.95 \\
28 & 5202800112 & 59747.048 & 676.0 & $632.7\pm1.0$ & $1.322\pm0.005$ & 21 & $3.8\pm2.0$ & $0.02\pm0.03$ & $1.37\pm0.38$ & 0.93 \\
29 & 5202800112 & 59747.492 & 1267.0 & $756.3\pm0.8$ & $1.322\pm0.004$ & 39 & $4.0\pm1.9$ & $0.16\pm0.03$ & $1.12\pm0.08$ & 1.05 \\
30 & 5202800113 & 59748.082 & 492.0 & $743.6\pm1.2$ & $1.34\pm0.01$ & 15 & $5.3\pm1.9$ & $0.08\pm0.05$ & $1.0\pm0.21$ & 0.88 \\
31 & 5202800113 & 59748.383 & 497.0 & $626.9\pm1.1$ & $1.08\pm0.01$ & 15 & $3.3\pm2.1$ & $-0.0\pm0.04$ & $0.23\pm5.71$ & 0.93 \\
32 & 5202800113 & 59748.392 & 565.0 & $619.7\pm1.0$ & $1.08\pm0.01$ & 17 & $3.6\pm2.1$ & $0.0\pm0.01$ & $1.91\pm0.91$ & 0.84 \\
33 & 5202800113 & 59748.785 & 1085.0 & $687.3\pm0.8$ & $1.08\pm0.01$ & 33 & $3.2\pm1.9$ & $0.08\pm0.03$ & $1.21\pm0.15$ & 0.98 \\
34 & 5202800114 & 59749.415 & 377.0 & $769.3\pm1.4$ & $1.075\pm0.004$ & 11 & $2.0\pm1.9$ & $0.07\pm0.05$ & $1.48\pm0.29$ & 0.79 \\
35 & 5202800115 & 59752.385 & 67.0 & $622.8\pm3.0$ & $1.2\pm0.01$ & 1 & \nodata & $-0.72\pm0.14$ & $0.33\pm0.1$ & 0.52 \\
36 & 5202800115 & 59752.393 & 114.0 & $613.4\pm2.3$ & $1.2\pm0.01$ & 3 & \nodata  & $-0.36\pm0.07$ & $0.46\pm0.1$ & 0.7 \\
39 & 5202800116 & 59753.353 & 46.0 & $765.8\pm4.1$ & $1.19\pm0.01$ & 1 & \nodata & $-0.76\pm0.14$ & $0.4\pm0.12$ & 0.48 \\
40 & 5202800116 & 59753.873 & 346.0 & $699.5\pm1.4$ & $1.179\pm0.004$ & 10 & $5.3\pm2.0$ & $0.04\pm0.04$ & $1.43\pm0.38$ & 0.97 \\
41 & 5202800117 & 59755.628 & 561.0 & $707.9\pm1.1$ & $1.03\pm0.01$ & 17 & $5.1\pm2.0$ & $0.11\pm0.05$ & $1.0\pm0.19$ & 0.81 \\
44 & 5202800118 & 59756.392 & 115.0 & $630.1\pm2.3$ & $1.39\pm0.01$ & 3 & $4.5\pm2.3$ & $-0.13\pm0.09$ & $0.29\pm0.45$ & 0.65 \\
46 & 5202800119 & 59757.158 & 805.0 & $715.1\pm0.9$ & $1.86\pm0.01$ & 24 & $5.2\pm1.9$ & $0.1\pm0.04$ & $1.17\pm0.14$ & 0.93 \\
47 & 5202800119 & 59757.293 & 343.0 & $692.9\pm1.4$ & $1.062\pm0.004$ & 10 & $3.0\pm2.0$ & \nodata & $0.0\pm0.59$ & 0.96 \\
49 & 5202800120 & 59758.457 & 87.0 & $651.7\pm2.7$ & $0.845\pm0.003$ & 2 & $7.1\pm2.4$ & $-0.37\pm0.1$ & $0.45\pm0.16$ & 0.69 \\
51 & 5202800120 & 59758.9 & 90.0 & $667.7\pm2.7$ & $1.128\pm0.004$ & 2 & $6.7\pm2.4$ & $-0.34\pm0.11$ & $0.7\pm0.14$ & 0.67 \\
52 & 5202800120 & 59758.964 & 206.0 & $631.9\pm1.8$ & $1.128\pm0.004$ & 6 & $5.6\pm2.2$ & $0.0\pm0.01$ & $2.04\pm2.88$ & 0.85 \\
53 & 5202800121 & 59759.029 & 327.0 & $607.2\pm1.4$ & $1.05\pm0.01$ & 10 & \nodata & $-0.15\pm0.04$ & $0.41\pm0.16$ & 1.03 \\
54 & 5202800121 & 59759.095 & 143.0 & $653.2\pm2.1$ & $1.0\pm0.01$ & 4 & $5.0\pm2.2$ & $-0.15\pm0.07$ & $0.58\pm0.23$ & 0.78 \\
55 & 5202800121 & 59759.759 & 697.0 & $636.8\pm1.0$ & $2.86\pm0.02$ & 21 & $4.4\pm2.0$ & $0.07\pm0.04$ & $1.17\pm0.2$ & 0.64 \\
56 & 5202800121 & 59759.808 & 423.0 & $592.4\pm1.2$ & $2.86\pm0.02$ & 13 & $3.9\pm2.2$ & $0.03\pm0.03$ & $1.6\pm0.34$ & 0.72 \\
57 & 5202800122 & 59760.648 & 406.0 & $666.4\pm1.3$ & $3.21\pm0.01$ & 12 & $2.4\pm2.0$ & $-0.03\pm0.04$ & $0.3\pm0.93$ & 0.94 \\
58 & 5202800122 & 59760.712 & 307.0 & $658.9\pm1.5$ & $3.22\pm0.01$ & 9 & $3.2\pm2.1$ & $0.01\pm0.02$ & $1.91\pm0.74$ & 0.68 \\
59 & 5202800123 & 59761.762 & 273.0 & $613.9\pm1.5$ & $0.88\pm0.01$ & 8 & $3.1\pm2.1$ & \nodata & $0.01\pm0.49$ & 0.73 \\
61 & 5202800124 & 59762.39 & 439.0 & $642.2\pm1.2$ & $1.24\pm0.01$ & 13 & $3.7\pm2.1$ & \nodata & $0.0\pm1.33$ & 0.92 \\
62 & 5202800124 & 59762.967 & 313.0 & $578.9\pm1.4$ & $1.24\pm0.01$ & 9 & $1.0\pm2.2$ & $0.03\pm0.05$ & $1.22\pm0.68$ & 0.75 \\
63 & 5202800125 & 59764.126 & 393.0 & $621.0\pm1.3$ & $1.36\pm0.01$ & 11 & $1.9\pm2.1$ & $0.02\pm0.03$ & $1.64\pm0.42$ & 0.88 \\
64 & 5202800125 & 59764.193 & 627.0 & $629.2\pm1.0$ & $1.38\pm0.01$ & 19 & $3.4\pm2.1$ & $0.07\pm0.04$ & $1.22\pm0.21$ & 0.75 \\
65 & 5202800126 & 59765.892 & 349.0 & $582.1\pm1.3$ & $2.7\pm0.03$ & 10 & $4.5\pm2.2$ & $0.03\pm0.05$ & $1.24\pm0.53$ & 0.71 \\
66 & 5202800126 & 59765.933 & 43.0 & $618.1\pm3.8$ & $2.73\pm0.03$ & 1 & \nodata & $-1.08\pm0.11$ & $0.41\pm0.05$ & 0.59 \\
67 & 5202800126 & 59765.934 & 140.0 & $588.6\pm2.1$ & $2.73\pm0.03$ & 3 & $5.0\pm2.4$ & $-0.31\pm0.08$ & $0.38\pm0.13$ & 0.69 \\
69 & 5202800127 & 59766.666 & 294.0 & $616.5\pm1.4$ & $0.93\pm0.01$ & 9 & $5.9\pm2.2$ & \nodata & $0.0\pm1.21$ & 0.96 \\
70 & 5202800127 & 59766.731 & 314.0 & $615.2\pm1.4$ & $0.93\pm0.01$ & 9 & $5.2\pm2.2$ & $0.04\pm0.05$ & $1.32\pm0.43$ & 0.83 \\
71 & 5202800128 & 59767.291 & 613.0 & $634.9\pm1.0$ & \nodata & 19 & \nodata & $0.04\pm0.04$ & $1.26\pm0.28$ & 0.89 \\
73 & 5202800128 & 59767.612 & 107.0 & $602.4\pm2.4$ & $2.11\pm0.01$ & 3 & \nodata & $-0.3\pm0.08$ & $0.46\pm0.08$ & 0.67 \\
74 & 5202800128 & 59767.614 & 864.0 & $585.9\pm0.8$ & \nodata & 26 & $3.3\pm2.1$ & $0.0\pm0.01$ & $1.85\pm0.75$ & 0.81 \\
75 & 5202800129 & 59768.149 & 328.0 & $661.1\pm1.4$ & $2.1\pm0.02$ & 9 & $1.9\pm2.1$ & $0.1\pm0.06$ & $1.18\pm0.22$ & 0.79 \\
76 & 5202800129 & 59768.195 & 634.0 & $683.5\pm1.0$ & $2.1\pm0.02$ & 19 & $3.1\pm2.0$ & $0.05\pm0.04$ & $1.23\pm0.23$ & 0.88 \\
77 & 5202800130 & 59769.035 & 334.0 & $599.7\pm1.3$ & $0.871\pm0.004$ & 10 & $3.9\pm2.2$ & $-0.06\pm0.07$ & $0.18\pm0.6$ & 1.08 \\
78 & 5202800130 & 59769.099 & 548.0 & $648.0\pm1.1$ & $0.862\pm0.003$ & 15 & $5.7\pm2.1$ & $0.08\pm0.05$ & $0.98\pm0.22$ & 0.87 \\
79 & 5202800130 & 59769.168 & 199.0 & $659.6\pm1.8$ & $0.862\pm0.005$ & 6 & $1.3\pm2.1$ & $-0.06\pm0.31$ & $0.11\pm1.11$ & 0.83 \\
80 & 5202800131 & 59771.233 & 372.0 & $603.7\pm1.3$ & $2.22\pm0.01$ & 11 & $0.8\pm2.1$ & \nodata & $0.0\pm0.95$ & 0.79 \\
81 & 5202800131 & 59771.297 & 375.0 & $548.7\pm1.2$ & $2.22\pm0.01$ & 11 & \nodata & \nodata & $0.01\pm0.6$ & 0.93 \\
82 & 5202800131 & 59771.362 & 373.0 & $608.2\pm1.3$ & $0.776\pm0.003$ & 11 & $3.9\pm2.1$ & $-0.03\pm0.04$ & $0.49\pm0.77$ & 0.95 \\
83 & 5202800132 & 59783.974 & 311.0 & $491.9\pm1.3$ & $1.04\pm0.02$ & 9 & $7.6\pm2.4$ & $0.03\pm0.05$ & $1.15\pm0.68$ & 0.75 \\
84 & 5202800133 & 59784.039 & 298.0 & $503.7\pm1.3$ & $1.04\pm0.02$ & 9 & \nodata & $-0.12\pm0.06$ & $0.2\pm0.28$ & 0.92 \\
85 & 5202800133 & 59784.426 & 359.0 & $522.3\pm1.2$ & $0.94\pm0.01$ & 11 & $4.9\pm2.3$ & $-0.03\pm0.08$ & $0.16\pm1.07$ & 0.99 \\
86 & 5202800134 & 59785.071 & 394.0 & $510.1\pm1.1$ & $1.03\pm0.01$ & 12 & $3.7\pm2.3$ & $-0.01\pm0.04$ & $0.26\pm1.86$ & 0.84 \\
87 & 5202800134 & 59785.393 & 318.0 & $499.4\pm1.3$ & $2.66\pm0.01$ & 9 & $4.2\pm2.4$ & $-0.03\pm0.05$ & $0.35\pm0.77$ & 0.63 \\
89 & 5202800135 & 59793.183 & 550.0 & $410.0\pm0.9$ & $1.0\pm0.02$ & 17 & $2.8\pm2.5$ & \nodata & $0.0\pm0.44$ & 1.21 \\
90 & 5202800136 & 59794.108 & 83.0 & $413.9\pm2.2$ & $1.02\pm0.02$ & 2 & $3.5\pm3.0$ & $-0.63\pm0.09$ & $0.38\pm0.08$ & 0.88 \\
91 & 5202800137 & 59795.721 & 43.0 & $409.9\pm3.1$ & $1.08\pm0.02$ & 1 & $0.3\pm3.5$ & $-1.3\pm0.11$ & $0.35\pm0.05$ & 0.8 \\
92 & 5202800138 & 59796.79 & 1159.0 & $392.9\pm0.6$ & $2.41\pm0.01$ & 36 & \nodata & $0.01\pm0.02$ & $1.24\pm0.78$ & 0.85 \\
93 & 5202800139 & 59797.563 & 443.0 & $368.4\pm0.9$ & $2.43\pm0.01$ & 13 & \nodata & $-0.01\pm0.02$ & $1.39\pm0.99$ & 0.64 \\
5202800140 & 5202800140 & 59804.593 & 2261.4 & $213.6\pm0.3$ & $0.66\pm0.02$ & 68 & $3.1\pm3.4$ & $0.01\pm0.02$ & $1.24\pm0.51$ & 0.82 \\
5202800141 & 5202800141 & 59805.496 & 1816.0 & $204.3\pm0.3$ & $0.6\pm0.02$ & 56 & \nodata & $-0.04\pm0.08$ & $0.11\pm0.46$ & 1.03 \\
5202800142 & 5202800142 & 59806.98 & 664.0 & $182.8\pm0.5$ & $0.62\pm0.03$ & 20 & $2.4\pm3.8$ & $-0.11\pm0.2$ & $0.1\pm0.31$ & 0.9 \\
5202800143 & 5202800143 & 59807.496 & 1745.0 & $162.5\pm0.3$ & $0.58\pm0.02$ & 54 & \nodata & $-0.03\pm0.02$ & $0.57\pm0.25$ & 0.99 \\
5202800144 & 5202800144 & 59808.657 & 1728.0 & $132.8\pm0.3$ & $0.68\pm0.02$ & 52 & $0.8\pm4.4$ & \nodata & $0.0\pm1.18$ & 1.02 \\
5202800145 & 5202800145 & 59809.753 & 1032.1 & $89.6\pm0.3$ & $1.62\pm0.04$ & 31 & $6.9\pm5.4$ & $0.06\pm0.03$ & $0.7\pm0.19$ & 1.0 \\
5202800146 & 5202800146 & 59810.721 & 1092.0 & $52.4\pm0.2$ & $1.45\pm0.04$ & 33 & $8.4\pm7.0$ & $-0.06\pm0.32$ & $0.07\pm0.57$ & 0.82 \\
5202800147 & 5202800147 & 59811.689 & 525.0 & $32.8\pm0.3$ & $1.17\pm0.05$ & 16 & \nodata & $-0.09\pm0.04$ & $0.58\pm0.18$ & 1.04 \\
5202800148 & 5202800148 & 59813.495 & 589.0 & $18.8\pm0.2$ & $0.78\pm0.04$ & 18 & $8.8\pm11.8$ & $-0.08\pm0.03$ & $0.63\pm0.16$ & 1.14 \\
5202800160 & 5202800160 & 59814.334 & 1726.0 & $9.3\pm0.1$ & $0.77\pm0.02$ & 75 & $20.9\pm14.1$ & $0.01\pm0.01$ & $1.53\pm0.27$ & 0.9 \\
\enddata
\end{deluxetable}

%% file: 1a1744.bbl
\begin{thebibliography}{}
\expandafter\ifx\csname natexlab\endcsname\relax\def\natexlab#1{#1}\fi
\providecommand{\url}[1]{\href{#1}{#1}}
\providecommand{\dodoi}[1]{doi:~\href{http://doi.org/#1}{\nolinkurl{#1}}}
\providecommand{\doeprint}[1]{\href{http://ascl.net/#1}{\nolinkurl{http://ascl.net/#1}}}
\providecommand{\doarXiv}[1]{\href{https://arxiv.org/abs/#1}{\nolinkurl{https://arxiv.org/abs/#1}}}

\bibitem[{{Akaike}(1974)}]{akaike74}
{Akaike}, H. 1974, IEEE Transactions on Automatic Control, 19, 716

\bibitem[{{Alpar} {et~al.}(1992){Alpar}, {Hasinger}, {Shaham}, \&
  {Yancopoulos}}]{alpar92}
{Alpar}, M.~A., {Hasinger}, G., {Shaham}, J., \& {Yancopoulos}, S. 1992, \aap,
  257, 627

\bibitem[{{Altamirano} {et~al.}(2008){Altamirano}, {van der Klis},
  {M{\'e}ndez}, {Jonker}, {Klein-Wolt}, \& {Lewin}}]{altamirano08}
{Altamirano}, D., {van der Klis}, M., {M{\'e}ndez}, M., {et~al.} 2008, \apj,
  685, 436, \dodoi{10.1086/590897}

\bibitem[{{Altamirano} {et~al.}(2010){Altamirano}, {Homan}, {Linares},
  {Patruno}, {Yang}, {Watts}, {Kalamkar}, {Casella}, {Armas-Padilla},
  {Cavecchi}, {Degenaar}, {Russell}, {Kaur}, {van der Klis}, {Rea}, \&
  {Wijnands}}]{altamirano10}
{Altamirano}, D., {Homan}, J., {Linares}, M., {et~al.} 2010, The Astronomer's
  Telegram, 2952, 1

\bibitem[{{Arcodia} {et~al.}(2018){Arcodia}, {Campana}, {Salvaterra}, \&
  {Ghisellini}}]{arcodia18}
{Arcodia}, R., {Campana}, S., {Salvaterra}, R., \& {Ghisellini}, G. 2018, \aap,
  616, A170, \dodoi{10.1051/0004-6361/201732322}

\bibitem[{{Arnaud}(1996)}]{arnaud96}
{Arnaud}, K.~A. 1996, in Astronomical Society of the Pacific Conference Series,
  Vol. 101, Astronomical Data Analysis Software and Systems V, ed. G.~H.
  {Jacoby} \& J.~{Barnes}, 17

\bibitem[{{Astropy Collaboration} {et~al.}(2013){Astropy Collaboration},
  {Robitaille}, {Tollerud}, {Greenfield}, {Droettboom}, {Bray}, {Aldcroft},
  {Davis}, {Ginsburg}, {Price-Whelan}, {Kerzendorf}, {Conley}, {Crighton},
  {Barbary}, {Muna}, {Ferguson}, {Grollier}, {Parikh}, {Nair}, {Unther},
  {Deil}, {Woillez}, {Conseil}, {Kramer}, {Turner}, {Singer}, {Fox}, {Weaver},
  {Zabalza}, {Edwards}, {Azalee Bostroem}, {Burke}, {Casey}, {Crawford},
  {Dencheva}, {Ely}, {Jenness}, {Labrie}, {Lim}, {Pierfederici}, {Pontzen},
  {Ptak}, {Refsdal}, {Servillat}, \& {Streicher}}]{astropy:2013}
{Astropy Collaboration}, {Robitaille}, T.~P., {Tollerud}, E.~J., {et~al.} 2013,
  \aap, 558, A33, \dodoi{10.1051/0004-6361/201322068}

\bibitem[{{Astropy Collaboration} {et~al.}(2018){Astropy Collaboration},
  {Price-Whelan}, {Sip{\H{o}}cz}, {G{\"u}nther}, {Lim}, {Crawford}, {Conseil},
  {Shupe}, {Craig}, {Dencheva}, {Ginsburg}, {Vand erPlas}, {Bradley},
  {P{\'e}rez-Su{\'a}rez}, {de Val-Borro}, {Aldcroft}, {Cruz}, {Robitaille},
  {Tollerud}, {Ardelean}, {Babej}, {Bach}, {Bachetti}, {Bakanov}, {Bamford},
  {Barentsen}, {Barmby}, {Baumbach}, {Berry}, {Biscani}, {Boquien}, {Bostroem},
  {Bouma}, {Brammer}, {Bray}, {Breytenbach}, {Buddelmeijer}, {Burke},
  {Calderone}, {Cano Rodr{\'\i}guez}, {Cara}, {Cardoso}, {Cheedella}, {Copin},
  {Corrales}, {Crichton}, {D'Avella}, {Deil}, {Depagne}, {Dietrich}, {Donath},
  {Droettboom}, {Earl}, {Erben}, {Fabbro}, {Ferreira}, {Finethy}, {Fox},
  {Garrison}, {Gibbons}, {Goldstein}, {Gommers}, {Greco}, {Greenfield},
  {Groener}, {Grollier}, {Hagen}, {Hirst}, {Homeier}, {Horton}, {Hosseinzadeh},
  {Hu}, {Hunkeler}, {Ivezi{\'c}}, {Jain}, {Jenness}, {Kanarek}, {Kendrew},
  {Kern}, {Kerzendorf}, {Khvalko}, {King}, {Kirkby}, {Kulkarni}, {Kumar},
  {Lee}, {Lenz}, {Littlefair}, {Ma}, {Macleod}, {Mastropietro}, {McCully},
  {Montagnac}, {Morris}, {Mueller}, {Mumford}, {Muna}, {Murphy}, {Nelson},
  {Nguyen}, {Ninan}, {N{\"o}the}, {Ogaz}, {Oh}, {Parejko}, {Parley}, {Pascual},
  {Patil}, {Patil}, {Plunkett}, {Prochaska}, {Rastogi}, {Reddy Janga},
  {Sabater}, {Sakurikar}, {Seifert}, {Sherbert}, {Sherwood-Taylor}, {Shih},
  {Sick}, {Silbiger}, {Singanamalla}, {Singer}, {Sladen}, {Sooley},
  {Sornarajah}, {Streicher}, {Teuben}, {Thomas}, {Tremblay}, {Turner},
  {Terr{\'o}n}, {van Kerkwijk}, {de la Vega}, {Watkins}, {Weaver}, {Whitmore},
  {Woillez}, {Zabalza}, \& {Astropy Contributors}}]{astropy:2018}
{Astropy Collaboration}, {Price-Whelan}, A.~M., {Sip{\H{o}}cz}, B.~M., {et~al.}
  2018, \aj, 156, 123, \dodoi{10.3847/1538-3881/aabc4f}

\bibitem[{{Bahramian} \& {Degenaar}(2022)}]{bahramian22}
{Bahramian}, A., \& {Degenaar}, N. 2022, arXiv e-prints, arXiv:2206.10053,
  \dodoi{10.48550/arXiv.2206.10053}

\bibitem[{Bahramian \& Rushton(2022)}]{bahramian22_lrlx}
Bahramian, A., \& Rushton, A. 2022, bersavosh/XRB-LrLx\_pub: update 20220908,
  v220908,  Zenodo, \dodoi{10.5281/zenodo.7059313}

\bibitem[{{Ba{\l}uci{\'n}ska-Church} {et~al.}(2011){Ba{\l}uci{\'n}ska-Church},
  {Schulz}, {Wilms}, {Gibiec}, {Hanke}, {Spencer}, {Rushton}, \&
  {Church}}]{balucinskachurch11}
{Ba{\l}uci{\'n}ska-Church}, M., {Schulz}, N.~S., {Wilms}, J., {et~al.} 2011,
  \aap, 530, A102, \dodoi{10.1051/0004-6361/201015931}

\bibitem[{{Belloni} {et~al.}(2002){Belloni}, {Psaltis}, \& {van der
  Klis}}]{belloni02}
{Belloni}, T., {Psaltis}, D., \& {van der Klis}, M. 2002, \apj, 572, 392,
  \dodoi{10.1086/340290}

\bibitem[{Bendat \& Piersol(2011)}]{bendat11}
Bendat, J.~S., \& Piersol, A.~G. 2011, Random data: analysis and measurement
  procedures (John Wiley \& Sons)

\bibitem[{{Bhattacharyya} {et~al.}(2006{\natexlab{a}}){Bhattacharyya},
  {Strohmayer}, {Markwardt}, \& {Swank}}]{bhattacharyya06a}
{Bhattacharyya}, S., {Strohmayer}, T.~E., {Markwardt}, C.~B., \& {Swank}, J.~H.
  2006{\natexlab{a}}, \apjl, 639, L31, \dodoi{10.1086/501438}

\bibitem[{{Bhattacharyya} {et~al.}(2006{\natexlab{b}}){Bhattacharyya},
  {Strohmayer}, {Swank}, \& {Markwardt}}]{bhattacharyya06b}
{Bhattacharyya}, S., {Strohmayer}, T.~E., {Swank}, J.~H., \& {Markwardt}, C.~B.
  2006{\natexlab{b}}, \apj, 652, 603, \dodoi{10.1086/507786}

\bibitem[{{Briggs}(1995)}]{briggs95}
{Briggs}, D.~S. 1995, in American Astronomical Society Meeting Abstracts, Vol.
  187, American Astronomical Society Meeting Abstracts, 112.02

\bibitem[{Bright {et~al.}(2020)}]{bright2020}
Bright, J.~S., {et~al.} 2020, Nature Astron., 4, 697,
  \dodoi{10.1038/s41550-020-1023-5}

\bibitem[{{Buisson} {et~al.}(2021){Buisson}, {Altamirano}, {Armas Padilla},
  {Arzoumanian}, {Bult}, {Castro Segura}, {Charles}, {Degenaar}, {D{\'\i}az
  Trigo}, {van den Eijnden}, {Fogantini}, {Gandhi}, {Gendreau}, {Hare},
  {Homan}, {Knigge}, {Malacaria}, {Mendez}, {Mu{\~n}oz Darias}, {Ng},
  {{\"O}zbey Arabac{\i}}, {Remillard}, {Strohmayer}, {Tombesi}, {Tomsick},
  {Vincentelli}, \& {Walton}}]{buisson21}
{Buisson}, D.~J.~K., {Altamirano}, D., {Armas Padilla}, M., {et~al.} 2021,
  \mnras, 503, 5600, \dodoi{10.1093/mnras/stab863}

\bibitem[{{Bult} {et~al.}(2021){Bult}, {Strohmayer}, {Malacaria}, {Ng}, \&
  {Wadiasingh}}]{bult21}
{Bult}, P., {Strohmayer}, T.~E., {Malacaria}, C., {Ng}, M., \& {Wadiasingh}, Z.
  2021, \apj, 912, 120, \dodoi{10.3847/1538-4357/abf13f}

\bibitem[{{Bult} \& {van der Klis}(2015)}]{bult15}
{Bult}, P., \& {van der Klis}, M. 2015, \apj, 806, 90,
  \dodoi{10.1088/0004-637X/806/1/90}

\bibitem[{{Burrows} {et~al.}(2005){Burrows}, {Hill}, {Nousek}, {Kennea},
  {Wells}, {Osborne}, {Abbey}, {Beardmore}, {Mukerjee}, {Short}, {Chincarini},
  {Campana}, {Citterio}, {Moretti}, {Pagani}, {Tagliaferri}, {Giommi},
  {Capalbi}, {Tamburelli}, {Angelini}, {Cusumano}, {Br{\"a}uninger}, {Burkert},
  \& {Hartner}}]{burrows05}
{Burrows}, D.~N., {Hill}, J.~E., {Nousek}, J.~A., {et~al.} 2005, \ssr, 120,
  165, \dodoi{10.1007/s11214-005-5097-2}

\bibitem[{{Carpenter} {et~al.}(1977){Carpenter}, {Eyles}, {Skinner}, {Wilson},
  \& {Willmore}}]{carpenter77}
{Carpenter}, G.~F., {Eyles}, C.~J., {Skinner}, G.~K., {Wilson}, A.~M., \&
  {Willmore}, A.~P. 1977, \mnras, 179, 27P, \dodoi{10.1093/mnras/179.1.27P}

\bibitem[{{CASA Team} {et~al.}(2022){CASA Team}, {Bean}, {Bhatnagar}, {Castro},
  {Donovan Meyer}, {Emonts}, {Garcia}, {Garwood}, {Golap}, {Gonzalez Villalba},
  {Harris}, {Hayashi}, {Hoskins}, {Hsieh}, {Jagannathan}, {Kawasaki},
  {Keimpema}, {Kettenis}, {Lopez}, {Marvil}, {Masters}, {McNichols},
  {Mehringer}, {Miel}, {Moellenbrock}, {Montesino}, {Nakazato}, {Ott}, {Petry},
  {Pokorny}, {Raba}, {Rau}, {Schiebel}, {Schweighart}, {Sekhar}, {Shimada},
  {Small}, {Steeb}, {Sugimoto}, {Suoranta}, {Tsutsumi}, {van Bemmel},
  {Verkouter}, {Wells}, {Xiong}, {Szomoru}, {Griffith}, {Glendenning}, \&
  {Kern}}]{casa22}
{CASA Team}, {Bean}, B., {Bhatnagar}, S., {et~al.} 2022, \pasp, 134, 114501,
  \dodoi{10.1088/1538-3873/ac9642}

\bibitem[{{Chakrabarty} {et~al.}(2003){Chakrabarty}, {Morgan}, {Muno},
  {Galloway}, {Wijnands}, {van der Klis}, \& {Markwardt}}]{chakrabarty03}
{Chakrabarty}, D., {Morgan}, E.~H., {Muno}, M.~P., {et~al.} 2003, \nat, 424,
  42, \dodoi{10.1038/nature01732}

\bibitem[{{Chakraborty} {et~al.}(2011){Chakraborty}, {Bhattacharyya}, \&
  {Mukherjee}}]{chakraborty11}
{Chakraborty}, M., {Bhattacharyya}, S., \& {Mukherjee}, A. 2011, \mnras, 418,
  490, \dodoi{10.1111/j.1365-2966.2011.19499.x}

\bibitem[{{Church} {et~al.}(1998){Church}, {Parmar}, {Balucinska-Church},
  {Oosterbroek}, {dal Fiume}, \& {Orlandini}}]{church98}
{Church}, M.~J., {Parmar}, A.~N., {Balucinska-Church}, M., {et~al.} 1998, \aap,
  338, 556, \dodoi{10.48550/arXiv.astro-ph/9806223}

\bibitem[{{Church} {et~al.}(2005){Church}, {Reed}, {Dotani},
  {Ba{\l}uci{\'n}ska-Church}, \& {Smale}}]{church05}
{Church}, M.~J., {Reed}, D., {Dotani}, T., {Ba{\l}uci{\'n}ska-Church}, M., \&
  {Smale}, A.~P. 2005, \mnras, 359, 1336,
  \dodoi{10.1111/j.1365-2966.2005.08728.x}

\bibitem[{{Coriat} {et~al.}(2011){Coriat}, {Corbel}, {Prat}, {Miller-Jones},
  {Cseh}, {Tzioumis}, {Brocksopp}, {Rodriguez}, {Fender}, \&
  {Sivakoff}}]{coriat2011}
{Coriat}, M., {Corbel}, S., {Prat}, L., {et~al.} 2011, \mnras, 414, 677,
  \dodoi{10.1111/j.1365-2966.2011.18433.x}

\bibitem[{da~Costa-Luis {et~al.}(2022)da~Costa-Luis, Larroque, Altendorf, Mary,
  richardsheridan, Korobov, Raphael, Ivanov, Bargull, Rodrigues, Chen, Lee,
  Newey, James, JC, Zugnoni, Pagel, mjstevens777, Dektyarev, Rothberg,
  Alexander, Panteleit, Dill, FichteFoll, Sturm, HeoHeo, van Kemenade,
  McCracken, \& MapleCCC}]{dacostaluis22}
da~Costa-Luis, C., Larroque, S.~K., Altendorf, K., {et~al.} 2022, {tqdm: A
  fast, Extensible Progress Bar for Python and CLI}, v4.64.0,  Zenodo,
  \dodoi{10.5281/zenodo.6412640}

\bibitem[{{Davison} {et~al.}(1976){Davison}, {Burnell}, {Ives}, {Wilson}, \&
  {Carpenter}}]{davison76}
{Davison}, P., {Burnell}, J., {Ives}, J., {Wilson}, A., \& {Carpenter}, G.
  1976, \iaucirc, 2925, 2

\bibitem[{{Degenaar} {et~al.}(2013){Degenaar}, {Miller}, {Wijnands},
  {Altamirano}, \& {Fabian}}]{degenaar13}
{Degenaar}, N., {Miller}, J.~M., {Wijnands}, R., {Altamirano}, D., \& {Fabian},
  A.~C. 2013, \apjl, 767, L37, \dodoi{10.1088/2041-8205/767/2/L37}

\bibitem[{{D{\'\i}az Trigo} {et~al.}(2006){D{\'\i}az Trigo}, {Parmar},
  {Boirin}, {M{\'e}ndez}, \& {Kaastra}}]{diaztrigo06}
{D{\'\i}az Trigo}, M., {Parmar}, A.~N., {Boirin}, L., {M{\'e}ndez}, M., \&
  {Kaastra}, J.~S. 2006, \aap, 445, 179, \dodoi{10.1051/0004-6361:20053586}

\bibitem[{{D{\'\i}az Trigo} {et~al.}(2009){D{\'\i}az Trigo}, {Parmar},
  {Boirin}, {Motch}, {Talavera}, \& {Balman}}]{diaztrigo09}
{D{\'\i}az Trigo}, M., {Parmar}, A.~N., {Boirin}, L., {et~al.} 2009, \aap, 493,
  145, \dodoi{10.1051/0004-6361:200810154}

\bibitem[{{D{\'\i}az Trigo} {et~al.}(2018){D{\'\i}az Trigo}, {Altamirano},
  {Din{\c{c}}er}, {Miller-Jones}, {Russell}, {Sanna}, {Bailyn}, {Lewis},
  {Migliari}, \& {Rahoui}}]{trigo2018}
{D{\'\i}az Trigo}, M., {Altamirano}, D., {Din{\c{c}}er}, T., {et~al.} 2018,
  \aap, 616, A23, \dodoi{10.1051/0004-6361/201832693}

\bibitem[{{Evans} {et~al.}(2007){Evans}, {Beardmore}, {Page}, {Tyler},
  {Osborne}, {Goad}, {O'Brien}, {Vetere}, {Racusin}, {Morris}, {Burrows},
  {Capalbi}, {Perri}, {Gehrels}, \& {Romano}}]{evans07}
{Evans}, P.~A., {Beardmore}, A.~P., {Page}, K.~L., {et~al.} 2007, \aap, 469,
  379, \dodoi{10.1051/0004-6361:20077530}

\bibitem[{{Evans} {et~al.}(2009){Evans}, {Beardmore}, {Page}, {Osborne},
  {O'Brien}, {Willingale}, {Starling}, {Burrows}, {Godet}, {Vetere}, {Racusin},
  {Goad}, {Wiersema}, {Angelini}, {Capalbi}, {Chincarini}, {Gehrels}, {Kennea},
  {Margutti}, {Morris}, {Mountford}, {Pagani}, {Perri}, {Romano}, \&
  {Tanvir}}]{evans09}
---. 2009, \mnras, 397, 1177, \dodoi{10.1111/j.1365-2966.2009.14913.x}

\bibitem[{{Fender}(2006)}]{fender2006}
{Fender}, R. 2006, in Compact stellar X-ray sources, Vol.~39, 381--419,
  \dodoi{10.48550/arXiv.astro-ph/0303339}

\bibitem[{{Fender} \& {Bright}(2019)}]{fender2019}
{Fender}, R., \& {Bright}, J. 2019, \mnras, 489, 4836,
  \dodoi{10.1093/mnras/stz2000}

\bibitem[{{Fender} {et~al.}(2016){Fender}, {Woudt}, {Corbel}, {Coriat},
  {Daigne}, {Falcke}, {Girard}, {Heywood}, {Horesh}, {Horrell}, {Jonker},
  {Joseph}, {Kamble}, {Knigge}, {K{\"o}rding}, {Kotze}, {Kouveliotou}, {Lynch},
  {Maccarone}, {Meintjes}, {Migliari}, {Murphy}, {Nagayama}, {Nelemans},
  {Nicholson}, {O'Brien}, {Oodendaal}, {Oozeer}, {Osborne}, {P{\'e}rez-Torres},
  {Ratcliffe}, {Ribeiro}, {Rol}, {Rushton}, {Scaife}, {Schurch}, {Sivakoff},
  {Staley}, {Steeghs}, {Stewart}, {Swinbank}, {Vergani}, {Warner}, {Wiersema},
  {Armstrong}, {Groot}, {McBride}, {Miller-Jones}, {Mooley}, {Stappers},
  {Wijers}, {Bietenholz}, {Blyth}, {B{\"o}ttcher}, {Buckley}, {Charles},
  {Chomiuk}, {Coppejans}, {de Blok}, {van der Heyden}, {van der Horst}, \& {van
  Soelen}}]{2016mks..confE..13F}
{Fender}, R., {Woudt}, P.~A., {Corbel}, S., {et~al.} 2016, in MeerKAT Science:
  On the Pathway to the SKA, 13, \dodoi{10.22323/1.277.0013}

\bibitem[{{Folkner} {et~al.}(2009){Folkner}, {Williams}, \&
  {Boggs}}]{folkner09}
{Folkner}, W.~M., {Williams}, J.~G., \& {Boggs}, D.~H. 2009, Interplanetary
  Network Progress Report, 42-178, 1

\bibitem[{{Fomalont} {et~al.}(2001){Fomalont}, {Geldzahler}, \&
  {Bradshaw}}]{fomalont01}
{Fomalont}, E.~B., {Geldzahler}, B.~J., \& {Bradshaw}, C.~F. 2001, \apj, 558,
  283, \dodoi{10.1086/322479}

\bibitem[{{Ford} \& {van der Klis}(1998)}]{ford98}
{Ford}, E.~C., \& {van der Klis}, M. 1998, \apjl, 506, L39,
  \dodoi{10.1086/311638}

\bibitem[{{Fortner} {et~al.}(1989){Fortner}, {Lamb}, \& {Miller}}]{fortner89}
{Fortner}, B., {Lamb}, F.~K., \& {Miller}, G.~S. 1989, \nat, 342, 775,
  \dodoi{10.1038/342775a0}

\bibitem[{{Frank} {et~al.}(1987){Frank}, {King}, \& {Lasota}}]{frank87}
{Frank}, J., {King}, A.~R., \& {Lasota}, J.~P. 1987, \aap, 178, 137

\bibitem[{{Fridriksson}(2011)}]{fridriksson11}
{Fridriksson}, J.~K. 2011, PhD thesis, MIT, Center for Space Research/Kavli
  Institute

\bibitem[{{Fridriksson} {et~al.}(2015){Fridriksson}, {Homan}, \&
  {Remillard}}]{fridriksson15}
{Fridriksson}, J.~K., {Homan}, J., \& {Remillard}, R.~A. 2015, \apj, 809, 52,
  \dodoi{10.1088/0004-637X/809/1/52}

\bibitem[{{Gambino} {et~al.}(2019){Gambino}, {Iaria}, {Di Salvo}, {Mazzola},
  {Marino}, {Burderi}, {Riggio}, {Sanna}, \& {D'Amico}}]{gambino19}
{Gambino}, A.~F., {Iaria}, R., {Di Salvo}, T., {et~al.} 2019, \aap, 625, A92,
  \dodoi{10.1051/0004-6361/201832984}

\bibitem[{{Gandhi} {et~al.}(2022){Gandhi}, {Kawamuro}, {D{\'\i}az Trigo},
  {Paice}, {Boorman}, {Cappi}, {Done}, {Fabian}, {Fukumura}, {Garc{\'\i}a},
  {Greenwell}, {Guainazzi}, {Makishima}, {Tashiro}, {Tomaru}, {Tombesi}, \&
  {Ueda}}]{gandhi22}
{Gandhi}, P., {Kawamuro}, T., {D{\'\i}az Trigo}, M., {et~al.} 2022, Nature
  Astronomy, 6, 1364, \dodoi{10.1038/s41550-022-01857-y}

\bibitem[{{Gavriil} {et~al.}(2012){Gavriil}, {Strohmayer}, \&
  {Bhattacharyya}}]{gavriil12}
{Gavriil}, F.~P., {Strohmayer}, T.~E., \& {Bhattacharyya}, S. 2012, \apj, 753,
  2, \dodoi{10.1088/0004-637X/753/1/2}

\bibitem[{{Gehrels} {et~al.}(2004){Gehrels}, {Chincarini}, {Giommi}, {Mason},
  {Nousek}, {Wells}, {White}, {Barthelmy}, {Burrows}, {Cominsky}, {Hurley},
  {Marshall}, {M{\'e}sz{\'a}ros}, {Roming}, {Angelini}, {Barbier}, {Belloni},
  {Campana}, {Caraveo}, {Chester}, {Citterio}, {Cline}, {Cropper}, {Cummings},
  {Dean}, {Feigelson}, {Fenimore}, {Frail}, {Fruchter}, {Garmire}, {Gendreau},
  {Ghisellini}, {Greiner}, {Hill}, {Hunsberger}, {Krimm}, {Kulkarni}, {Kumar},
  {Lebrun}, {Lloyd-Ronning}, {Markwardt}, {Mattson}, {Mushotzky}, {Norris},
  {Osborne}, {Paczynski}, {Palmer}, {Park}, {Parsons}, {Paul}, {Rees},
  {Reynolds}, {Rhoads}, {Sasseen}, {Schaefer}, {Short}, {Smale}, {Smith},
  {Stella}, {Tagliaferri}, {Takahashi}, {Tashiro}, {Townsley}, {Tueller},
  {Turner}, {Vietri}, {Voges}, {Ward}, {Willingale}, {Zerbi}, \&
  {Zhang}}]{gehrels04}
{Gehrels}, N., {Chincarini}, G., {Giommi}, P., {et~al.} 2004, \apj, 611, 1005,
  \dodoi{10.1086/422091}

\bibitem[{{Gendreau} {et~al.}(2016){Gendreau}, {Arzoumanian}, {Adkins},
  {Albert}, {Anders}, {Aylward}, {Baker}, {Balsamo}, {Bamford}, {Benegalrao},
  {Berry}, {Bhalwani}, {Black}, {Blaurock}, {Bronke}, {Brown}, {Budinoff},
  {Cantwell}, {Cazeau}, {Chen}, {Clement}, {Colangelo}, {Coleman},
  {Coopersmith}, {Dehaven}, {Doty}, {Egan}, {Enoto}, {Fan}, {Ferro}, {Foster},
  {Galassi}, {Gallo}, {Green}, {Grosh}, {Ha}, {Hasouneh}, {Heefner}, {Hestnes},
  {Hoge}, {Jacobs}, {J{\o}rgensen}, {Kaiser}, {Kellogg}, {Kenyon}, {Koenecke},
  {Kozon}, {LaMarr}, {Lambertson}, {Larson}, {Lentine}, {Lewis}, {Lilly},
  {Liu}, {Malonis}, {Manthripragada}, {Markwardt}, {Matonak}, {Mcginnis},
  {Miller}, {Mitchell}, {Mitchell}, {Mohammed}, {Monroe}, {Montt de Garcia},
  {Mul{\'e}}, {Nagao}, {Ngo}, {Norris}, {Norwood}, {Novotka}, {Okajima},
  {Olsen}, {Onyeachu}, {Orosco}, {Peterson}, {Pevear}, {Pham}, {Pollard},
  {Pope}, {Powers}, {Powers}, {Price}, {Prigozhin}, {Ramirez}, {Reid},
  {Remillard}, {Rogstad}, {Rosecrans}, {Rowe}, {Sager}, {Sanders}, {Savadkin},
  {Saylor}, {Schaeffer}, {Schweiss}, {Semper}, {Serlemitsos}, {Shackelford},
  {Soong}, {Struebel}, {Vezie}, {Villasenor}, {Winternitz}, {Wofford},
  {Wright}, {Yang}, \& {Yu}}]{gendreau16}
{Gendreau}, K.~C., {Arzoumanian}, Z., {Adkins}, P.~W., {et~al.} 2016, in
  Society of Photo-Optical Instrumentation Engineers (SPIE) Conference Series,
  Vol. 9905, Space Telescopes and Instrumentation 2016: Ultraviolet to Gamma
  Ray, ed. J.-W.~A. {den Herder}, T.~{Takahashi}, \& M.~{Bautz}, 99051H,
  \dodoi{10.1117/12.2231304}

\bibitem[{{Gladstone} {et~al.}(2007){Gladstone}, {Done}, \&
  {Gierli{\'n}ski}}]{gladstone07}
{Gladstone}, J., {Done}, C., \& {Gierli{\'n}ski}, M. 2007, \mnras, 378, 13,
  \dodoi{10.1111/j.1365-2966.2007.11675.x}

\bibitem[{{Harrison} {et~al.}(2013){Harrison}, {Craig}, {Christensen},
  {Hailey}, {Zhang}, {Boggs}, {Stern}, {Cook}, {Forster}, {Giommi},
  {Grefenstette}, {Kim}, {Kitaguchi}, {Koglin}, {Madsen}, {Mao}, {Miyasaka},
  {Mori}, {Perri}, {Pivovaroff}, {Puccetti}, {Rana}, {Westergaard}, {Willis},
  {Zoglauer}, {An}, {Bachetti}, {Barri{\`e}re}, {Bellm}, {Bhalerao},
  {Brejnholt}, {Fuerst}, {Liebe}, {Markwardt}, {Nynka}, {Vogel}, {Walton},
  {Wik}, {Alexander}, {Cominsky}, {Hornschemeier}, {Hornstrup}, {Kaspi},
  {Madejski}, {Matt}, {Molendi}, {Smith}, {Tomsick}, {Ajello}, {Ballantyne},
  {Balokovi{\'c}}, {Barret}, {Bauer}, {Blandford}, {Brandt}, {Brenneman},
  {Chiang}, {Chakrabarty}, {Chenevez}, {Comastri}, {Dufour}, {Elvis}, {Fabian},
  {Farrah}, {Fryer}, {Gotthelf}, {Grindlay}, {Helfand}, {Krivonos}, {Meier},
  {Miller}, {Natalucci}, {Ogle}, {Ofek}, {Ptak}, {Reynolds}, {Rigby},
  {Tagliaferri}, {Thorsett}, {Treister}, \& {Urry}}]{harrison13}
{Harrison}, F.~A., {Craig}, W.~W., {Christensen}, F.~E., {et~al.} 2013, \apj,
  770, 103, \dodoi{10.1088/0004-637X/770/2/103}

\bibitem[{{Hasinger} \& {van der Klis}(1989)}]{hasinger89}
{Hasinger}, G., \& {van der Klis}, M. 1989, \aap, 225, 79

\bibitem[{{Heger} {et~al.}(2007){Heger}, {Cumming}, \& {Woosley}}]{heger07}
{Heger}, A., {Cumming}, A., \& {Woosley}, S.~E. 2007, \apj, 665, 1311,
  \dodoi{10.1086/517491}

\bibitem[{{Heywood}(2020)}]{2020ascl.soft09003H}
{Heywood}, I. 2020, {oxkat: Semi-automated imaging of MeerKAT observations},
  Astrophysics Source Code Library, record ascl:2009.003.
\newblock \doeprint{2009.003}

\bibitem[{{Heywood} {et~al.}(2022){Heywood}, {Jarvis}, {Hale}, {Whittam},
  {Bester}, {Hugo}, {Kenyon}, {Prescott}, {Smirnov}, {Tasse}, {Afonso}, {Best},
  {Collier}, {Deane}, {Frank}, {Hardcastle}, {Knowles}, {Maddox}, {Murphy},
  {Prandoni}, {Randriamampandry}, {Santos}, {Sekhar}, {Tabatabaei}, {Taylor},
  \& {Thorat}}]{heywood22}
{Heywood}, I., {Jarvis}, M.~J., {Hale}, C.~L., {et~al.} 2022, \mnras, 509,
  2150, \dodoi{10.1093/mnras/stab3021}

\bibitem[{{HI4PI Collaboration} {et~al.}(2016){HI4PI Collaboration}, {Ben
  Bekhti}, {Fl{\"o}er}, {Keller}, {Kerp}, {Lenz}, {Winkel}, {Bailin},
  {Calabretta}, {Dedes}, {Ford}, {Gibson}, {Haud}, {Janowiecki}, {Kalberla},
  {Lockman}, {McClure-Griffiths}, {Murphy}, {Nakanishi}, {Pisano}, \&
  {Staveley-Smith}}]{hi4pi16}
{HI4PI Collaboration}, {Ben Bekhti}, N., {Fl{\"o}er}, L., {et~al.} 2016, \aap,
  594, A116, \dodoi{10.1051/0004-6361/201629178}

\bibitem[{{Hjellming} \& {Rupen}(1995)}]{hjellming1995}
{Hjellming}, R.~M., \& {Rupen}, M.~P. 1995, \nat, 375, 464,
  \dodoi{10.1038/375464a0}

\bibitem[{{Homan}(2012)}]{homan12}
{Homan}, J. 2012, \apjl, 760, L30, \dodoi{10.1088/2041-8205/760/2/L30}

\bibitem[{{Homan} {et~al.}(2007{\natexlab{a}}){Homan}, {Belloni}, {Wijnands},
  {van der Klis}, {Swank}, {Smith}, {Pereira}, \& {Markwardt}}]{homan07b}
{Homan}, J., {Belloni}, T., {Wijnands}, R., {et~al.} 2007{\natexlab{a}}, The
  Astronomer's Telegram, 1144, 1

\bibitem[{{Homan} {et~al.}(2015){Homan}, {Fridriksson}, \&
  {Remillard}}]{homan15}
{Homan}, J., {Fridriksson}, J.~K., \& {Remillard}, R.~A. 2015, \apj, 812, 80,
  \dodoi{10.1088/0004-637X/812/1/80}

\bibitem[{{Homan} {et~al.}(1999){Homan}, {Jonker}, {Wijnands}, {van der Klis},
  \& {van Paradijs}}]{homan99}
{Homan}, J., {Jonker}, P.~G., {Wijnands}, R., {van der Klis}, M., \& {van
  Paradijs}, J. 1999, \apjl, 516, L91, \dodoi{10.1086/312000}

\bibitem[{{Homan} {et~al.}(2007{\natexlab{b}}){Homan}, {Wijnands},
  {Altamirano}, \& {Belloni}}]{homan07c}
{Homan}, J., {Wijnands}, R., {Altamirano}, D., \& {Belloni}, T.
  2007{\natexlab{b}}, The Astronomer's Telegram, 1165, 1

\bibitem[{{Homan} {et~al.}(2004){Homan}, {Wijnands}, {Rupen}, {Fender},
  {Hjellming}, {di Salvo}, \& {van der Klis}}]{homan04}
{Homan}, J., {Wijnands}, R., {Rupen}, M.~P., {et~al.} 2004, \aap, 418, 255,
  \dodoi{10.1051/0004-6361:20034258}

\bibitem[{{Homan} {et~al.}(2010){Homan}, {van der Klis}, {Fridriksson},
  {Remillard}, {Wijnands}, {M{\'e}ndez}, {Lin}, {Altamirano}, {Casella},
  {Belloni}, \& {Lewin}}]{homan10}
{Homan}, J., {van der Klis}, M., {Fridriksson}, J.~K., {et~al.} 2010, \apj,
  719, 201, \dodoi{10.1088/0004-637X/719/1/201}

\bibitem[{{Hughes} {et~al.}(2022){Hughes}, {Sivakoff}, {Fender}, {Woudt},
  {Miller-Jones}, \& {van den Eijnden}}]{hughes22}
{Hughes}, A.~K., {Sivakoff}, G.~R., {Fender}, R., {et~al.} 2022, The
  Astronomer's Telegram, 15410, 1

\bibitem[{{Hugo} {et~al.}(2022){Hugo}, {Perkins}, {Merry}, {Mauch}, \&
  {Smirnov}}]{hugo22}
{Hugo}, B.~V., {Perkins}, S., {Merry}, B., {Mauch}, T., \& {Smirnov}, O.~M.
  2022, in Astronomical Society of the Pacific Conference Series, Vol. 532,
  Astronomical Society of the Pacific Conference Series, ed. J.~E. {Ruiz},
  F.~{Pierfedereci}, \& P.~{Teuben}, 541.
\newblock \doarXiv{2206.09179}

\bibitem[{{Hunter}(2007)}]{hunter07}
{Hunter}, J.~D. 2007, Computing in Science and Engineering, 9, 90,
  \dodoi{10.1109/MCSE.2007.55}

\bibitem[{{Huppenkothen} \& {Bachetti}(2018)}]{huppenkothen18}
{Huppenkothen}, D., \& {Bachetti}, M. 2018, \apjs, 236, 13,
  \dodoi{10.3847/1538-4365/aabe38}

\bibitem[{{Huppenkothen} {et~al.}(2019{\natexlab{a}}){Huppenkothen},
  {Bachetti}, {Stevens}, {Migliari}, {Balm}, {Hammad}, {Khan}, {Mishra},
  {Rashid}, {Sharma}, {Ribeiro}, \& {Blanco}}]{huppenkothen19a}
{Huppenkothen}, D., {Bachetti}, M., {Stevens}, A., {et~al.} 2019{\natexlab{a}},
  The Journal of Open Source Software, 4, 1393, \dodoi{10.21105/joss.01393}

\bibitem[{{Huppenkothen} {et~al.}(2019{\natexlab{b}}){Huppenkothen},
  {Bachetti}, {Stevens}, {Migliari}, {Balm}, {Hammad}, {Khan}, {Mishra},
  {Rashid}, {Sharma}, {Martinez Ribeiro}, \& {Valles Blanco}}]{huppenkothen19b}
{Huppenkothen}, D., {Bachetti}, M., {Stevens}, A.~L., {et~al.}
  2019{\natexlab{b}}, \apj, 881, 39, \dodoi{10.3847/1538-4357/ab258d}

\bibitem[{{Hyodo} {et~al.}(2009){Hyodo}, {Ueda}, {Yuasa}, {Maeda}, {Makishima},
  \& {Koyama}}]{hyodo09}
{Hyodo}, Y., {Ueda}, Y., {Yuasa}, T., {et~al.} 2009, \pasj, 61, S99,
  \dodoi{10.1093/pasj/61.sp1.S99}

\bibitem[{{Iaria} {et~al.}(2007{\natexlab{a}}){Iaria}, {di Salvo}, {Lavagetto},
  {D'A{\'\i}}, \& {Robba}}]{iaria07b}
{Iaria}, R., {di Salvo}, T., {Lavagetto}, G., {D'A{\'\i}}, A., \& {Robba},
  N.~R. 2007{\natexlab{a}}, \aap, 464, 291, \dodoi{10.1051/0004-6361:20065644}

\bibitem[{{Iaria} {et~al.}(2007{\natexlab{b}}){Iaria}, {Lavagetto},
  {D'A{\'\i}}, {di Salvo}, \& {Robba}}]{iaria07a}
{Iaria}, R., {Lavagetto}, G., {D'A{\'\i}}, A., {di Salvo}, T., \& {Robba},
  N.~R. 2007{\natexlab{b}}, \aap, 463, 289, \dodoi{10.1051/0004-6361:20065862}

\bibitem[{{Ingram} \& {Done}(2010)}]{ingram10}
{Ingram}, A., \& {Done}, C. 2010, \mnras, 405, 2447,
  \dodoi{10.1111/j.1365-2966.2010.16614.x}

\bibitem[{{Jia} {et~al.}(2023){Jia}, {Qu}, {Lu}, {Zhang}, {Zhang}, {Song},
  {Zhang}, {Huang}, {Ma}, {Tao}, {Liu}, \& {Yu}}]{jia23}
{Jia}, S.~M., {Qu}, J.~L., {Lu}, F.~J., {et~al.} 2023, \mnras, 521, 4792,
  \dodoi{10.1093/mnras/stad876}

\bibitem[{{Jimenez-Garate} {et~al.}(2002){Jimenez-Garate}, {Raymond}, \&
  {Liedahl}}]{jimenezgarate02}
{Jimenez-Garate}, M.~A., {Raymond}, J.~C., \& {Liedahl}, D.~A. 2002, \apj, 581,
  1297, \dodoi{10.1086/344364}

\bibitem[{{Jonker} {et~al.}(2000){Jonker}, {van der Klis}, {Homan}, {Wijnands},
  {van Paradijs}, {M{\'e}ndez}, {Kuulkers}, \& {Ford}}]{jonker00}
{Jonker}, P.~G., {van der Klis}, M., {Homan}, J., {et~al.} 2000, \apj, 531,
  453, \dodoi{10.1086/308471}

\bibitem[{{Jonker} {et~al.}(1999){Jonker}, {van der Klis}, \&
  {Wijnands}}]{jonker99}
{Jonker}, P.~G., {van der Klis}, M., \& {Wijnands}, R. 1999, \apjl, 511, L41,
  \dodoi{10.1086/311840}

\bibitem[{{Kaastra} \& {Bleeker}(2016)}]{kaastra16}
{Kaastra}, J.~S., \& {Bleeker}, J.~A.~M. 2016, \aap, 587, A151,
  \dodoi{10.1051/0004-6361/201527395}

\bibitem[{{Kenyon} {et~al.}(2018){Kenyon}, {Smirnov}, {Grobler}, \&
  {Perkins}}]{kenyon18}
{Kenyon}, J.~S., {Smirnov}, O.~M., {Grobler}, T.~L., \& {Perkins}, S.~J. 2018,
  \mnras, 478, 2399, \dodoi{10.1093/mnras/sty1221}

\bibitem[{{Ko} \& {Kallman}(1991)}]{ko91}
{Ko}, Y.-K., \& {Kallman}, T.~R. 1991, \apj, 374, 721, \dodoi{10.1086/170156}

\bibitem[{{Kobayashi} {et~al.}(2022){Kobayashi}, {Negoro}, {Nakajima},
  {Tanaka}, {Soejima}, {Mihara}, {Kawamuro}, {Yamada}, {Tamagawa}, {Matsuoka},
  {Sakamoto}, {Serino}, {Sugita}, {Hiramatsu}, {Yoshida}, {Tsuboi}, {Iwakiri},
  {Kohara}, {Shidatsu}, {Iwasaki}, {Kawai}, {Niwano}, {Hosokawa}, {Imai},
  {Ito}, {Takamatsu}, {Nakahira}, {Ueno}, {Tomida}, {Ishikawa}, {Tominaga},
  {Nagatsuka}, {Kurihara}, {Ueda}, {Ogawa}, {Setoguchi}, {Yoshitake}, {Inaba},
  {Tsunemi}, {Yamauchi}, {Nonaka}, {Sato}, {Hatsuda}, {Fukuoka}, {Yamaoka},
  {Kawakubo}, \& {Sugizaki}}]{2022ATel15407....1K}
{Kobayashi}, K., {Negoro}, H., {Nakajima}, M., {et~al.} 2022, The Astronomer's
  Telegram, 15407, 1

\bibitem[{{Kong} {et~al.}(2006){Kong}, {Charles}, {Homer}, {Kuulkers}, \&
  {O'Donoghue}}]{kong06}
{Kong}, A.~K.~H., {Charles}, P.~A., {Homer}, L., {Kuulkers}, E., \&
  {O'Donoghue}, D. 2006, \mnras, 368, 781,
  \dodoi{10.1111/j.1365-2966.2006.10157.x}

\bibitem[{{Kubota} {et~al.}(1998){Kubota}, {Tanaka}, {Makishima}, {Ueda},
  {Dotani}, {Inoue}, \& {Yamaoka}}]{kubota98}
{Kubota}, A., {Tanaka}, Y., {Makishima}, K., {et~al.} 1998, \pasj, 50, 667,
  \dodoi{10.1093/pasj/50.6.667}

\bibitem[{{Kuulkers} {et~al.}(2003){Kuulkers}, {den Hartog}, {in't Zand},
  {Verbunt}, {Harris}, \& {Cocchi}}]{Kuulkers2003}
{Kuulkers}, E., {den Hartog}, P.~R., {in't Zand}, J.~J.~M., {et~al.} 2003,
  \aap, 399, 663, \dodoi{10.1051/0004-6361:20021781}

\bibitem[{{LaMarr} {et~al.}(2016){LaMarr}, {Prigozhin}, {Remillard}, {Malonis},
  {Gendreau}, {Arzoumanian}, {Markwardt}, \& {Baumgartner}}]{lamarr16}
{LaMarr}, B., {Prigozhin}, G., {Remillard}, R., {et~al.} 2016, in Society of
  Photo-Optical Instrumentation Engineers (SPIE) Conference Series, Vol. 9905,
  Space Telescopes and Instrumentation 2016: Ultraviolet to Gamma Ray, ed.
  J.-W.~A. {den Herder}, T.~{Takahashi}, \& M.~{Bautz}, 99054W,
  \dodoi{10.1117/12.2232784}

\bibitem[{{Liddle}(2007)}]{liddle07}
{Liddle}, A.~R. 2007, \mnras, 377, L74,
  \dodoi{10.1111/j.1745-3933.2007.00306.x}

\bibitem[{{Lin} {et~al.}(2007){Lin}, {Remillard}, \& {Homan}}]{lin07}
{Lin}, D., {Remillard}, R.~A., \& {Homan}, J. 2007, \apj, 667, 1073,
  \dodoi{10.1086/521181}

\bibitem[{{Lin} {et~al.}(2009){Lin}, {Remillard}, \& {Homan}}]{lin09}
---. 2009, \apj, 696, 1257, \dodoi{10.1088/0004-637X/696/2/1257}

\bibitem[{{Ludlam} {et~al.}(2022){Ludlam}, {Cackett}, {Garc{\'\i}a}, {Miller},
  {Stevens}, {Fabian}, {Homan}, {Ng}, {Guillot}, {Buisson}, \&
  {Chakrabarty}}]{ludlam22}
{Ludlam}, R.~M., {Cackett}, E.~M., {Garc{\'\i}a}, J.~A., {et~al.} 2022, \apj,
  927, 112, \dodoi{10.3847/1538-4357/ac5028}

\bibitem[{{Mancuso} {et~al.}(2023){Mancuso}, {Altamirano}, {Bult}, {Chenevez},
  {Guillot}, {G{\"u}ver}, {Jaisawal}, {Malacaria}, {Ng}, {Sanna}, \&
  {Strohmayer}}]{mancuso23}
{Mancuso}, G.~C., {Altamirano}, D., {Bult}, P., {et~al.} 2023, \mnras, 521,
  5616, \dodoi{10.1093/mnras/stad949}

\bibitem[{{Marino} {et~al.}(2022){Marino}, {Anitra}, {Mazzola}, {Di Salvo},
  {Sanna}, {Bult}, {Guillot}, {Mancuso}, {Ng}, {Riggio}, {Albayati},
  {Altamirano}, {Arzoumanian}, {Burderi}, {Cabras}, {Chakrabarty}, {Deiosso},
  {Gendreau}, {Iaria}, {Manca}, \& {Strohmayer}}]{marino22}
{Marino}, A., {Anitra}, A., {Mazzola}, S.~M., {et~al.} 2022, \mnras, 515, 3838,
  \dodoi{10.1093/mnras/stac2038}

\bibitem[{{Matsuoka} {et~al.}(2009){Matsuoka}, {Kawasaki}, {Ueno}, {Tomida},
  {Kohama}, {Suzuki}, {Adachi}, {Ishikawa}, {Mihara}, {Sugizaki}, {Isobe},
  {Nakagawa}, {Tsunemi}, {Miyata}, {Kawai}, {Kataoka}, {Morii}, {Yoshida},
  {Negoro}, {Nakajima}, {Ueda}, {Chujo}, {Yamaoka}, {Yamazaki}, {Nakahira},
  {You}, {Ishiwata}, {Miyoshi}, {Eguchi}, {Hiroi}, {Katayama}, \&
  {Ebisawa}}]{matsuoka09}
{Matsuoka}, M., {Kawasaki}, K., {Ueno}, S., {et~al.} 2009, \pasj, 61, 999,
  \dodoi{10.1093/pasj/61.5.999}

\bibitem[{{M{\'e}ndez} {et~al.}(1997){M{\'e}ndez}, {van der Klis}, {van
  Paradijs}, {Lewin}, {Lamb}, {Vaughan}, {Kuulkers}, \& {Psaltis}}]{mendez97}
{M{\'e}ndez}, M., {van der Klis}, M., {van Paradijs}, J., {et~al.} 1997, \apjl,
  485, L37, \dodoi{10.1086/310803}

\bibitem[{{Middleditch} \& {Priedhorsky}(1986)}]{middleditch86}
{Middleditch}, J., \& {Priedhorsky}, W.~C. 1986, \apj, 306, 230,
  \dodoi{10.1086/164335}

\bibitem[{{Migliari} \& {Fender}(2006)}]{migliari06}
{Migliari}, S., \& {Fender}, R.~P. 2006, \mnras, 366, 79,
  \dodoi{10.1111/j.1365-2966.2005.09777.x}

\bibitem[{{Migliari} {et~al.}(2004){Migliari}, {Fender}, {Rupen}, {Wachter},
  {Jonker}, {Homan}, \& {van der Klis}}]{migliari2004}
{Migliari}, S., {Fender}, R.~P., {Rupen}, M., {et~al.} 2004, \mnras, 351, 186,
  \dodoi{10.1111/j.1365-2966.2004.07768.x}

\bibitem[{{Migliari} {et~al.}(2007){Migliari}, {Miller-Jones}, {Fender},
  {Homan}, {Di Salvo}, {Rothschild}, {Rupen}, {Tomsick}, {Wijnands}, \& {van
  der Klis}}]{migliari07}
{Migliari}, S., {Miller-Jones}, J.~C.~A., {Fender}, R.~P., {et~al.} 2007, \apj,
  671, 706, \dodoi{10.1086/522516}

\bibitem[{{Migliari} {et~al.}(2010){Migliari}, {Tomsick}, {Miller-Jones},
  {Heinz}, {Hynes}, {Fender}, {Gallo}, {Jonker}, \& {Maccarone}}]{migliari2010}
{Migliari}, S., {Tomsick}, J.~A., {Miller-Jones}, J.~C.~A., {et~al.} 2010,
  \apj, 710, 117, \dodoi{10.1088/0004-637X/710/1/117}

\bibitem[{{Miller-Jones} {et~al.}(2010){Miller-Jones}, {Sivakoff},
  {Altamirano}, {Tudose}, {Migliari}, {Dhawan}, {Fender}, {Garrett}, {Heinz},
  {K{\"o}rding}, {Krimm}, {Linares}, {Maitra}, {Markoff}, {Paragi},
  {Remillard}, {Rupen}, {Rushton}, {Russell}, {Sarazin}, \&
  {Spencer}}]{millerjones2010}
{Miller-Jones}, J.~C.~A., {Sivakoff}, G.~R., {Altamirano}, D., {et~al.} 2010,
  \apjl, 716, L109, \dodoi{10.1088/2041-8205/716/2/L109}

\bibitem[{{Miller-Jones} {et~al.}(2012){Miller-Jones}, {Moin}, {Tingay},
  {Reynolds}, {Phillips}, {Tzioumis}, {Fender}, {McCallum}, {Nicolson}, \&
  {Tudose}}]{millerjones2012}
{Miller-Jones}, J.~C.~A., {Moin}, A., {Tingay}, S.~J., {et~al.} 2012, \mnras,
  419, L49, \dodoi{10.1111/j.1745-3933.2011.01176.x}

\bibitem[{{Mondal} {et~al.}(2023){Mondal}, {Raychaudhuri}, \&
  {Dewangan}}]{mondal23}
{Mondal}, A.~S., {Raychaudhuri}, B., \& {Dewangan}, G.~C. 2023, arXiv e-prints,
  arXiv:2309.12637.
\newblock \doarXiv{2309.12637}

\bibitem[{{Motta} \& {Fender}(2019)}]{motta19}
{Motta}, S.~E., \& {Fender}, R.~P. 2019, \mnras, 483, 3686,
  \dodoi{10.1093/mnras/sty3331}

\bibitem[{{Motta} {et~al.}(2017){Motta}, {Rouco Escorial}, {Kuulkers},
  {Mu{\~n}oz-Darias}, \& {Sanna}}]{motta17}
{Motta}, S.~E., {Rouco Escorial}, A., {Kuulkers}, E., {Mu{\~n}oz-Darias}, T.,
  \& {Sanna}, A. 2017, \mnras, 468, 2311, \dodoi{10.1093/mnras/stx570}

\bibitem[{{Mu{\~n}oz-Darias} {et~al.}(2014){Mu{\~n}oz-Darias}, {Fender},
  {Motta}, \& {Belloni}}]{munoz14}
{Mu{\~n}oz-Darias}, T., {Fender}, R.~P., {Motta}, S.~E., \& {Belloni}, T.~M.
  2014, \mnras, 443, 3270, \dodoi{10.1093/mnras/stu1334}

\bibitem[{{Muno} {et~al.}(2002{\natexlab{a}}){Muno}, {{\"O}zel}, \&
  {Chakrabarty}}]{muno02b}
{Muno}, M.~P., {{\"O}zel}, F., \& {Chakrabarty}, D. 2002{\natexlab{a}}, \apj,
  581, 550, \dodoi{10.1086/344152}

\bibitem[{{Muno} {et~al.}(2002{\natexlab{b}}){Muno}, {Remillard}, \&
  {Chakrabarty}}]{muno02a}
{Muno}, M.~P., {Remillard}, R.~A., \& {Chakrabarty}, D. 2002{\natexlab{b}},
  \apjl, 568, L35, \dodoi{10.1086/340269}

\bibitem[{{NASA High Energy Astrophysics Science Archive Research Center
  (HEASARC)}(2014)}]{heasoft}
{NASA High Energy Astrophysics Science Archive Research Center (HEASARC)}.
  2014, {HEAsoft: Unified Release of FTOOLS and XANADU}, Astrophysics Source
  Code Library, record ascl:1408.004.
\newblock \doeprint{1408.004}

\bibitem[{{Neilsen} \& {Lee}(2009)}]{neilsen09}
{Neilsen}, J., \& {Lee}, J.~C. 2009, \nat, 458, 481,
  \dodoi{10.1038/nature07680}

\bibitem[{{Ng} {et~al.}(2022){Ng}, {Remillard}, {Steiner}, {Chakrabarty}, \&
  {Pasham}}]{ngmason22}
{Ng}, M., {Remillard}, R.~A., {Steiner}, J.~F., {Chakrabarty}, D., \& {Pasham},
  D.~R. 2022, \apj, 940, 138, \dodoi{10.3847/1538-4357/ac9965}

\bibitem[{{Offringa} {et~al.}(2014){Offringa}, {McKinley}, {Hurley-Walker},
  {Briggs}, {Wayth}, {Kaplan}, {Bell}, {Feng}, {Neben}, {Hughes}, {Rhee},
  {Murphy}, {Bhat}, {Bernardi}, {Bowman}, {Cappallo}, {Corey}, {Deshpande},
  {Emrich}, {Ewall-Wice}, {Gaensler}, {Goeke}, {Greenhill}, {Hazelton},
  {Hindson}, {Johnston-Hollitt}, {Jacobs}, {Kasper}, {Kratzenberg}, {Lenc},
  {Lonsdale}, {Lynch}, {McWhirter}, {Mitchell}, {Morales}, {Morgan},
  {Kudryavtseva}, {Oberoi}, {Ord}, {Pindor}, {Procopio}, {Prabu}, {Riding},
  {Roshi}, {Shankar}, {Srivani}, {Subrahmanyan}, {Tingay}, {Waterson},
  {Webster}, {Whitney}, {Williams}, \& {Williams}}]{offringa14}
{Offringa}, A.~R., {McKinley}, B., {Hurley-Walker}, N., {et~al.} 2014, \mnras,
  444, 606, \dodoi{10.1093/mnras/stu1368}

\bibitem[{{Oosterbroek} {et~al.}(1995){Oosterbroek}, {van der Klis},
  {Kuulkers}, {van Paradijs}, \& {Lewin}}]{oosterbroek95}
{Oosterbroek}, T., {van der Klis}, M., {Kuulkers}, E., {van Paradijs}, J., \&
  {Lewin}, W.~H.~G. 1995, \aap, 297, 141

\bibitem[{{Parmar} {et~al.}(2002){Parmar}, {Oosterbroek}, {Boirin}, \&
  {Lumb}}]{parmar02}
{Parmar}, A.~N., {Oosterbroek}, T., {Boirin}, L., \& {Lumb}, D. 2002, \aap,
  386, 910, \dodoi{10.1051/0004-6361:20020281}

\bibitem[{{Parmar} {et~al.}(1986){Parmar}, {White}, {Giommi}, \&
  {Gottwald}}]{parmar86}
{Parmar}, A.~N., {White}, N.~E., {Giommi}, P., \& {Gottwald}, M. 1986, \apj,
  308, 199, \dodoi{10.1086/164490}

\bibitem[{{Penninx} {et~al.}(1988){Penninx}, {Lewin}, {Zijlstra}, {Mitsuda}, \&
  {van Paradijs}}]{penninx88}
{Penninx}, W., {Lewin}, W. H.~G., {Zijlstra}, A.~A., {Mitsuda}, K., \& {van
  Paradijs}, J. 1988, \nat, 336, 146, \dodoi{10.1038/336146a0}

\bibitem[{{Perez} \& {Granger}(2007)}]{perez07}
{Perez}, F., \& {Granger}, B.~E. 2007, Computing in Science and Engineering, 9,
  21, \dodoi{10.1109/MCSE.2007.53}

\bibitem[{{Pike} {et~al.}(2022){Pike}, {Jaodand}, {Ludlam}, \&
  {Tomsick}}]{pike22}
{Pike}, S.~N., {Jaodand}, A., {Ludlam}, R.~M., \& {Tomsick}, J.~A. 2022, The
  Astronomer's Telegram, 15429, 1

\bibitem[{{Piraino} {et~al.}(2002){Piraino}, {Santangelo}, \&
  {Kaaret}}]{piraino02}
{Piraino}, S., {Santangelo}, A., \& {Kaaret}, P. 2002, \apj, 567, 1091,
  \dodoi{10.1086/338037}

\bibitem[{{Prigozhin} {et~al.}(2016){Prigozhin}, {Gendreau}, {Doty}, {Foster},
  {Remillard}, {Malonis}, {LaMarr}, {Vezie}, {Egan}, {Villasenor},
  {Arzoumanian}, {Baumgartner}, {Scholze}, {Laubis}, {Krumrey}, \&
  {Huber}}]{prigozhin16}
{Prigozhin}, G., {Gendreau}, K., {Doty}, J.~P., {et~al.} 2016, in Society of
  Photo-Optical Instrumentation Engineers (SPIE) Conference Series, Vol. 9905,
  Space Telescopes and Instrumentation 2016: Ultraviolet to Gamma Ray, ed.
  J.-W.~A. {den Herder}, T.~{Takahashi}, \& M.~{Bautz}, 99051I,
  \dodoi{10.1117/12.2231718}

\bibitem[{{Raman} {et~al.}(2018){Raman}, {Maitra}, \& {Paul}}]{raman18}
{Raman}, G., {Maitra}, C., \& {Paul}, B. 2018, \mnras, 477, 5358,
  \dodoi{10.1093/mnras/sty918}

\bibitem[{{Ransom} {et~al.}(2002){Ransom}, {Eikenberry}, \&
  {Middleditch}}]{ransom02}
{Ransom}, S.~M., {Eikenberry}, S.~S., \& {Middleditch}, J. 2002, \aj, 124,
  1788, \dodoi{10.1086/342285}

\bibitem[{{Reig} {et~al.}(2004){Reig}, {van Straaten}, \& {van der
  Klis}}]{reig04}
{Reig}, P., {van Straaten}, S., \& {van der Klis}, M. 2004, \apj, 602, 918,
  \dodoi{10.1086/381143}

\bibitem[{{Remillard} {et~al.}(2022){Remillard}, {Loewenstein}, {Steiner},
  {Prigozhin}, {LaMarr}, {Enoto}, {Gendreau}, {Arzoumanian}, {Markwardt},
  {Basak}, {Stevens}, {Ray}, {Altamirano}, \& {Buisson}}]{remillard22}
{Remillard}, R.~A., {Loewenstein}, M., {Steiner}, J.~F., {et~al.} 2022, \aj,
  163, 130, \dodoi{10.3847/1538-3881/ac4ae6}

\bibitem[{{Revnivtsev} {et~al.}(2001){Revnivtsev}, {Churazov}, {Gilfanov}, \&
  {Sunyaev}}]{revnivtsev01}
{Revnivtsev}, M., {Churazov}, E., {Gilfanov}, M., \& {Sunyaev}, R. 2001, \aap,
  372, 138, \dodoi{10.1051/0004-6361:20010434}

\bibitem[{{Rhodes} {et~al.}(2022){Rhodes}, {Fender}, {Motta}, {van den
  Eijnden}, {Williams}, {Bright}, \& {Sivakoff}}]{rhodes22}
{Rhodes}, L., {Fender}, R.~P., {Motta}, S., {et~al.} 2022, \mnras, 513, 2708,
  \dodoi{10.1093/mnras/stac954}

\bibitem[{{Russell} {et~al.}(2013){Russell}, {Markoff}, {Casella}, {Cantrell},
  {Chatterjee}, {Fender}, {Gallo}, {Gandhi}, {Homan}, {Maitra}, {Miller-Jones},
  {O'Brien}, \& {Shahbaz}}]{russell2013}
{Russell}, D.~M., {Markoff}, S., {Casella}, P., {et~al.} 2013, \mnras, 429,
  815, \dodoi{10.1093/mnras/sts377}

\bibitem[{{Russell} {et~al.}(2021){Russell}, {Degenaar}, {van den Eijnden},
  {Del Santo}, {Segreto}, {Altamirano}, {Beri}, {D{\'\i}az Trigo}, \&
  {Miller-Jones}}]{Russell2021}
{Russell}, T.~D., {Degenaar}, N., {van den Eijnden}, J., {et~al.} 2021, \mnras,
  508, L6, \dodoi{10.1093/mnrasl/slab087}

\bibitem[{{Schnerr} {et~al.}(2003){Schnerr}, {Reerink}, {van der Klis},
  {Homan}, {M{\'e}ndez}, {Fender}, \& {Kuulkers}}]{schnerr03}
{Schnerr}, R.~S., {Reerink}, T., {van der Klis}, M., {et~al.} 2003, \aap, 406,
  221, \dodoi{10.1051/0004-6361:20030682}

\bibitem[{{Shimura} \& {Takahara}(1995)}]{shimura95}
{Shimura}, T., \& {Takahara}, F. 1995, \apj, 445, 780, \dodoi{10.1086/175740}

\bibitem[{{Shirey} {et~al.}(1999){Shirey}, {Bradt}, \& {Levine}}]{shirey99}
{Shirey}, R.~E., {Bradt}, H.~V., \& {Levine}, A.~M. 1999, \apj, 517, 472,
  \dodoi{10.1086/307188}

\bibitem[{{Shirey} {et~al.}(1998){Shirey}, {Bradt}, {Levine}, \&
  {Morgan}}]{shirey98}
{Shirey}, R.~E., {Bradt}, H.~V., {Levine}, A.~M., \& {Morgan}, E.~H. 1998,
  \apj, 506, 374, \dodoi{10.1086/306247}

\bibitem[{{Smale} {et~al.}(2001){Smale}, {Church}, \&
  {Ba{\l}uci{\'n}ska-Church}}]{smale01}
{Smale}, A.~P., {Church}, M.~J., \& {Ba{\l}uci{\'n}ska-Church}, M. 2001, \apj,
  550, 962, \dodoi{10.1086/319800}

\bibitem[{{Smale} {et~al.}(2002){Smale}, {Church}, \&
  {Ba{\l}uci{\'n}ska-Church}}]{smale02}
---. 2002, \apj, 581, 1286, \dodoi{10.1086/344339}

\bibitem[{{Spencer} {et~al.}(2013){Spencer}, {Rushton},
  {Ba{\l}uci{\'n}ska-Church}, {Paragi}, {Schulz}, {Wilms}, {Pooley}, \&
  {Church}}]{spencer2013}
{Spencer}, R.~E., {Rushton}, A.~P., {Ba{\l}uci{\'n}ska-Church}, M., {et~al.}
  2013, \mnras, 435, L48, \dodoi{10.1093/mnrasl/slt090}

\bibitem[{{Stella} \& {Vietri}(1998)}]{stella98}
{Stella}, L., \& {Vietri}, M. 1998, \apjl, 492, L59, \dodoi{10.1086/311075}

\bibitem[{{Tan} \& {Biswas}(2012)}]{tan12}
{Tan}, M.~Y.~J., \& {Biswas}, R. 2012, \mnras, 419, 3292,
  \dodoi{10.1111/j.1365-2966.2011.19969.x}

\bibitem[{{Tarana} {et~al.}(2006){Tarana}, {Bazzano}, {Ubertini}, {Cocchi},
  {G{\"o}tz}, {Capitanio}, {Bird}, \& {Fiocchi}}]{tarana06}
{Tarana}, A., {Bazzano}, A., {Ubertini}, P., {et~al.} 2006, \aap, 448, 335,
  \dodoi{10.1051/0004-6361:20053917}

\bibitem[{{Tashiro} {et~al.}(2018){Tashiro}, {Maejima}, {Toda}, {Kelley},
  {Reichenthal}, {Lobell}, {Petre}, {Guainazzi}, {Costantini}, {Edison},
  {Fujimoto}, {Grim}, {Hayashida}, {den Herder}, {Ishisaki}, {Paltani},
  {Matsushita}, {Mori}, {Sneiderman}, {Takei}, {Terada}, {Tomida}, {Akamatsu},
  {Angelini}, {Arai}, {Awaki}, {Babyk}, {Bamba}, {Barfknecht}, {Barnstable},
  {Bialas}, {Blagojevic}, {Bonafede}, {Brambora}, {Brenneman}, {Brown},
  {Brown}, {Burns}, {Canavan}, {Carnahan}, {Chiao}, {Comber}, {Corrales}, {de
  Vries}, {Dercksen}, {Diaz-Trigo}, {Dillard}, {DiPirro}, {Done}, {Dotani},
  {Ebisawa}, {Eckart}, {Enoto}, {Ezoe}, {Ferrigno}, {Fukazawa}, {Fujita},
  {Furuzawa}, {Gallo}, {Graham}, {Gu}, {Hagino}, {Hamaguchi}, {Hatsukade},
  {Hawes}, {Hayashi}, {Hegarty}, {Hell}, {Hiraga}, {Hodges-Kluck}, {Holland},
  {Hornschemeier}, {Hoshino}, {Ichinohe}, {Iizuka}, {Ishibashi}, {Ishida},
  {Ishikawa}, {Ishimura}, {James}, {Kallman}, {Kara}, {Katsuda}, {Kenyon},
  {Kilbourne}, {Kimball}, {Kitaguti}, {Kitamoto}, {Kobayashi}, {Kohmura},
  {Koyama}, {Kubota}, {Leutenegger}, {Lockard}, {Loewenstein}, {Maeda},
  {Marbley}, {Markevitch}, {Matsumoto}, {Matsuzaki}, {McCammon}, {McNamara},
  {Miko}, {Miller}, {Miller}, {Minesugi}, {Mitsuishi}, {Mizuno}, {Mori},
  {Mukai}, {Murakami}, {Mushotzky}, {Nakajima}, {Nakamura}, {Nakashima},
  {Nakazawa}, {Natsukari}, {Nigo}, {Nishioka}, {Nobukawa}, {Nobukawa}, {Noda},
  {Odaka}, {Ogawa}, {Ohashi}, {Ohno}, {Ohta}, {Okajima}, {Okamoto}, {Onizuka},
  {Ota}, {Ozaki}, {Plucinsky}, {Porter}, {Pottschmidt}, {Sato}, {Sato},
  {Sawada}, {Seta}, {Shelton}, {Shibano}, {Shida}, {Shidatsu}, {Shirron},
  {Simionescu}, {Smith}, {Someya}, {Soong}, {Suagawara}, {Szymkowiak},
  {Takahashi}, {Tamagawa}, {Tamura}, {Tanaka}, {Terashima}, {Tsuboi},
  {Tsujimoto}, {Tsunemi}, {Tsuru}, {Uchida}, {Uchiyama}, {Ueda}, {Uno},
  {Walsh}, {Watanabe}, {Williams}, {Wolfs}, {Wright}, {Yamada}, {Yamaguchi},
  {Yamaoka}, {Yamasaki}, {Yamauchi}, {Yamauchi}, {Yanagase}, {Yaqoob},
  {Yasuda}, {Yoshioka}, {Zabala}, \& {Irina}}]{tashiro18}
{Tashiro}, M., {Maejima}, H., {Toda}, K., {et~al.} 2018, in Society of
  Photo-Optical Instrumentation Engineers (SPIE) Conference Series, Vol. 10699,
  Space Telescopes and Instrumentation 2018: Ultraviolet to Gamma Ray, ed.
  J.-W.~A. {den Herder}, S.~{Nikzad}, \& K.~{Nakazawa}, 1069922,
  \dodoi{10.1117/12.2309455}

\bibitem[{{Tobrej} {et~al.}(2023){Tobrej}, {Rai}, {Ghising}, {Tamang}, \&
  {Paul}}]{tobrej23}
{Tobrej}, M., {Rai}, B., {Ghising}, M., {Tamang}, R., \& {Paul}, B.~C. 2023,
  arXiv e-prints, arXiv:2309.11817.
\newblock \doarXiv{2309.11817}

\bibitem[{{Trueba} {et~al.}(2020){Trueba}, {Miller}, {Fabian}, {Kaastra},
  {Kallman}, {Lohfink}, {Proga}, {Raymond}, {Reynolds}, {Reynolds}, \&
  {Zoghbi}}]{trueba20}
{Trueba}, N., {Miller}, J.~M., {Fabian}, A.~C., {et~al.} 2020, \apjl, 899, L16,
  \dodoi{10.3847/2041-8213/aba9de}

\bibitem[{{Trueba} {et~al.}(2022){Trueba}, {Miller}, {Fabian}, {Kaastra},
  {Kallman}, {Lohfink}, {Ludlam}, {Proga}, {Raymond}, {Reynolds}, {Reynolds},
  \& {Zoghbi}}]{trueba22}
---. 2022, \apj, 925, 113, \dodoi{10.3847/1538-4357/ac3766}

\bibitem[{{van den Eijnden} {et~al.}(2022){van den Eijnden}, {Fender},
  {Miller-Jones}, {Russell}, {Saikia}, {Sivakoff}, \& {Carotenuto}}]{vde2022}
{van den Eijnden}, J., {Fender}, R., {Miller-Jones}, J.~C.~A., {et~al.} 2022,
  \mnras, 516, 2641, \dodoi{10.1093/mnras/stac2392}

\bibitem[{{van den Eijnden} {et~al.}(2018){van den Eijnden}, {Degenaar},
  {Pinto}, {Patruno}, {Wette}, {Messenger}, {Hern{\'a}ndez Santisteban},
  {Wijnands}, {Miller}, {Altamirano}, {Paerels}, {Chakrabarty}, \&
  {Fabian}}]{vandeneijnden17}
{van den Eijnden}, J., {Degenaar}, N., {Pinto}, C., {et~al.} 2018, \mnras, 475,
  2027, \dodoi{10.1093/mnras/stx3224}

\bibitem[{{van den Eijnden} {et~al.}(2021){van den Eijnden}, {Degenaar},
  {Russell}, {Wijnands}, {Bahramian}, {Miller-Jones}, {Hern{\'a}ndez
  Santisteban}, {Gallo}, {Atri}, {Plotkin}, {Maccarone}, {Sivakoff}, {Miller},
  {Reynolds}, {Russell}, {Maitra}, {Heinke}, {Armas Padilla}, \&
  {Shaw}}]{vandeneijnden2021}
{van den Eijnden}, J., {Degenaar}, N., {Russell}, T.~D., {et~al.} 2021, \mnras,
  507, 3899, \dodoi{10.1093/mnras/stab1995}

\bibitem[{{van der Klis}(1989)}]{vdk89}
{van der Klis}, M. 1989, \araa, 27, 517,
  \dodoi{10.1146/annurev.aa.27.090189.002505}

\bibitem[{{van der Klis}(1995)}]{vdk95}
{van der Klis}, M. 1995, in X-ray Binaries, 252--307

\bibitem[{{van der Klis}(2000)}]{vdk00}
---. 2000, \araa, 38, 717, \dodoi{10.1146/annurev.astro.38.1.717}

\bibitem[{{van der Klis}(2004)}]{vdk04}
---. 2004, arXiv e-prints, astro, \dodoi{10.48550/arXiv.astro-ph/0410551}

\bibitem[{{van der Klis}(2006)}]{vdk06}
---. 2006, in Compact stellar X-ray sources, Vol.~39, 39--112

\bibitem[{{van der Klis} {et~al.}(1996){van der Klis}, {Swank}, {Zhang},
  {Jahoda}, {Morgan}, {Lewin}, {Vaughan}, \& {van Paradijs}}]{vdk96}
{van der Klis}, M., {Swank}, J.~H., {Zhang}, W., {et~al.} 1996, \apjl, 469, L1,
  \dodoi{10.1086/310251}

\bibitem[{{van Paradijs} {et~al.}(1988){van Paradijs}, {Hasinger}, {Lewin},
  {van der Klis}, {Sztajno}, {Schulz}, \& {Jansen}}]{vanparadijs88}
{van Paradijs}, J., {Hasinger}, G., {Lewin}, W.~H.~G., {et~al.} 1988, \mnras,
  231, 379, \dodoi{10.1093/mnras/231.2.379}

\bibitem[{{van Straaten} {et~al.}(2005){van Straaten}, {van der Klis}, \&
  {Wijnands}}]{vanstraaten05}
{van Straaten}, S., {van der Klis}, M., \& {Wijnands}, R. 2005, \apj, 619, 455,
  \dodoi{10.1086/426183}

\bibitem[{{Vaughan} {et~al.}(1994){Vaughan}, {van der Klis}, {Wood}, {Norris},
  {Hertz}, {Michelson}, {van Paradijs}, {Lewin}, {Mitsuda}, \&
  {Penninx}}]{vaughan94}
{Vaughan}, B.~A., {van der Klis}, M., {Wood}, K.~S., {et~al.} 1994, \apj, 435,
  362, \dodoi{10.1086/174818}

\bibitem[{Virtanen {et~al.}(2020)Virtanen, Gommers, Oliphant, Haberland, Reddy,
  Cournapeau, Burovski, Peterson, Weckesser, Bright, {van der Walt}, Brett,
  Wilson, Millman, Mayorov, Nelson, Jones, Kern, Larson, Carey, Polat, Feng,
  Moore, {VanderPlas}, Laxalde, Perktold, Cimrman, Henriksen, Quintero, Harris,
  Archibald, Ribeiro, Pedregosa, {van Mulbregt}, \& {SciPy 1.0
  Contributors}}]{virtanen20}
Virtanen, P., Gommers, R., Oliphant, T.~E., {et~al.} 2020, Nature Methods, 17,
  261, \dodoi{10.1038/s41592-019-0686-2}

\bibitem[{{Watts}(2012)}]{watts12}
{Watts}, A.~L. 2012, \araa, 50, 609,
  \dodoi{10.1146/annurev-astro-040312-132617}

\bibitem[{{Wijnands} {et~al.}(2017){Wijnands}, {Parikh}, {Altamirano}, {Homan},
  \& {Degenaar}}]{wijnands17}
{Wijnands}, R., {Parikh}, A.~S., {Altamirano}, D., {Homan}, J., \& {Degenaar},
  N. 2017, \mnras, 472, 559, \dodoi{10.1093/mnras/stx2006}

\bibitem[{{Wijnands} \& {van der Klis}(1999)}]{wijnands99b}
{Wijnands}, R., \& {van der Klis}, M. 1999, \apj, 522, 965,
  \dodoi{10.1086/307698}

\bibitem[{{Wijnands} \& {van der Klis}(2001)}]{wijnands01}
---. 2001, \mnras, 321, 537, \dodoi{10.1046/j.1365-8711.2001.04058.x}

\bibitem[{{Wijnands} {et~al.}(1999){Wijnands}, {van der Klis}, \&
  {Rijkhorst}}]{wijnands99a}
{Wijnands}, R., {van der Klis}, M., \& {Rijkhorst}, E.-J. 1999, \apjl, 512,
  L39, \dodoi{10.1086/311872}

\bibitem[{{Wijnands} {et~al.}(1996){Wijnands}, {van der Klis}, {Psaltis},
  {Lamb}, {Kuulkers}, {Dieters}, {van Paradijs}, \& {Lewin}}]{wijnands96}
{Wijnands}, R.~A.~D., {van der Klis}, M., {Psaltis}, D., {et~al.} 1996, \apjl,
  469, L5, \dodoi{10.1086/310257}

\bibitem[{{Wilms} {et~al.}(2000){Wilms}, {Allen}, \& {McCray}}]{wilms00}
{Wilms}, J., {Allen}, A., \& {McCray}, R. 2000, \apj, 542, 914,
  \dodoi{10.1086/317016}

\end{thebibliography}
